%% file: main.tex
\documentclass[sigconf,nonacm]{acmart} 

\renewcommand\footnotetextcopyrightpermission[1]{} 

\acmConference[ESEC/FSE 2022]{The 30th ACM Joint European Software Engineering Conference and Symposium on the Foundations of Software Engineering}{14 - 18 November, 2022}{Singapore}

\usepackage{enumitem}
\usepackage[tikz]{bclogo}
\usepackage{tikz}
\usetikzlibrary{calc}
\usepackage{pgfplots}
\pgfplotsset{compat=1.17}
\usetikzlibrary{positioning}
\usepackage{amsmath}
\usepackage{booktabs}
\usepackage{listings}
\usepackage{xcolor}
\usepackage{tikzscale}
\usepackage{hyperref}
\usepackage{graphicx}
\usepackage{float}
\usepackage{subcaption}
\usepackage{multirow}
\usepackage{soul}
\usepackage{xspace}
\usepackage{fancyvrb}
\usepackage[many]{tcolorbox}
\usepackage{listings}
\usepackage{highlighter}
\usepackage{seqsplit}
\usepackage{fontawesome}
\usepackage{soul}
\usepackage{lipsum}

\definecolor{edgeblue}{RGB}{0, 0, 200}
\definecolor{blue}{RGB}{0, 0, 255}

\theoremstyle{definition}

\renewenvironment{quote}{%
  \list{}{%
    \leftmargin0.2cm   
    \rightmargin\leftmargin
  }
  \item\relax
}
{\endlist}

\definecolor{lst-gray}{rgb}{0.98,0.98,0.98}
\definecolor{lst-blue}{RGB}{40,0.0,255}
\definecolor{lst-green}{RGB}{65,128,95}
\definecolor{lst-red}{RGB}{200,0,85}

\definecolor{lightmauve}{rgb}{0.86, 0.82, 1.0}
\definecolor{lightcornflowerblue}{rgb}{0.6, 0.81, 0.93}
\definecolor{lightgreen}{rgb}{0.56, 0.93, 0.56}

\colorlet{codehighlight}{lightmauve!40!white}

\newcommand{\hlc}[2][yellow]{{%
    \colorlet{foo}{#1!40!white}%
    \sethlcolor{foo}\hl{#2}}%
}

\usepackage{listings}
\input{latex-listings-powershell.tex}
\title{AutoTSG: Learning and Synthesis for Incident Troubleshooting}

\author{Manish Shetty, Chetan Bansal, Sai Pramod Upadhyayula,\\ Arjun Radhakrishna, Anurag Gupta}
\email{{t-mamola, chetanb, saupa, arradha, anugup}@microsoft.com}
\affiliation{%
   \institution{Microsoft}
  \streetaddress{}
   \country{}
}

\begin{document}

\newcommand{\tool}{\texttt{AutoTSG}}
\newcommand{\CompanyX}{Microsoft}
\newcommand{\Paragraph}[1]{\smallskip\noindent{\bf #1}}
\newcommand{\Subparagraph}[1]{\smallskip\noindent{\emph{#1}}}

\newcommand{\todo}[1]{\textcolor{red}{TODO: #1}}
\newcommand{\del}[1]{\textcolor{red}{\st{#1}}}
\newcommand{\add}[1]{\textbf{\textcolor{blue}{#1}}}
\newcommand{\verify}[1]{\textbf{\textcolor{magenta}{#1}}}
\newcommand{\addRef}{\textbf{\textcolor{red}{[REF]}}}
\newcommand{\reduceVSpace}{\vspace{-3mm}}
\newcommand{\str}[1]{\ensuremath{\mathtt{``#1"}}}

\newcommand{\manish}[1]{\textcolor{blue}{\\ \fbox{MS:} {\textit{#1}}}}
\newcommand{\arjun}[1]{\textcolor{red}{\\ \fbox{AR:} {\textit{#1}}}}

\input{abstract}

\maketitle

\input{introduction}
\input{empirical-study}
\input{tool}
\input{evaluation}
\input{related-work}
\input{conclusion}

\bibliographystyle{ACM-Reference-Format}
\bibliography{references}

\end{document}

%% file: latex-listings-powershell.tex
\lstdefinelanguage{PowerShell}{
    morekeywords={
        StormEvents, Get-TransportRule, Update-GridTenantProvisioningStamp, variable, command, parameters
    },
	morekeywords={
		Add-Content,Add-PSSnapin,Clear-Content,Clear-History,Clear-Host,Clear-Item,Clear-ItemProperty,Clear-Variable,Compare-Object,Connect-PSSession,ConvertFrom-String,Convert-Path,Copy-Item,Copy-ItemProperty,Disable-PSBreakpoint,Disconnect-PSSession,Enable-PSBreakpoint,Enter-PSSession,Exit-PSSession,Export-Alias,Export-Csv,Export-PSSession,ForEach-Object,Format-Custom,Format-Hex,,Format-Table,Format-Wide,Get-Alias,Get-ChildItem,Get-Clipboard,Get-Command,Get-ComputerInfo,Get-Content,Get-History,Get-Item,Get-ItemProperty,Get-ItemPropertyValue,Get-Job,Get-Location,Get-Member,Get-Module,Get-Process,Get-PSBreakpoint,Get-PSCallStack,Get-PSDrive,Get-PSSession,Get-PSSnapin,Get-Service,Get-TimeZone,Get-Unique,Get-Variable,Get-WmiObject,Group-Object,help,Import-Alias,Import-Csv,Import-Module,Import-PSSession,Invoke-Command,Invoke-Expression,Invoke-History,Invoke-Item,Invoke-RestMethod,Invoke-WebRequest,Invoke-WmiMethod,Measure-Object,mkdir,Move-Item,Move-ItemProperty,New-Alias,New-Item,New-Module,New-PSDrive,New-PSSession,New-PSSessionConfigurationFile,New-Variable,Out-GridView,Out-Host,Out-Printer,Pop-Location,powershell_ise.exe,Push-Location,Receive-Job,Receive-PSSession,Remove-Item,Remove-ItemProperty,Remove-Job,Remove-Module,Remove-PSBreakpoint,Remove-PSDrive,Remove-PSSession,Remove-PSSnapin,Remove-Variable,Remove-WmiObject,Rename-Item,Rename-ItemProperty,Resolve-Path,Resume-Job,Select-Object,Select-String,Set-Alias,Set-Clipboard,Set-Content,Set-Item,Set-ItemProperty,Set-Location,Set-PSBreakpoint,Set-TimeZone,Set-Variable,Set-WmiInstance,Show-Command,Sort-Object,Start-Job,Start-Process,Start-Service,Start-Sleep,Stop-Job,Stop-Process,Stop-Service,Suspend-Job,Tee-Object,Trace-Command,Wait-Job,Where-Object,Write-Output
	},
	morekeywords={
		Add-AppxPackage,Add-AppxProvisionedPackage,Add-AppxVolume,Add-BitsFile,Add-CertificateEnrollmentPolicyServer,Add-Computer,Add-Content,Add-History,Add-JobTrigger,Add-KdsRootKey,Add-LocalGroupMember,Add-Member,Add-PSSnapin,Add-Type,Add-WindowsCapability,Add-WindowsDriver,Add-WindowsImage,Add-WindowsPackage,Checkpoint-Computer,Clear-Content,Clear-EventLog,Clear-History,Clear-Item,Clear-ItemProperty,Clear-KdsCache,Clear-RecycleBin,Clear-Tpm,Clear-Variable,Clear-WindowsCorruptMountPoint,Compare-Object,Complete-BitsTransfer,Complete-DtiagnosticTransaction,Complete-Transaction,Confirm-SecureBootUEFI,Connect-PSSession,Connect-WSMan,ConvertFrom-Csv,ConvertFrom-Json,ConvertFrom-SecureString,ConvertFrom-String,ConvertFrom-StringData,Convert-Path,Convert-String,ConvertTo-Csv,ConvertTo-Html,ConvertTo-Json,ConvertTo-ProcessMitigationPolicy,ConvertTo-SecureString,ConvertTo-TpmOwnerAuth,ConvertTo-Xml,Copy-Item,Copy-ItemProperty,Debug-Job,Debug-Process,Debug-Runspace,Disable-AppBackgroundTaskDiagnosticLog,Disable-ComputerRestore,Disable-JobTrigger,Disable-LocalUser,Disable-PSBreakpoint,Disable-PSRemoting,Disable-PSSessionConfiguration,Disable-RunspaceDebug,Disable-ScheduledJob,Disable-TlsCipherSuite,Disable-TlsEccCurve,Disable-TlsSessionTicketKey,Disable-TpmAutoProvisioning,Disable-WindowsErrorReporting,Disable-WindowsOptionalFeature,Disable-WSManCredSSP,Disconnect-PSSession,Disconnect-WSMan,Dismount-AppxVolume,Dismount-WindowsImage,Enable-AppBackgroundTaskDiagnosticLog,Enable-ComputerRestore,Enable-JobTrigger,Enable-LocalUser,Enable-PSBreakpoint,Enable-PSRemoting,Enable-PSSessionConfiguration,Enable-RunspaceDebug,Enable-ScheduledJob,Enable-TlsCipherSuite,Enable-TlsEccCurve,Enable-TlsSessionTicketKey,Enable-TpmAutoProvisioning,Enable-WindowsErrorReporting,Enable-WindowsOptionalFeature,Enable-WSManCredSSP,Enter-PSHostProcess,Enter-PSSession,Exit-PSHostProcess,Exit-PSSession,Expand-WindowsCustomDataImage,Expand-WindowsImage,Export-Alias,Export-BinaryMiLog,Export-Certificate,Export-Clixml,Export-Console,Export-Counter,Export-Csv,Export-FormatData,Export-ModuleMember,Export-PfxCertificate,Export-ProvisioningPackage,Export-PSSession,Export-StartLayout,Export-StartLayoutEdgeAssets,Export-TlsSessionTicketKey,Export-Trace,Export-WindowsCapabilitySource,Export-WindowsDriver,Export-WindowsImage,Find-Package,Find-PackageProvider,ForEach-Object,Format-Custom,Format-SecureBootUEFI,Format-Table,Format-Wide,Get-Acl,Get-Alias,Get-AppxDefaultVolume,Get-AppxPackage,Get-AppxPackageManifest,Get-AppxProvisionedPackage,Get-AppxVolume,Get-AuthenticodeSignature,Get-BitsTransfer,Get-Certificate,Get-CertificateAutoEnrollmentPolicy,Get-CertificateEnrollmentPolicyServer,Get-CertificateNotificationTask,Get-ChildItem,Get-CimAssociatedInstance,Get-CimClass,Get-CimInstance,Get-CimSession,Get-Clipboard,Get-CmsMessage,Get-Command,Get-ComputerInfo,Get-ComputerRestorePoint,Get-Content,Get-ControlPanelItem,Get-Counter,Get-Credential,Get-Culture,Get-DAPolicyChange,Get-Date,Get-DeliveryOptimizationLog,Get-DeliveryOptimizationPerfSnap,Get-DeliveryOptimizationPerfSnapThisMonth,Get-DeliveryOptimizationStatus,Get-DODownloadMode,Get-DOPercentageMaxBackgroundBandwidth,Get-DOPercentageMaxForegroundBandwidth,Get-Event,Get-EventLog,Get-EventSubscriber,Get-ExecutionPolicy,Get-FormatData,Get-Help,Get-History,Get-Host,Get-HotFix,Get-Item,Get-ItemProperty,Get-ItemPropertyValue,Get-Job,Get-JobTrigger,Get-KdsConfiguration,Get-KdsRootKey,Get-LocalGroup,Get-LocalGroupMember,Get-LocalUser,Get-Location,Get-Member,Get-Module,Get-Package,Get-PackageProvider,Get-PackageSource,Get-PfxCertificate,Get-PfxData,Get-PmemDisk,Get-PmemPhysicalDevice,Get-PmemUnusedRegion,Get-Process,Get-ProcessMitigation,Get-ProvisioningPackage,Get-PSBreakpoint,Get-PSCallStack,Get-PSDrive,Get-PSHostProcessInfo,Get-PSProvider,Get-PSReadlineKeyHandler,Get-PSReadlineOption,Get-PSSession,Get-PSSessionCapability,Get-PSSessionConfiguration,Get-PSSnapin,Get-Random,Get-Runspace,Get-RunspaceDebug,Get-ScheduledJob,Get-ScheduledJobOption,Get-SecureBootPolicy,Get-SecureBootUEFI,Get-Service,Get-TimeZone,Get-TlsCipherSuite,Get-TlsEccCurve,Get-Tpm,Get-TpmEndorsementKeyInfo,Get-TpmSupportedFeature,Get-TraceSource,Get-Transaction,Get-TroubleshootingPack,Get-TrustedProvisioningCertificate,Get-TypeData,Get-UICulture,Get-Unique,Get-Variable,Get-WIMBootEntry,Get-WinAcceptLanguageFromLanguageListOptOut,Get-WinCultureFromLanguageListOptOut,Get-WinDefaultInputMethodOverride,Get-WindowsCapability,Get-WindowsDeveloperLicense,Get-WindowsDriver,Get-WindowsEdition,Get-WindowsErrorReporting,Get-WindowsImage,Get-WindowsImageContent,Get-WindowsOptionalFeature,Get-WindowsPackage,Get-WindowsSearchSetting,Get-WinEvent,Get-WinHomeLocation,Get-WinLanguageBarOption,Get-WinSystemLocale,Get-WinUILanguageOverride,Get-WinUserLanguageList,Get-WmiObject,Get-WSManCredSSP,Get-WSManInstance,Group-Object,Import-Alias,Import-BinaryMiLog,Import-Certificate,Import-Clixml,Import-Counter,Import-Csv,Import-LocalizedData,Import-Module,Import-PackageProvider,Import-PfxCertificate,Import-PSSession,Import-StartLayout,Import-TpmOwnerAuth,Initialize-PmemPhysicalDevice,Initialize-Tpm,Install-Package,Install-PackageProvider,Install-ProvisioningPackage,Install-TrustedProvisioningCertificate,Invoke-CimMethod,Invoke-Command,Invoke-CommandInDesktopPackage,Invoke-DscResource,Invoke-Expression,Invoke-History,Invoke-Item,Invoke-RestMethod,Invoke-TroubleshootingPack,Invoke-WebRequest,Invoke-WmiMethod,Invoke-WSManAction,Join-DtiagnosticResourceManager,Join-Path,Limit-EventLog,Measure-Command,Measure-Object,Mount-AppxVolume,Mount-WindowsImage,Move-AppxPackage,Move-Item,Move-ItemProperty,New-Alias,New-CertificateNotificationTask,New-CimInstance,New-CimSession,New-CimSessionOption,New-DtiagnosticTransaction,New-Event,New-EventLog,New-FileCatalog,New-Item,New-ItemProperty,New-JobTrigger,New-LocalGroup,New-LocalUser,New-Module,New-ModuleManifest,New-NetIPsecAuthProposal,New-NetIPsecMainModeCryptoProposal,New-NetIPsecQuickModeCryptoProposal,New-Object,New-PmemDisk,New-ProvisioningRepro,New-PSDrive,New-PSRoleCapabilityFile,New-PSSession,New-PSSessionConfigurationFile,New-PSSessionOption,New-PSTransportOption,New-PSWorkflowExecutionOption,New-ScheduledJobOption,New-SelfSignedCertificate,New-Service,New-TimeSpan,New-TlsSessionTicketKey,New-Variable,New-WebServiceProxy,New-WindowsCustomImage,New-WindowsImage,New-WinEvent,New-WinUserLanguageList,New-WSManInstance,New-WSManSessionOption,Optimize-AppxProvisionedPackages,Optimize-WindowsImage,Out-Default,Out-File,Out-GridView,Out-Host,Out-Null,Out-Printer,Out-String,Pop-Location,Protect-CmsMessage,Publish-DscConfiguration,Push-Location,Read-Host,Receive-DtiagnosticTransaction,Receive-Job,Receive-PSSession,Register-ArgumentCompleter,Register-CimIndicationEvent,Register-EngineEvent,Register-ObjectEvent,Register-PackageSource,Register-PSSessionConfiguration,Register-ScheduledJob,Register-WmiEvent,Remove-AppxPackage,Remove-AppxProvisionedPackage,Remove-AppxVolume,Remove-BitsTransfer,Remove-CertificateEnrollmentPolicyServer,Remove-CertificateNotificationTask,Remove-CimInstance,Remove-CimSession,Remove-Computer,Remove-Event,Remove-EventLog,Remove-Item,Remove-ItemProperty,Remove-Job,Remove-JobTrigger,Remove-LocalGroup,Remove-LocalGroupMember,Remove-LocalUser,Remove-Module,Remove-PmemDisk,Remove-PSBreakpoint,Remove-PSDrive,Remove-PSReadlineKeyHandler,Remove-PSSession,Remove-PSSnapin,Remove-TypeData,Remove-Variable,Remove-WindowsCapability,Remove-WindowsDriver,Remove-WindowsImage,Remove-WindowsPackage,Remove-WmiObject,Remove-WSManInstance,Rename-Computer,Rename-Item,Rename-ItemProperty,Rename-LocalGroup,Rename-LocalUser,Repair-WindowsImage,Reset-ComputerMachinePassword,Resolve-DnsName,Resolve-Path,Restart-Computer,Restart-Service,Restore-Computer,Resume-BitsTransfer,Resume-Job,Resume-ProvisioningSession,Resume-Service,Save-Help,Save-Package,Save-WindowsImage,Select-Object,Select-String,Select-Xml,Send-DtiagnosticTransaction,Send-MailMessage,Set-Acl,Set-Alias,Set-AppBackgroundTaskResourcePolicy,Set-AppxDefaultVolume,Set-AppXProvisionedDataFile,Set-AuthenticodeSignature,Set-BitsTransfer,Set-CertificateAutoEnrollmentPolicy,Set-CimInstance,Set-Clipboard,Set-Content,Set-Culture,Set-Date,Set-DODownloadMode,Set-DOPercentageMaxBackgroundBandwidth,Set-DOPercentageMaxForegroundBandwidth,Set-DscLocalConfigurationManager,Set-ExecutionPolicy,Set-Item,Set-ItemProperty,Set-JobTrigger,Set-KdsConfiguration,Set-LocalGroup,Set-LocalUser,Set-Location,Set-PackageSource,Set-ProcessMitigation,Set-PSBreakpoint,Set-PSDebug,Set-PSReadlineKeyHandler,Set-PSReadlineOption,Set-PSSessionConfiguration,Set-ScheduledJob,Set-ScheduledJobOption,Set-SecureBootUEFI,Set-Service,Set-StrictMode,Set-TimeZone,Set-TpmOwnerAuth,Set-TraceSource,Set-Variable,Set-WinAcceptLanguageFromLanguageListOptOut,Set-WinCultureFromLanguageListOptOut,Set-WinDefaultInputMethodOverride,Set-WindowsEdition,Set-WindowsProductKey,Set-WindowsSearchSetting,Set-WinHomeLocation,Set-WinLanguageBarOption,Set-WinSystemLocale,Set-WinUILanguageOverride,Set-WinUserLanguageList,Set-WmiInstance,Set-WSManInstance,Set-WSManQuickConfig,Show-Command,Show-ControlPanelItem,Show-EventLog,Show-WindowsDeveloperLicenseRegistration,Sort-Object,Split-Path,Split-WindowsImage,Start-BitsTransfer,Start-DscConfiguration,Start-DtiagnosticResourceManager,Start-Job,Start-Process,Start-Service,Start-Sleep,Start-Transaction,Start-Transcript,Stop-Computer,Stop-DtiagnosticResourceManager,Stop-Job,Stop-Process,Stop-Service,Stop-Transcript,Suspend-BitsTransfer,Suspend-Job,Suspend-Service,Switch-Certificate,Tee-Object,Test-Certificate,Test-ComputerSecureChannel,Test-Connection,Test-DscConfiguration,Test-FileCatalog,Test-KdsRootKey,Test-ModuleManifest,Test-Path,Test-PSSessionConfigurationFile,Test-WSMan,Trace-Command,Unblock-File,Unblock-Tpm,Undo-DtiagnosticTransaction,Undo-Transaction,Uninstall-Package,Uninstall-ProvisioningPackage,Uninstall-TrustedProvisioningCertificate,Unprotect-CmsMessage,Unregister-Event,Unregister-PackageSource,Unregister-PSSessionConfiguration,Unregister-ScheduledJob,Unregister-WindowsDeveloperLicense,Update-FormatData,Update-Help,Update-List,Update-TypeData,Update-WIMBootEntry,Use-Transaction,Use-WindowsUnattend,Wait-Debugger,Wait-Event,Wait-Job,Wait-Process,Where-Object,Write-Debug,Write-Error,Write-EventLog,Write-Host,Write-Information,Write-Output,Write-Progress,Write-Verbose,Write-Warning
	},
	morekeywords={
		Add-BitLockerKeyProtector,Add-DnsClientNrptRule,Add-DtcClusterTMMapping,Add-EtwTraceProvider,Add-InitiatorIdToMaskingSet,Add-MpPreference,Add-NetEventNetworkAdapter,Add-NetEventPacketCaptureProvider,Add-NetEventProvider,Add-NetEventVFPProvider,Add-NetEventVmNetworkAdapter,Add-NetEventVmSwitch,Add-NetEventVmSwitchProvider,Add-NetEventWFPCaptureProvider,Add-NetIPHttpsCertBinding,Add-NetLbfoTeamMember,Add-NetLbfoTeamNic,Add-NetNatExternalAddress,Add-NetNatStaticMapping,Add-NetSwitchTeamMember,Add-Odbsn,Add-PartitionAccessPath,Add-PhysicalDisk,Add-Printer,Add-PrinterDriver,Add-PrinterPort,Add-StorageFaultDomain,Add-TargetPortToMaskingSet,Add-VirtualDiskToMaskingSet,Add-VpnConnection,Add-VpnConnectionRoute,Add-VpnConnectionTriggerApplication,Add-VpnConnectionTriggerDnsConfiguration,Add-VpnConnectionTriggerTrustedNetwork,AfterAll,AfterEach,Assert-MockCalled,Assert-VerifiableMocks,Backup-BitLockerKeyProtector,BackupToAAD-BitLockerKeyProtector,BeforeAll,BeforeEach,Block-FileShareAccess,Block-SmbShareAccess,Clear-BitLockerAutoUnlock,Clear-Disk,Clear-DnsClientCache,Clear-FileStorageTier,Clear-Host,Clear-PcsvDeviceLog,Clear-StorageDiagnosticInfo,Close-SmbOpenFile,Close-SmbSession,Compress-Archive,Configuration,Connect-IscsiTarget,Connect-VirtualDisk,Context,convert,ConvertFrom-SddlString,Copy-NetFirewallRule,Copy-NetIPsecMainModeCryptoSet,Copy-NetIPsecMainModeRule,Copy-NetIPsecPhase1AuthSet,Copy-NetIPsecPhase2AuthSet,Copy-NetIPsecQuickModeCryptoSet,Copy-NetIPsecRule,Debug-FileShare,Debug-MMAppPrelaunch,Debug-StorageSubSystem,Debug-Volume,Describe,Disable-BitLocker,Disable-BitLockerAutoUnlock,Disable-DAManualEntryPointSelection,Disable-Dsebug,Disable-MMAgent,Disable-NetAdapter,Disable-NetAdapterBinding,Disable-NetAdapterChecksumOffload,Disable-NetAdapterEncapsulatedPacketTaskOffload,Disable-NetAdapterIPsecOffload,Disable-NetAdapterLso,Disable-NetAdapterPacketDirect,Disable-NetAdapterPowerManagement,Disable-NetAdapterQos,Disable-NetAdapterRdma,Disable-NetAdapterRsc,Disable-NetAdapterRss,Disable-NetAdapterSriov,Disable-NetAdapterVmq,Disable-NetDnsTransitionConfiguration,Disable-NetFirewallRule,Disable-NetIPHttpsProfile,Disable-NetIPsecMainModeRule,Disable-NetIPsecRule,Disable-NetNatTransitionConfiguration,Disable-NetworkSwitchEthernetPort,Disable-NetworkSwitchFeature,Disable-NetworkSwitchVlan,Disable-OdbcPerfCounter,Disable-PhysicalDiskIdentification,Disable-PnpDevice,Disable-PSTrace,Disable-PSWSManCombinedTrace,Disable-ScheduledTask,Disable-SmbDelegation,Disable-StorageEnclosureIdentification,Disable-StorageEnclosurePower,Disable-StorageHighAvailability,Disable-StorageMaintenanceMode,Disable-WdacBidTrace,Disable-WSManTrace,Disconnect-IscsiTarget,Disconnect-VirtualDisk,Dismount-DiskImage,Enable-BitLocker,Enable-BitLockerAutoUnlock,Enable-DAManualEntryPointSelection,Enable-Dsebug,Enable-MMAgent,Enable-NetAdapter,Enable-NetAdapterBinding,Enable-NetAdapterChecksumOffload,Enable-NetAdapterEncapsulatedPacketTaskOffload,Enable-NetAdapterIPsecOffload,Enable-NetAdapterLso,Enable-NetAdapterPacketDirect,Enable-NetAdapterPowerManagement,Enable-NetAdapterQos,Enable-NetAdapterRdma,Enable-NetAdapterRsc,Enable-NetAdapterRss,Enable-NetAdapterSriov,Enable-NetAdapterVmq,Enable-NetDnsTransitionConfiguration,Enable-NetFirewallRule,Enable-NetIPHttpsProfile,Enable-NetIPsecMainModeRule,Enable-NetIPsecRule,Enable-NetNatTransitionConfiguration,Enable-NetworkSwitchEthernetPort,Enable-NetworkSwitchFeature,Enable-NetworkSwitchVlan,Enable-OdbcPerfCounter,Enable-PhysicalDiskIdentification,Enable-PnpDevice,Enable-PSTrace,Enable-PSWSManCombinedTrace,Enable-ScheduledTask,Enable-SmbDelegation,Enable-StorageEnclosureIdentification,Enable-StorageEnclosurePower,Enable-StorageHighAvailability,Enable-StorageMaintenanceMode,Enable-WdacBidTrace,Enable-WSManTrace,Expand-Archive,Export-ODataEndpointProxy,Export-ScheduledTask,Find-Command,Find-DscResource,Find-Module,Find-NetIPsecRule,Find-NetRoute,Find-RoleCapability,Find-Script,Flush-EtwTraceSession,Format-Hex,Format-Volume,Get-AppBackgroundTask,Get-AppxLastError,Get-AppxLog,Get-AutologgerConfig,Get-BitLockerVolume,Get-ClusteredScheduledTask,Get-DAClientExperienceConfiguration,Get-DAConnectionStatus,Get-DAEntryPointTableItem,Get-DedupProperties,Get-Disk,Get-DiskImage,Get-DiskStorageNodeView,Get-DnsClient,Get-DnsClientCache,Get-DnsClientGlobalSetting,Get-DnsClientNrptGlobal,Get-DnsClientNrptPolicy,Get-DnsClientNrptRule,Get-DnsClientServerAddress,Get-DscConfiguration,Get-DscConfigurationStatus,Get-DscLocalConfigurationManager,Get-DscResource,Get-Dtc,Get-DtcAdvancedHostSetting,Get-DtcAdvancedSetting,Get-DtcClusterDefault,Get-DtcClusterTMMapping,Get-Dtefault,Get-DtcLog,Get-DtcNetworkSetting,Get-DtcTransaction,Get-DtcTransactionsStatistics,Get-DtcTransactionsTraceSession,Get-DtcTransactionsTraceSetting,Get-EtwTraceProvider,Get-EtwTraceSession,Get-FileHash,Get-FileIntegrity,Get-FileShare,Get-FileShareAccessControlEntry,Get-FileStorageTier,Get-InitiatorId,Get-InitiatorPort,Get-InstalledModule,Get-InstalledScript,Get-IscsiConnection,Get-IscsiSession,Get-IscsiTarget,Get-IscsiTargetPortal,Get-IseSnippet,Get-LogProperties,Get-MaskingSet,Get-MMAgent,Get-MockDynamicParameters,Get-MpComputerStatus,Get-MpPreference,Get-MpThreat,Get-MpThreatCatalog,Get-MpThreatDetection,Get-NCSIPolicyConfiguration,Get-Net6to4Configuration,Get-NetAdapter,Get-NetAdapterAdvancedProperty,Get-NetAdapterBinding,Get-NetAdapterChecksumOffload,Get-NetAdapterEncapsulatedPacketTaskOffload,Get-NetAdapterHardwareInfo,Get-NetAdapterIPsecOffload,Get-NetAdapterLso,Get-NetAdapterPacketDirect,Get-NetAdapterPowerManagement,Get-NetAdapterQos,Get-NetAdapterRdma,Get-NetAdapterRsc,Get-NetAdapterRss,Get-NetAdapterSriov,Get-NetAdapterSriovVf,Get-NetAdapterStatistics,Get-NetAdapterVmq,Get-NetAdapterVMQQueue,Get-NetAdapterVPort,Get-NetCompartment,Get-NetConnectionProfile,Get-NetDnsTransitionConfiguration,Get-NetDnsTransitionMonitoring,Get-NetEventNetworkAdapter,Get-NetEventPacketCaptureProvider,Get-NetEventProvider,Get-NetEventSession,Get-NetEventVFPProvider,Get-NetEventVmNetworkAdapter,Get-NetEventVmSwitch,Get-NetEventVmSwitchProvider,Get-NetEventWFPCaptureProvider,Get-NetFirewallAddressFilter,Get-NetFirewallApplicationFilter,Get-NetFirewallInterfaceFilter,Get-NetFirewallInterfaceTypeFilter,Get-NetFirewallPortFilter,Get-NetFirewallProfile,Get-NetFirewallRule,Get-NetFirewallSecurityFilter,Get-NetFirewallServiceFilter,Get-NetFirewallSetting,Get-NetIPAddress,Get-NetIPConfiguration,Get-NetIPHttpsConfiguration,Get-NetIPHttpsState,Get-NetIPInterface,Get-NetIPseospSetting,Get-NetIPsecMainModeCryptoSet,Get-NetIPsecMainModeRule,Get-NetIPsecMainModeSA,Get-NetIPsecPhase1AuthSet,Get-NetIPsecPhase2AuthSet,Get-NetIPsecQuickModeCryptoSet,Get-NetIPsecQuickModeSA,Get-NetIPsecRule,Get-NetIPv4Protocol,Get-NetIPv6Protocol,Get-NetIsatapConfiguration,Get-NetLbfoTeam,Get-NetLbfoTeamMember,Get-NetLbfoTeamNic,Get-NetNat,Get-NetNatExternalAddress,Get-NetNatGlobal,Get-NetNatSession,Get-NetNatStaticMapping,Get-NetNatTransitionConfiguration,Get-NetNatTransitionMonitoring,Get-NetNeighbor,Get-NetOffloadGlobalSetting,Get-NetPrefixPolicy,Get-NetQosPolicy,Get-NetRoute,Get-NetSwitchTeam,Get-NetSwitchTeamMember,Get-NetTCPConnection,Get-NetTCPSetting,Get-NetTeredoConfiguration,Get-NetTeredoState,Get-NetTransportFilter,Get-NetUDPEndpoint,Get-NetUDPSetting,Get-NetworkSwitchEthernetPort,Get-NetworkSwitchFeature,Get-NetworkSwitchGlobalData,Get-NetworkSwitchVlan,Get-Odbriver,Get-Odbsn,Get-OdbcPerfCounter,Get-OffloadDataTransferSetting,Get-OperationValidation,Get-Partition,Get-PartitionSupportedSize,Get-PcsvDevice,Get-PcsvDeviceLog,Get-PhysicalDisk,Get-PhysicalDiskStorageNodeView,Get-PhysicalExtent,Get-PhysicalExtentAssociation,Get-PnpDevice,Get-PnpDeviceProperty,Get-PrintConfiguration,Get-Printer,Get-PrinterDriver,Get-PrinterPort,Get-PrinterProperty,Get-PrintJob,Get-PSRepository,Get-ResiliencySetting,Get-ScheduledTask,Get-ScheduledTaskInfo,Get-SmbBandWidthLimit,Get-SmbClientConfiguration,Get-SmbClientNetworkInterface,Get-SmbConnection,Get-SmbDelegation,Get-SmbGlobalMapping,Get-SmbMapping,Get-SmbMultichannelConnection,Get-SmbMultichannelConstraint,Get-SmbOpenFile,Get-SmbServerConfiguration,Get-SmbServerNetworkInterface,Get-SmbSession,Get-SmbShare,Get-SmbShareAccess,Get-SmbWitnessClient,Get-StartApps,Get-StorageAdvancedProperty,Get-StorageDiagnosticInfo,Get-StorageEnclosure,Get-StorageEnclosureStorageNodeView,Get-StorageEnclosureVendorData,Get-StorageExtendedStatus,Get-StorageFaultDomain,Get-StorageFileServer,Get-StorageFirmwareInformation,Get-StorageHealthAction,Get-StorageHealthReport,Get-StorageHealthSetting,Get-StorageJob,Get-StorageNode,Get-StoragePool,Get-StorageProvider,Get-StorageReliabilityCounter,Get-StorageSetting,Get-StorageSubSystem,Get-StorageTier,Get-StorageTierSupportedSize,Get-SupportedClusterSizes,Get-SupportedFileSystems,Get-TargetPort,Get-TargetPortal,Get-TestDriveItem,Get-Verb,Get-VirtualDisk,Get-VirtualDiskSupportedSize,Get-Volume,Get-VolumeCorruptionCount,Get-VolumeScrubPolicy,Get-VpnConnection,Get-VpnConnectionTrigger,Get-WdacBidTrace,Get-WindowsUpdateLog,Get-WUAVersion,Get-WUIsPendingReboot,Get-WULastInstallationDate,Get-WULastScanSuccessDate,Grant-FileShareAccess,Grant-SmbShareAccess,help,Hide-VirtualDisk,Import-IseSnippet,Import-PowerShellDataFile,ImportSystemModules,In,Initialize-Disk,InModuleScope,Install-Dtc,Install-Module,Install-Script,Install-WUUpdates,Invoke-AsWorkflow,Invoke-Mock,Invoke-OperationValidation,Invoke-Pester,It,Lock-BitLocker,mkdir,Mock,more,Mount-DiskImage,Move-SmbWitnessClient,New-AutologgerConfig,New-DAEntryPointTableItem,New-DscChecksum,New-EapConfiguration,New-EtwTraceSession,New-FileShare,New-Fixture,New-Guid,New-IscsiTargetPortal,New-IseSnippet,New-MaskingSet,New-NetAdapterAdvancedProperty,New-NetEventSession,New-NetFirewallRule,New-NetIPAddress,New-NetIPHttpsConfiguration,New-NetIPseospSetting,New-NetIPsecMainModeCryptoSet,New-NetIPsecMainModeRule,New-NetIPsecPhase1AuthSet,New-NetIPsecPhase2AuthSet,New-NetIPsecQuickModeCryptoSet,New-NetIPsecRule,New-NetLbfoTeam,New-NetNat,New-NetNatTransitionConfiguration,New-NetNeighbor,New-NetQosPolicy,New-NetRoute,New-NetSwitchTeam,New-NetTransportFilter,New-NetworkSwitchVlan,New-Partition,New-PesterOption,New-PSWorkflowSession,New-ScheduledTask,New-ScheduledTaskAction,New-ScheduledTaskPrincipal,New-ScheduledTaskSettingsSet,New-ScheduledTaskTrigger,New-ScriptFileInfo,New-SmbGlobalMapping,New-SmbMapping,New-SmbMultichannelConstraint,New-SmbShare,New-StorageFileServer,New-StoragePool,New-StorageSubsystemVirtualDisk,New-StorageTier,New-TemporaryFile,New-VirtualDisk,New-VirtualDiskClone,New-VirtualDiskSnapshot,New-Volume,New-VpnServerAddress,Open-NetGPO,Optimize-StoragePool,Optimize-Volume,oss,Pause,prompt,PSConsoleHostReadline,Publish-Module,Publish-Script,Read-PrinterNfcTag,Register-ClusteredScheduledTask,Register-DnsClient,Register-IscsiSession,Register-PSRepository,Register-ScheduledTask,Register-StorageSubsystem,Remove-AutologgerConfig,Remove-BitLockerKeyProtector,Remove-DAEntryPointTableItem,Remove-DnsClientNrptRule,Remove-DscConfigurationDocument,Remove-DtcClusterTMMapping,Remove-EtwTraceProvider,Remove-FileShare,Remove-InitiatorId,Remove-InitiatorIdFromMaskingSet,Remove-IscsiTargetPortal,Remove-MaskingSet,Remove-MpPreference,Remove-MpThreat,Remove-NetAdapterAdvancedProperty,Remove-NetEventNetworkAdapter,Remove-NetEventPacketCaptureProvider,Remove-NetEventProvider,Remove-NetEventSession,Remove-NetEventVFPProvider,Remove-NetEventVmNetworkAdapter,Remove-NetEventVmSwitch,Remove-NetEventVmSwitchProvider,Remove-NetEventWFPCaptureProvider,Remove-NetFirewallRule,Remove-NetIPAddress,Remove-NetIPHttpsCertBinding,Remove-NetIPHttpsConfiguration,Remove-NetIPseospSetting,Remove-NetIPsecMainModeCryptoSet,Remove-NetIPsecMainModeRule,Remove-NetIPsecMainModeSA,Remove-NetIPsecPhase1AuthSet,Remove-NetIPsecPhase2AuthSet,Remove-NetIPsecQuickModeCryptoSet,Remove-NetIPsecQuickModeSA,Remove-NetIPsecRule,Remove-NetLbfoTeam,Remove-NetLbfoTeamMember,Remove-NetLbfoTeamNic,Remove-NetNat,Remove-NetNatExternalAddress,Remove-NetNatStaticMapping,Remove-NetNatTransitionConfiguration,Remove-NetNeighbor,Remove-NetQosPolicy,Remove-NetRoute,Remove-NetSwitchTeam,Remove-NetSwitchTeamMember,Remove-NetTransportFilter,Remove-NetworkSwitchEthernetPortIPAddress,Remove-NetworkSwitchVlan,Remove-Odbsn,Remove-Partition,Remove-PartitionAccessPath,Remove-PhysicalDisk,Remove-Printer,Remove-PrinterDriver,Remove-PrinterPort,Remove-PrintJob,Remove-SmbBandwidthLimit,Remove-SmbGlobalMapping,Remove-SmbMapping,Remove-SmbMultichannelConstraint,Remove-SmbShare,Remove-StorageFaultDomain,Remove-StorageFileServer,Remove-StorageHealthIntent,Remove-StorageHealthSetting,Remove-StoragePool,Remove-StorageTier,Remove-TargetPortFromMaskingSet,Remove-VirtualDisk,Remove-VirtualDiskFromMaskingSet,Remove-VpnConnection,Remove-VpnConnectionRoute,Remove-VpnConnectionTriggerApplication,Remove-VpnConnectionTriggerDnsConfiguration,Remove-VpnConnectionTriggerTrustedNetwork,Rename-DAEntryPointTableItem,Rename-MaskingSet,Rename-NetAdapter,Rename-NetFirewallRule,Rename-NetIPHttpsConfiguration,Rename-NetIPsecMainModeCryptoSet,Rename-NetIPsecMainModeRule,Rename-NetIPsecPhase1AuthSet,Rename-NetIPsecPhase2AuthSet,Rename-NetIPsecQuickModeCryptoSet,Rename-NetIPsecRule,Rename-NetLbfoTeam,Rename-NetSwitchTeam,Rename-Printer,Repair-FileIntegrity,Repair-VirtualDisk,Repair-Volume,Reset-DAClientExperienceConfiguration,Reset-DAEntryPointTableItem,Reset-DtcLog,Reset-NCSIPolicyConfiguration,Reset-Net6to4Configuration,Reset-NetAdapterAdvancedProperty,Reset-NetDnsTransitionConfiguration,Reset-NetIPHttpsConfiguration,Reset-NetIsatapConfiguration,Reset-NetTeredoConfiguration,Reset-PhysicalDisk,Reset-StorageReliabilityCounter,Resize-Partition,Resize-StorageTier,Resize-VirtualDisk,Restart-NetAdapter,Restart-PcsvDevice,Restart-PrintJob,Restore-DscConfiguration,Restore-NetworkSwitchConfiguration,Resume-BitLocker,Resume-PrintJob,Revoke-FileShareAccess,Revoke-SmbShareAccess,SafeGetCommand,Save-EtwTraceSession,Save-Module,Save-NetGPO,Save-NetworkSwitchConfiguration,Save-Script,Send-EtwTraceSession,Set-AutologgerConfig,Set-ClusteredScheduledTask,Set-DAClientExperienceConfiguration,Set-DAEntryPointTableItem,Set-Disk,Set-DnsClient,Set-DnsClientGlobalSetting,Set-DnsClientNrptGlobal,Set-DnsClientNrptRule,Set-DnsClientServerAddress,Set-DtcAdvancedHostSetting,Set-DtcAdvancedSetting,Set-DtcClusterDefault,Set-DtcClusterTMMapping,Set-Dtefault,Set-DtcLog,Set-DtcNetworkSetting,Set-DtcTransaction,Set-DtcTransactionsTraceSession,Set-DtcTransactionsTraceSetting,Set-DynamicParameterVariables,Set-EtwTraceProvider,Set-FileIntegrity,Set-FileShare,Set-FileStorageTier,Set-InitiatorPort,Set-IscsiChapSecret,Set-LogProperties,Set-MMAgent,Set-MpPreference,Set-NCSIPolicyConfiguration,Set-Net6to4Configuration,Set-NetAdapter,Set-NetAdapterAdvancedProperty,Set-NetAdapterBinding,Set-NetAdapterChecksumOffload,Set-NetAdapterEncapsulatedPacketTaskOffload,Set-NetAdapterIPsecOffload,Set-NetAdapterLso,Set-NetAdapterPacketDirect,Set-NetAdapterPowerManagement,Set-NetAdapterQos,Set-NetAdapterRdma,Set-NetAdapterRsc,Set-NetAdapterRss,Set-NetAdapterSriov,Set-NetAdapterVmq,Set-NetConnectionProfile,Set-NetDnsTransitionConfiguration,Set-NetEventPacketCaptureProvider,Set-NetEventProvider,Set-NetEventSession,Set-NetEventVFPProvider,Set-NetEventVmSwitchProvider,Set-NetEventWFPCaptureProvider,Set-NetFirewallAddressFilter,Set-NetFirewallApplicationFilter,Set-NetFirewallInterfaceFilter,Set-NetFirewallInterfaceTypeFilter,Set-NetFirewallPortFilter,Set-NetFirewallProfile,Set-NetFirewallRule,Set-NetFirewallSecurityFilter,Set-NetFirewallServiceFilter,Set-NetFirewallSetting,Set-NetIPAddress,Set-NetIPHttpsConfiguration,Set-NetIPInterface,Set-NetIPseospSetting,Set-NetIPsecMainModeCryptoSet,Set-NetIPsecMainModeRule,Set-NetIPsecPhase1AuthSet,Set-NetIPsecPhase2AuthSet,Set-NetIPsecQuickModeCryptoSet,Set-NetIPsecRule,Set-NetIPv4Protocol,Set-NetIPv6Protocol,Set-NetIsatapConfiguration,Set-NetLbfoTeam,Set-NetLbfoTeamMember,Set-NetLbfoTeamNic,Set-NetNat,Set-NetNatGlobal,Set-NetNatTransitionConfiguration,Set-NetNeighbor,Set-NetOffloadGlobalSetting,Set-NetQosPolicy,Set-NetRoute,Set-NetTCPSetting,Set-NetTeredoConfiguration,Set-NetUDPSetting,Set-NetworkSwitchEthernetPortIPAddress,Set-NetworkSwitchPortMode,Set-NetworkSwitchPortProperty,Set-NetworkSwitchVlanProperty,Set-Odbriver,Set-Odbsn,Set-Partition,Set-PcsvDeviceBootConfiguration,Set-PcsvDeviceNetworkConfiguration,Set-PcsvDeviceUserPassword,Set-PhysicalDisk,Set-PrintConfiguration,Set-Printer,Set-PrinterProperty,Set-PSRepository,Set-ResiliencySetting,Set-ScheduledTask,Set-SmbBandwidthLimit,Set-SmbClientConfiguration,Set-SmbPathAcl,Set-SmbServerConfiguration,Set-SmbShare,Set-StorageFileServer,Set-StorageHealthSetting,Set-StoragePool,Set-StorageProvider,Set-StorageSetting,Set-StorageSubSystem,Set-StorageTier,Set-TestInconclusive,Setup,Set-VirtualDisk,Set-Volume,Set-VolumeScrubPolicy,Set-VpnConnection,Set-VpnConnectionIPsecConfiguration,Set-VpnConnectionProxy,Set-VpnConnectionTriggerDnsConfiguration,Set-VpnConnectionTriggerTrustedNetwork,Should,Show-NetFirewallRule,Show-NetIPsecRule,Show-VirtualDisk,Start-AppBackgroundTask,Start-AutologgerConfig,Start-Dtc,Start-DtcTransactionsTraceSession,Start-EtwTraceSession,Start-MpScan,Start-MpWDOScan,Start-NetEventSession,Start-PcsvDevice,Start-ScheduledTask,Start-StorageDiagnosticLog,Start-Trace,Start-WUScan,Stop-DscConfiguration,Stop-Dtc,Stop-DtcTransactionsTraceSession,Stop-EtwTraceSession,Stop-NetEventSession,Stop-PcsvDevice,Stop-ScheduledTask,Stop-StorageDiagnosticLog,Stop-StorageJob,Stop-Trace,Suspend-BitLocker,Suspend-PrintJob,Sync-NetIPsecRule,TabExpansion2,Test-Dtc,Test-NetConnection,Test-ScriptFileInfo,Unblock-FileShareAccess,Unblock-SmbShareAccess,Uninstall-Dtc,Uninstall-Module,Uninstall-Script,Unlock-BitLocker,Unregister-AppBackgroundTask,Unregister-ClusteredScheduledTask,Unregister-IscsiSession,Unregister-PSRepository,Unregister-ScheduledTask,Unregister-StorageSubsystem,Update-Disk,Update-DscConfiguration,Update-EtwTraceSession,Update-HostStorageCache,Update-IscsiTarget,Update-IscsiTargetPortal,Update-Module,Update-ModuleManifest,Update-MpSignature,Update-NetIPsecRule,Update-Script,Update-ScriptFileInfo,Update-SmbMultichannelConnection,Update-StorageFirmware,Update-StoragePool,Update-StorageProviderCache,Write-DtcTransactionsTraceSession,Write-PrinterNfcTag,Write-VolumeCache
	},
	morekeywords={Do,Else,For,ForEach,Function,If,In,Until,While},
	alsodigit={-},
	sensitive=false,
	morecomment=[l]{\#},
	morecomment=[n]{<\#}{\#>},
	morestring=[b]{"},
	morestring=[b]{'},
	morestring=[s]{@'}{'@},
	morestring=[s]{@"}{"@}
}

%% file: abstract.tex
\begin{abstract}
Incident management is a key aspect of operating large-scale cloud services. To aid with faster and efficient resolution of incidents, engineering teams document frequent troubleshooting steps in the form of Troubleshooting Guides (TSGs), to be used by on-call engineers (OCEs). However, TSGs are siloed, unstructured, and often incomplete, requiring developers to manually understand and execute necessary steps. This results in a plethora of issues such as on-call fatigue, reduced productivity, and human errors. In this work, we conduct a large-scale empirical study of over 4K+ TSGs mapped to \iftrue 1000s of \fi 
incidents and find that TSGs are widely used and help significantly reduce mitigation efforts. We then analyze feedback on TSGs provided by 400+ OCEs and propose a taxonomy of issues that highlights significant gaps in TSG quality. To alleviate these gaps, we investigate the automation of TSGs and propose \tool{} -- a novel framework for automation of TSGs to executable workflows by combining machine learning and program synthesis. Our evaluation of \tool{} on 50 TSGs shows the effectiveness in both identifying TSG statements (accuracy 0.89) and parsing them for execution (precision 0.94 and recall 0.91). Lastly, we survey ten \CompanyX{} engineers and show the importance of TSG automation and the usefulness of \tool{}. 

\end{abstract}


%% file: introduction.tex
\section{Introduction}

At \CompanyX{}, we operate services at a massive scale with 1000+ internal and external services built and operated by tens of thousands of engineers spread across the world and deployed in over 200 data centers worldwide. At such a large scale, we need effective incident management processes to minimize the impact of service incidents. Today, most software companies have on-call duty, which requires engineers building services to be responsible for handling (i.e., acknowledge, diagnose and mitigate) incidents 24x7 on a rotating basis. To standardize these incident management workflows, engineering teams document these steps as \textit{Troubleshooting Guides} (TSGs) which are then referred to and followed by the on-call engineers while handling production incidents. These TSGs help with knowledge sharing and avoid the challenges with tribal knowledge, especially when new engineers join the team. 

In \CompanyX{}, we have more than 50,000 TSGs which are used regularly for incident resolution by over 60,000 engineers every month. The TSGs are authored by the engineers and can contain various components such as commands and scripts for troubleshooting, big data queries for fetching diagnostics logs, natural language instructions and even screenshots. At the time of incident handling, the on-call engineer manually tries to follow the instructions described in the TSGs. The manual execution can consume a significant amount of effort since commands/queries must be copy-pasted and executed, and instructions must be parsed and understood. Further, similar to other kinds of software documentation \cite{aghajani2020software}, TSGs are also prone to various issues such as lack of readability, fragmentation, etc. These issues have significant detrimental impact on both engineering productivity and service health because it leads to increased effort by the on-call engineers and, also, higher customer impact due to increased incident resolution time. Further, with manual TSGs, there is a significant risk of outages\footnote{\url{https://www.wsj.com/articles/amazon-finds-the-cause-of-its-aws-outage-a-typo-1488490506}} due to human errors, on-call fatigue, and knowledge gaps. Hence, there is a need to automate the TSGs into executable workflows, which can help mitigate incidents with minimal human intervention.


\input{Figures/tsg-usage}

In this work, we conducted a large-scale empirical study to understand the usage and challenges of TSGs better. We find that incidents linked with TSGs have reduced mitigation time showing the effectiveness of TSGs. At the same time, TSGs are also prone to completeness, validation and maintenance issues. To mitigate these problems and reduce the manual effort and human error involved in executing TSGs, we investigate the problem of automating TSGs by converting them to executable workflows such as Jupyter notebooks. We propose \textbf{\tool{}}, a novel framework for TSG automation at scale. Executable TSGs help minimize the manual effort for the on-call engineer while also improving the maintainability with automated testing and validation. With \tool{}, our goal is to assist developers with automation of 50,000+ TSGs at \CompanyX{}. To overcome the unique challenges in this task, such as lack of labelled data and heterogeneity of information embedded in TSGs, we combine meta-learning and program synthesis for extraction of components (i.e., code, big data queries, natural language instructions, etc.) and parsing of components into constituents such as variables, commands in code and conditional, action statements in natural language. Our evaluation shows that \tool{} has high accuracy while also being useful based on the survey of on-call engineers. To summarize, we make the following main contributions
in this work:

\begin{enumerate}
    \item We do a large-scale empirical study on the usage and effectiveness of TSGs for incident resolution at \CompanyX{}.
    \item We analyze feedback provided by 400+ on-call engineers at \CompanyX{} to propose the first taxonomy of TSG quality issues. 
    \item We design and build \tool{}, a novel framework which combines machine learning and program synthesis to aid with the automation of TSGs at scale.
    \item We do a quantitative evaluation on 50 TSGs and survey 10 \CompanyX{} engineers to show the effectiveness of \tool{}.
\end{enumerate}
The rest of the paper is organized as follows: In Section \ref{sec:empirical-analysis}, we present insights from the empirical study about the usage of TSGs and motivate the need for automation. In Section 3, we provide an overview and the implementation details for \tool{}. In Section 4, we describe the experimental evaluation and user study for \tool{}. In Section 5, we discuss the related work followed by conclusion.

%% file: Figures/tsg-usage.tex
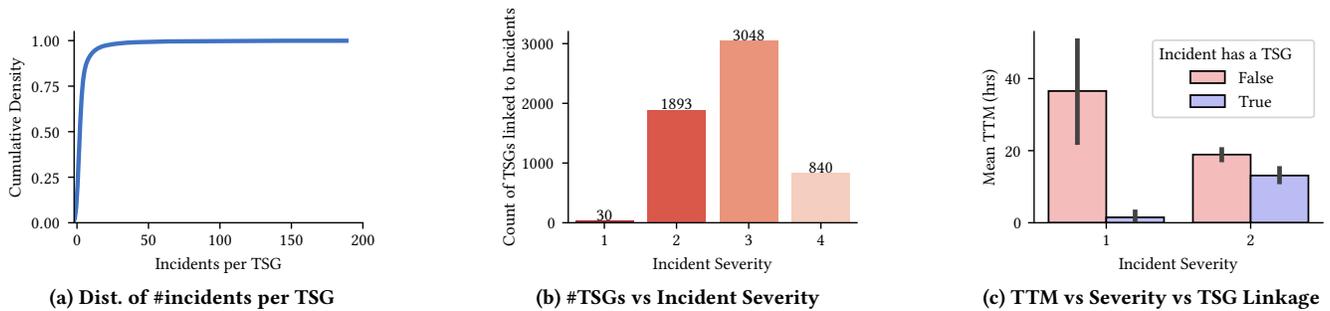
\begin{figure*}[t]
\centering
\subcaptionbox{Dist. of \#incidents per TSG\label{fig:emp-1}}
    {\scalebox{.65}{\input{Figures/incidents-per-tsg.pgf}}\vspace{-2mm}}%
\hfill 
\subcaptionbox{\#TSGs vs Incident Severity\label{fig:emp-2}}
    {\scalebox{.65}{\input{Figures/tsg-per-severity.pgf}}\vspace{-2mm}}%
\hfill 
\subcaptionbox{TTM vs Severity vs TSG Linkage\label{fig:emp-3}}
    {\scalebox{.65}{\input{Figures/ttm-severity.pgf}}\vspace{-2mm}}%
\caption{Analysis of TSG Usage}
\end{figure*}

%% file: Figures/incidents-per-tsg.pgf
\begingroup%
\makeatletter%
\begin{pgfpicture}%
\pgfpathrectangle{\pgfpointorigin}{\pgfqpoint{3.152316in}{2.139691in}}%
\pgfusepath{use as bounding box, clip}%
\begin{pgfscope}%
\pgfsetbuttcap%
\pgfsetmiterjoin%
\definecolor{currentfill}{rgb}{1.000000,1.000000,1.000000}%
\pgfsetfillcolor{currentfill}%
\pgfsetlinewidth{0.000000pt}%
\definecolor{currentstroke}{rgb}{1.000000,1.000000,1.000000}%
\pgfsetstrokecolor{currentstroke}%
\pgfsetdash{}{0pt}%
\pgfpathmoveto{\pgfqpoint{0.000000in}{0.000000in}}%
\pgfpathlineto{\pgfqpoint{3.152316in}{0.000000in}}%
\pgfpathlineto{\pgfqpoint{3.152316in}{2.139691in}}%
\pgfpathlineto{\pgfqpoint{0.000000in}{2.139691in}}%
\pgfpathclose%
\pgfusepath{fill}%
\end{pgfscope}%
\begin{pgfscope}%
\pgfsetbuttcap%
\pgfsetmiterjoin%
\definecolor{currentfill}{rgb}{1.000000,1.000000,1.000000}%
\pgfsetfillcolor{currentfill}%
\pgfsetlinewidth{0.000000pt}%
\definecolor{currentstroke}{rgb}{0.000000,0.000000,0.000000}%
\pgfsetstrokecolor{currentstroke}%
\pgfsetstrokeopacity{0.000000}%
\pgfsetdash{}{0pt}%
\pgfpathmoveto{\pgfqpoint{0.623149in}{0.499691in}}%
\pgfpathlineto{\pgfqpoint{2.948149in}{0.499691in}}%
\pgfpathlineto{\pgfqpoint{2.948149in}{2.039691in}}%
\pgfpathlineto{\pgfqpoint{0.623149in}{2.039691in}}%
\pgfpathclose%
\pgfusepath{fill}%
\end{pgfscope}%
\begin{pgfscope}%
\pgfsetbuttcap%
\pgfsetroundjoin%
\definecolor{currentfill}{rgb}{0.000000,0.000000,0.000000}%
\pgfsetfillcolor{currentfill}%
\pgfsetlinewidth{0.803000pt}%
\definecolor{currentstroke}{rgb}{0.000000,0.000000,0.000000}%
\pgfsetstrokecolor{currentstroke}%
\pgfsetdash{}{0pt}%
\pgfsys@defobject{currentmarker}{\pgfqpoint{0.000000in}{-0.048611in}}{\pgfqpoint{0.000000in}{0.000000in}}{%
\pgfpathmoveto{\pgfqpoint{0.000000in}{0.000000in}}%
\pgfpathlineto{\pgfqpoint{0.000000in}{-0.048611in}}%
\pgfusepath{stroke,fill}%
}%
\begin{pgfscope}%
\pgfsys@transformshift{0.646169in}{0.499691in}%
\pgfsys@useobject{currentmarker}{}%
\end{pgfscope}%
\end{pgfscope}%
\begin{pgfscope}%
\definecolor{textcolor}{rgb}{0.000000,0.000000,0.000000}%
\pgfsetstrokecolor{textcolor}%
\pgfsetfillcolor{textcolor}%
\pgftext[x=0.646169in,y=0.402469in,,top]{\color{textcolor}\rmfamily\fontsize{10.000000}{12.000000}\selectfont \(\displaystyle {0}\)}%
\end{pgfscope}%
\begin{pgfscope}%
\pgfsetbuttcap%
\pgfsetroundjoin%
\definecolor{currentfill}{rgb}{0.000000,0.000000,0.000000}%
\pgfsetfillcolor{currentfill}%
\pgfsetlinewidth{0.803000pt}%
\definecolor{currentstroke}{rgb}{0.000000,0.000000,0.000000}%
\pgfsetstrokecolor{currentstroke}%
\pgfsetdash{}{0pt}%
\pgfsys@defobject{currentmarker}{\pgfqpoint{0.000000in}{-0.048611in}}{\pgfqpoint{0.000000in}{0.000000in}}{%
\pgfpathmoveto{\pgfqpoint{0.000000in}{0.000000in}}%
\pgfpathlineto{\pgfqpoint{0.000000in}{-0.048611in}}%
\pgfusepath{stroke,fill}%
}%
\begin{pgfscope}%
\pgfsys@transformshift{1.221664in}{0.499691in}%
\pgfsys@useobject{currentmarker}{}%
\end{pgfscope}%
\end{pgfscope}%
\begin{pgfscope}%
\definecolor{textcolor}{rgb}{0.000000,0.000000,0.000000}%
\pgfsetstrokecolor{textcolor}%
\pgfsetfillcolor{textcolor}%
\pgftext[x=1.221664in,y=0.402469in,,top]{\color{textcolor}\rmfamily\fontsize{10.000000}{12.000000}\selectfont \(\displaystyle {50}\)}%
\end{pgfscope}%
\begin{pgfscope}%
\pgfsetbuttcap%
\pgfsetroundjoin%
\definecolor{currentfill}{rgb}{0.000000,0.000000,0.000000}%
\pgfsetfillcolor{currentfill}%
\pgfsetlinewidth{0.803000pt}%
\definecolor{currentstroke}{rgb}{0.000000,0.000000,0.000000}%
\pgfsetstrokecolor{currentstroke}%
\pgfsetdash{}{0pt}%
\pgfsys@defobject{currentmarker}{\pgfqpoint{0.000000in}{-0.048611in}}{\pgfqpoint{0.000000in}{0.000000in}}{%
\pgfpathmoveto{\pgfqpoint{0.000000in}{0.000000in}}%
\pgfpathlineto{\pgfqpoint{0.000000in}{-0.048611in}}%
\pgfusepath{stroke,fill}%
}%
\begin{pgfscope}%
\pgfsys@transformshift{1.797159in}{0.499691in}%
\pgfsys@useobject{currentmarker}{}%
\end{pgfscope}%
\end{pgfscope}%
\begin{pgfscope}%
\definecolor{textcolor}{rgb}{0.000000,0.000000,0.000000}%
\pgfsetstrokecolor{textcolor}%
\pgfsetfillcolor{textcolor}%
\pgftext[x=1.797159in,y=0.402469in,,top]{\color{textcolor}\rmfamily\fontsize{10.000000}{12.000000}\selectfont \(\displaystyle {100}\)}%
\end{pgfscope}%
\begin{pgfscope}%
\pgfsetbuttcap%
\pgfsetroundjoin%
\definecolor{currentfill}{rgb}{0.000000,0.000000,0.000000}%
\pgfsetfillcolor{currentfill}%
\pgfsetlinewidth{0.803000pt}%
\definecolor{currentstroke}{rgb}{0.000000,0.000000,0.000000}%
\pgfsetstrokecolor{currentstroke}%
\pgfsetdash{}{0pt}%
\pgfsys@defobject{currentmarker}{\pgfqpoint{0.000000in}{-0.048611in}}{\pgfqpoint{0.000000in}{0.000000in}}{%
\pgfpathmoveto{\pgfqpoint{0.000000in}{0.000000in}}%
\pgfpathlineto{\pgfqpoint{0.000000in}{-0.048611in}}%
\pgfusepath{stroke,fill}%
}%
\begin{pgfscope}%
\pgfsys@transformshift{2.372654in}{0.499691in}%
\pgfsys@useobject{currentmarker}{}%
\end{pgfscope}%
\end{pgfscope}%
\begin{pgfscope}%
\definecolor{textcolor}{rgb}{0.000000,0.000000,0.000000}%
\pgfsetstrokecolor{textcolor}%
\pgfsetfillcolor{textcolor}%
\pgftext[x=2.372654in,y=0.402469in,,top]{\color{textcolor}\rmfamily\fontsize{10.000000}{12.000000}\selectfont \(\displaystyle {150}\)}%
\end{pgfscope}%
\begin{pgfscope}%
\pgfsetbuttcap%
\pgfsetroundjoin%
\definecolor{currentfill}{rgb}{0.000000,0.000000,0.000000}%
\pgfsetfillcolor{currentfill}%
\pgfsetlinewidth{0.803000pt}%
\definecolor{currentstroke}{rgb}{0.000000,0.000000,0.000000}%
\pgfsetstrokecolor{currentstroke}%
\pgfsetdash{}{0pt}%
\pgfsys@defobject{currentmarker}{\pgfqpoint{0.000000in}{-0.048611in}}{\pgfqpoint{0.000000in}{0.000000in}}{%
\pgfpathmoveto{\pgfqpoint{0.000000in}{0.000000in}}%
\pgfpathlineto{\pgfqpoint{0.000000in}{-0.048611in}}%
\pgfusepath{stroke,fill}%
}%
\begin{pgfscope}%
\pgfsys@transformshift{2.948149in}{0.499691in}%
\pgfsys@useobject{currentmarker}{}%
\end{pgfscope}%
\end{pgfscope}%
\begin{pgfscope}%
\definecolor{textcolor}{rgb}{0.000000,0.000000,0.000000}%
\pgfsetstrokecolor{textcolor}%
\pgfsetfillcolor{textcolor}%
\pgftext[x=2.948149in,y=0.402469in,,top]{\color{textcolor}\rmfamily\fontsize{10.000000}{12.000000}\selectfont \(\displaystyle {200}\)}%
\end{pgfscope}%
\begin{pgfscope}%
\definecolor{textcolor}{rgb}{0.000000,0.000000,0.000000}%
\pgfsetstrokecolor{textcolor}%
\pgfsetfillcolor{textcolor}%
\pgftext[x=1.785649in,y=0.223457in,,top]{\color{textcolor}\rmfamily\fontsize{10.000000}{12.000000}\selectfont Incidents per TSG}%
\end{pgfscope}%
\begin{pgfscope}%
\pgfsetbuttcap%
\pgfsetroundjoin%
\definecolor{currentfill}{rgb}{0.000000,0.000000,0.000000}%
\pgfsetfillcolor{currentfill}%
\pgfsetlinewidth{0.803000pt}%
\definecolor{currentstroke}{rgb}{0.000000,0.000000,0.000000}%
\pgfsetstrokecolor{currentstroke}%
\pgfsetdash{}{0pt}%
\pgfsys@defobject{currentmarker}{\pgfqpoint{-0.048611in}{0.000000in}}{\pgfqpoint{-0.000000in}{0.000000in}}{%
\pgfpathmoveto{\pgfqpoint{-0.000000in}{0.000000in}}%
\pgfpathlineto{\pgfqpoint{-0.048611in}{0.000000in}}%
\pgfusepath{stroke,fill}%
}%
\begin{pgfscope}%
\pgfsys@transformshift{0.623149in}{0.499691in}%
\pgfsys@useobject{currentmarker}{}%
\end{pgfscope}%
\end{pgfscope}%
\begin{pgfscope}%
\definecolor{textcolor}{rgb}{0.000000,0.000000,0.000000}%
\pgfsetstrokecolor{textcolor}%
\pgfsetfillcolor{textcolor}%
\pgftext[x=0.279012in, y=0.451466in, left, base]{\color{textcolor}\rmfamily\fontsize{10.000000}{12.000000}\selectfont \(\displaystyle {0.00}\)}%
\end{pgfscope}%
\begin{pgfscope}%
\pgfsetbuttcap%
\pgfsetroundjoin%
\definecolor{currentfill}{rgb}{0.000000,0.000000,0.000000}%
\pgfsetfillcolor{currentfill}%
\pgfsetlinewidth{0.803000pt}%
\definecolor{currentstroke}{rgb}{0.000000,0.000000,0.000000}%
\pgfsetstrokecolor{currentstroke}%
\pgfsetdash{}{0pt}%
\pgfsys@defobject{currentmarker}{\pgfqpoint{-0.048611in}{0.000000in}}{\pgfqpoint{-0.000000in}{0.000000in}}{%
\pgfpathmoveto{\pgfqpoint{-0.000000in}{0.000000in}}%
\pgfpathlineto{\pgfqpoint{-0.048611in}{0.000000in}}%
\pgfusepath{stroke,fill}%
}%
\begin{pgfscope}%
\pgfsys@transformshift{0.623149in}{0.866627in}%
\pgfsys@useobject{currentmarker}{}%
\end{pgfscope}%
\end{pgfscope}%
\begin{pgfscope}%
\definecolor{textcolor}{rgb}{0.000000,0.000000,0.000000}%
\pgfsetstrokecolor{textcolor}%
\pgfsetfillcolor{textcolor}%
\pgftext[x=0.279012in, y=0.818402in, left, base]{\color{textcolor}\rmfamily\fontsize{10.000000}{12.000000}\selectfont \(\displaystyle {0.25}\)}%
\end{pgfscope}%
\begin{pgfscope}%
\pgfsetbuttcap%
\pgfsetroundjoin%
\definecolor{currentfill}{rgb}{0.000000,0.000000,0.000000}%
\pgfsetfillcolor{currentfill}%
\pgfsetlinewidth{0.803000pt}%
\definecolor{currentstroke}{rgb}{0.000000,0.000000,0.000000}%
\pgfsetstrokecolor{currentstroke}%
\pgfsetdash{}{0pt}%
\pgfsys@defobject{currentmarker}{\pgfqpoint{-0.048611in}{0.000000in}}{\pgfqpoint{-0.000000in}{0.000000in}}{%
\pgfpathmoveto{\pgfqpoint{-0.000000in}{0.000000in}}%
\pgfpathlineto{\pgfqpoint{-0.048611in}{0.000000in}}%
\pgfusepath{stroke,fill}%
}%
\begin{pgfscope}%
\pgfsys@transformshift{0.623149in}{1.233563in}%
\pgfsys@useobject{currentmarker}{}%
\end{pgfscope}%
\end{pgfscope}%
\begin{pgfscope}%
\definecolor{textcolor}{rgb}{0.000000,0.000000,0.000000}%
\pgfsetstrokecolor{textcolor}%
\pgfsetfillcolor{textcolor}%
\pgftext[x=0.279012in, y=1.185338in, left, base]{\color{textcolor}\rmfamily\fontsize{10.000000}{12.000000}\selectfont \(\displaystyle {0.50}\)}%
\end{pgfscope}%
\begin{pgfscope}%
\pgfsetbuttcap%
\pgfsetroundjoin%
\definecolor{currentfill}{rgb}{0.000000,0.000000,0.000000}%
\pgfsetfillcolor{currentfill}%
\pgfsetlinewidth{0.803000pt}%
\definecolor{currentstroke}{rgb}{0.000000,0.000000,0.000000}%
\pgfsetstrokecolor{currentstroke}%
\pgfsetdash{}{0pt}%
\pgfsys@defobject{currentmarker}{\pgfqpoint{-0.048611in}{0.000000in}}{\pgfqpoint{-0.000000in}{0.000000in}}{%
\pgfpathmoveto{\pgfqpoint{-0.000000in}{0.000000in}}%
\pgfpathlineto{\pgfqpoint{-0.048611in}{0.000000in}}%
\pgfusepath{stroke,fill}%
}%
\begin{pgfscope}%
\pgfsys@transformshift{0.623149in}{1.600500in}%
\pgfsys@useobject{currentmarker}{}%
\end{pgfscope}%
\end{pgfscope}%
\begin{pgfscope}%
\definecolor{textcolor}{rgb}{0.000000,0.000000,0.000000}%
\pgfsetstrokecolor{textcolor}%
\pgfsetfillcolor{textcolor}%
\pgftext[x=0.279012in, y=1.552274in, left, base]{\color{textcolor}\rmfamily\fontsize{10.000000}{12.000000}\selectfont \(\displaystyle {0.75}\)}%
\end{pgfscope}%
\begin{pgfscope}%
\pgfsetbuttcap%
\pgfsetroundjoin%
\definecolor{currentfill}{rgb}{0.000000,0.000000,0.000000}%
\pgfsetfillcolor{currentfill}%
\pgfsetlinewidth{0.803000pt}%
\definecolor{currentstroke}{rgb}{0.000000,0.000000,0.000000}%
\pgfsetstrokecolor{currentstroke}%
\pgfsetdash{}{0pt}%
\pgfsys@defobject{currentmarker}{\pgfqpoint{-0.048611in}{0.000000in}}{\pgfqpoint{-0.000000in}{0.000000in}}{%
\pgfpathmoveto{\pgfqpoint{-0.000000in}{0.000000in}}%
\pgfpathlineto{\pgfqpoint{-0.048611in}{0.000000in}}%
\pgfusepath{stroke,fill}%
}%
\begin{pgfscope}%
\pgfsys@transformshift{0.623149in}{1.967436in}%
\pgfsys@useobject{currentmarker}{}%
\end{pgfscope}%
\end{pgfscope}%
\begin{pgfscope}%
\definecolor{textcolor}{rgb}{0.000000,0.000000,0.000000}%
\pgfsetstrokecolor{textcolor}%
\pgfsetfillcolor{textcolor}%
\pgftext[x=0.279012in, y=1.919210in, left, base]{\color{textcolor}\rmfamily\fontsize{10.000000}{12.000000}\selectfont \(\displaystyle {1.00}\)}%
\end{pgfscope}%
\begin{pgfscope}%
\definecolor{textcolor}{rgb}{0.000000,0.000000,0.000000}%
\pgfsetstrokecolor{textcolor}%
\pgfsetfillcolor{textcolor}%
\pgftext[x=0.223457in,y=1.269691in,,bottom,rotate=90.000000]{\color{textcolor}\rmfamily\fontsize{10.000000}{12.000000}\selectfont Cumulative Density}%
\end{pgfscope}%
\begin{pgfscope}%
\pgfpathrectangle{\pgfqpoint{0.623149in}{0.499691in}}{\pgfqpoint{2.325000in}{1.540000in}}%
\pgfusepath{clip}%
\pgfsetrectcap%
\pgfsetroundjoin%
\pgfsetlinewidth{2.509375pt}%
\definecolor{currentstroke}{rgb}{0.250980,0.447059,0.768627}%
\pgfsetstrokecolor{currentstroke}%
\pgfsetdash{}{0pt}%
\pgfpathmoveto{\pgfqpoint{0.613149in}{0.503659in}}%
\pgfpathlineto{\pgfqpoint{0.617474in}{0.506191in}}%
\pgfpathlineto{\pgfqpoint{0.628574in}{0.535279in}}%
\pgfpathlineto{\pgfqpoint{0.639674in}{0.623050in}}%
\pgfpathlineto{\pgfqpoint{0.650774in}{0.803353in}}%
\pgfpathlineto{\pgfqpoint{0.672974in}{1.320176in}}%
\pgfpathlineto{\pgfqpoint{0.684074in}{1.519944in}}%
\pgfpathlineto{\pgfqpoint{0.695174in}{1.647255in}}%
\pgfpathlineto{\pgfqpoint{0.706275in}{1.723392in}}%
\pgfpathlineto{\pgfqpoint{0.717375in}{1.770513in}}%
\pgfpathlineto{\pgfqpoint{0.728475in}{1.802055in}}%
\pgfpathlineto{\pgfqpoint{0.739575in}{1.825238in}}%
\pgfpathlineto{\pgfqpoint{0.750675in}{1.843764in}}%
\pgfpathlineto{\pgfqpoint{0.761775in}{1.859221in}}%
\pgfpathlineto{\pgfqpoint{0.772875in}{1.872199in}}%
\pgfpathlineto{\pgfqpoint{0.783975in}{1.883072in}}%
\pgfpathlineto{\pgfqpoint{0.795075in}{1.892178in}}%
\pgfpathlineto{\pgfqpoint{0.817276in}{1.906302in}}%
\pgfpathlineto{\pgfqpoint{0.839476in}{1.916610in}}%
\pgfpathlineto{\pgfqpoint{0.861676in}{1.924201in}}%
\pgfpathlineto{\pgfqpoint{0.894976in}{1.931696in}}%
\pgfpathlineto{\pgfqpoint{0.939377in}{1.939258in}}%
\pgfpathlineto{\pgfqpoint{0.983777in}{1.944699in}}%
\pgfpathlineto{\pgfqpoint{1.050378in}{1.950757in}}%
\pgfpathlineto{\pgfqpoint{1.128079in}{1.954535in}}%
\pgfpathlineto{\pgfqpoint{1.383381in}{1.961009in}}%
\pgfpathlineto{\pgfqpoint{2.304690in}{1.966052in}}%
\pgfpathlineto{\pgfqpoint{2.815295in}{1.966358in}}%
\pgfpathlineto{\pgfqpoint{2.815295in}{1.966358in}}%
\pgfusepath{stroke}%
\end{pgfscope}%
\begin{pgfscope}%
\pgfsetrectcap%
\pgfsetmiterjoin%
\pgfsetlinewidth{0.803000pt}%
\definecolor{currentstroke}{rgb}{0.000000,0.000000,0.000000}%
\pgfsetstrokecolor{currentstroke}%
\pgfsetdash{}{0pt}%
\pgfpathmoveto{\pgfqpoint{0.623149in}{0.499691in}}%
\pgfpathlineto{\pgfqpoint{0.623149in}{2.039691in}}%
\pgfusepath{stroke}%
\end{pgfscope}%
\begin{pgfscope}%
\pgfsetrectcap%
\pgfsetmiterjoin%
\pgfsetlinewidth{0.803000pt}%
\definecolor{currentstroke}{rgb}{0.000000,0.000000,0.000000}%
\pgfsetstrokecolor{currentstroke}%
\pgfsetdash{}{0pt}%
\pgfpathmoveto{\pgfqpoint{0.623149in}{0.499691in}}%
\pgfpathlineto{\pgfqpoint{2.948149in}{0.499691in}}%
\pgfusepath{stroke}%
\end{pgfscope}%
\end{pgfpicture}%
\makeatother%
\endgroup%

%% file: Figures/tsg-per-severity.pgf
\begingroup%
\makeatletter%
\begin{pgfpicture}%
\pgfpathrectangle{\pgfpointorigin}{\pgfqpoint{3.079013in}{2.416471in}}%
\pgfusepath{use as bounding box, clip}%
\begin{pgfscope}%
\pgfsetbuttcap%
\pgfsetmiterjoin%
\definecolor{currentfill}{rgb}{1.000000,1.000000,1.000000}%
\pgfsetfillcolor{currentfill}%
\pgfsetlinewidth{0.000000pt}%
\definecolor{currentstroke}{rgb}{1.000000,1.000000,1.000000}%
\pgfsetstrokecolor{currentstroke}%
\pgfsetdash{}{0pt}%
\pgfpathmoveto{\pgfqpoint{0.000000in}{0.000000in}}%
\pgfpathlineto{\pgfqpoint{3.079013in}{0.000000in}}%
\pgfpathlineto{\pgfqpoint{3.079013in}{2.416471in}}%
\pgfpathlineto{\pgfqpoint{0.000000in}{2.416471in}}%
\pgfpathclose%
\pgfusepath{fill}%
\end{pgfscope}%
\begin{pgfscope}%
\pgfsetbuttcap%
\pgfsetmiterjoin%
\definecolor{currentfill}{rgb}{1.000000,1.000000,1.000000}%
\pgfsetfillcolor{currentfill}%
\pgfsetlinewidth{0.000000pt}%
\definecolor{currentstroke}{rgb}{0.000000,0.000000,0.000000}%
\pgfsetstrokecolor{currentstroke}%
\pgfsetstrokeopacity{0.000000}%
\pgfsetdash{}{0pt}%
\pgfpathmoveto{\pgfqpoint{0.654013in}{0.499691in}}%
\pgfpathlineto{\pgfqpoint{2.979013in}{0.499691in}}%
\pgfpathlineto{\pgfqpoint{2.979013in}{2.039691in}}%
\pgfpathlineto{\pgfqpoint{0.654013in}{2.039691in}}%
\pgfpathclose%
\pgfusepath{fill}%
\end{pgfscope}%
\begin{pgfscope}%
\pgfpathrectangle{\pgfqpoint{0.654013in}{0.499691in}}{\pgfqpoint{2.325000in}{1.540000in}}%
\pgfusepath{clip}%
\pgfsetbuttcap%
\pgfsetmiterjoin%
\definecolor{currentfill}{rgb}{0.654412,0.162059,0.177941}%
\pgfsetfillcolor{currentfill}%
\pgfsetlinewidth{0.000000pt}%
\definecolor{currentstroke}{rgb}{0.000000,0.000000,0.000000}%
\pgfsetstrokecolor{currentstroke}%
\pgfsetstrokeopacity{0.000000}%
\pgfsetdash{}{0pt}%
\pgfpathmoveto{\pgfqpoint{0.712138in}{0.499691in}}%
\pgfpathlineto{\pgfqpoint{1.177138in}{0.499691in}}%
\pgfpathlineto{\pgfqpoint{1.177138in}{0.514127in}}%
\pgfpathlineto{\pgfqpoint{0.712138in}{0.514127in}}%
\pgfpathclose%
\pgfusepath{fill}%
\end{pgfscope}%
\begin{pgfscope}%
\pgfpathrectangle{\pgfqpoint{0.654013in}{0.499691in}}{\pgfqpoint{2.325000in}{1.540000in}}%
\pgfusepath{clip}%
\pgfsetbuttcap%
\pgfsetmiterjoin%
\definecolor{currentfill}{rgb}{0.852843,0.344020,0.289902}%
\pgfsetfillcolor{currentfill}%
\pgfsetlinewidth{0.000000pt}%
\definecolor{currentstroke}{rgb}{0.000000,0.000000,0.000000}%
\pgfsetstrokecolor{currentstroke}%
\pgfsetstrokeopacity{0.000000}%
\pgfsetdash{}{0pt}%
\pgfpathmoveto{\pgfqpoint{1.293388in}{0.499691in}}%
\pgfpathlineto{\pgfqpoint{1.758388in}{0.499691in}}%
\pgfpathlineto{\pgfqpoint{1.758388in}{1.410583in}}%
\pgfpathlineto{\pgfqpoint{1.293388in}{1.410583in}}%
\pgfpathclose%
\pgfusepath{fill}%
\end{pgfscope}%
\begin{pgfscope}%
\pgfpathrectangle{\pgfqpoint{0.654013in}{0.499691in}}{\pgfqpoint{2.325000in}{1.540000in}}%
\pgfusepath{clip}%
\pgfsetbuttcap%
\pgfsetmiterjoin%
\definecolor{currentfill}{rgb}{0.915980,0.581275,0.487157}%
\pgfsetfillcolor{currentfill}%
\pgfsetlinewidth{0.000000pt}%
\definecolor{currentstroke}{rgb}{0.000000,0.000000,0.000000}%
\pgfsetstrokecolor{currentstroke}%
\pgfsetstrokeopacity{0.000000}%
\pgfsetdash{}{0pt}%
\pgfpathmoveto{\pgfqpoint{1.874638in}{0.499691in}}%
\pgfpathlineto{\pgfqpoint{2.339638in}{0.499691in}}%
\pgfpathlineto{\pgfqpoint{2.339638in}{1.966358in}}%
\pgfpathlineto{\pgfqpoint{1.874638in}{1.966358in}}%
\pgfpathclose%
\pgfusepath{fill}%
\end{pgfscope}%
\begin{pgfscope}%
\pgfpathrectangle{\pgfqpoint{0.654013in}{0.499691in}}{\pgfqpoint{2.325000in}{1.540000in}}%
\pgfusepath{clip}%
\pgfsetbuttcap%
\pgfsetmiterjoin%
\definecolor{currentfill}{rgb}{0.955980,0.805980,0.743627}%
\pgfsetfillcolor{currentfill}%
\pgfsetlinewidth{0.000000pt}%
\definecolor{currentstroke}{rgb}{0.000000,0.000000,0.000000}%
\pgfsetstrokecolor{currentstroke}%
\pgfsetstrokeopacity{0.000000}%
\pgfsetdash{}{0pt}%
\pgfpathmoveto{\pgfqpoint{2.455888in}{0.499691in}}%
\pgfpathlineto{\pgfqpoint{2.920888in}{0.499691in}}%
\pgfpathlineto{\pgfqpoint{2.920888in}{0.903891in}}%
\pgfpathlineto{\pgfqpoint{2.455888in}{0.903891in}}%
\pgfpathclose%
\pgfusepath{fill}%
\end{pgfscope}%
\begin{pgfscope}%
\pgfsetbuttcap%
\pgfsetroundjoin%
\definecolor{currentfill}{rgb}{0.000000,0.000000,0.000000}%
\pgfsetfillcolor{currentfill}%
\pgfsetlinewidth{0.803000pt}%
\definecolor{currentstroke}{rgb}{0.000000,0.000000,0.000000}%
\pgfsetstrokecolor{currentstroke}%
\pgfsetdash{}{0pt}%
\pgfsys@defobject{currentmarker}{\pgfqpoint{0.000000in}{-0.048611in}}{\pgfqpoint{0.000000in}{0.000000in}}{%
\pgfpathmoveto{\pgfqpoint{0.000000in}{0.000000in}}%
\pgfpathlineto{\pgfqpoint{0.000000in}{-0.048611in}}%
\pgfusepath{stroke,fill}%
}%
\begin{pgfscope}%
\pgfsys@transformshift{0.944638in}{0.499691in}%
\pgfsys@useobject{currentmarker}{}%
\end{pgfscope}%
\end{pgfscope}%
\begin{pgfscope}%
\definecolor{textcolor}{rgb}{0.000000,0.000000,0.000000}%
\pgfsetstrokecolor{textcolor}%
\pgfsetfillcolor{textcolor}%
\pgftext[x=0.944638in,y=0.402469in,,top]{\color{textcolor}\rmfamily\fontsize{10.000000}{12.000000}\selectfont 1}%
\end{pgfscope}%
\begin{pgfscope}%
\pgfsetbuttcap%
\pgfsetroundjoin%
\definecolor{currentfill}{rgb}{0.000000,0.000000,0.000000}%
\pgfsetfillcolor{currentfill}%
\pgfsetlinewidth{0.803000pt}%
\definecolor{currentstroke}{rgb}{0.000000,0.000000,0.000000}%
\pgfsetstrokecolor{currentstroke}%
\pgfsetdash{}{0pt}%
\pgfsys@defobject{currentmarker}{\pgfqpoint{0.000000in}{-0.048611in}}{\pgfqpoint{0.000000in}{0.000000in}}{%
\pgfpathmoveto{\pgfqpoint{0.000000in}{0.000000in}}%
\pgfpathlineto{\pgfqpoint{0.000000in}{-0.048611in}}%
\pgfusepath{stroke,fill}%
}%
\begin{pgfscope}%
\pgfsys@transformshift{1.525888in}{0.499691in}%
\pgfsys@useobject{currentmarker}{}%
\end{pgfscope}%
\end{pgfscope}%
\begin{pgfscope}%
\definecolor{textcolor}{rgb}{0.000000,0.000000,0.000000}%
\pgfsetstrokecolor{textcolor}%
\pgfsetfillcolor{textcolor}%
\pgftext[x=1.525888in,y=0.402469in,,top]{\color{textcolor}\rmfamily\fontsize{10.000000}{12.000000}\selectfont 2}%
\end{pgfscope}%
\begin{pgfscope}%
\pgfsetbuttcap%
\pgfsetroundjoin%
\definecolor{currentfill}{rgb}{0.000000,0.000000,0.000000}%
\pgfsetfillcolor{currentfill}%
\pgfsetlinewidth{0.803000pt}%
\definecolor{currentstroke}{rgb}{0.000000,0.000000,0.000000}%
\pgfsetstrokecolor{currentstroke}%
\pgfsetdash{}{0pt}%
\pgfsys@defobject{currentmarker}{\pgfqpoint{0.000000in}{-0.048611in}}{\pgfqpoint{0.000000in}{0.000000in}}{%
\pgfpathmoveto{\pgfqpoint{0.000000in}{0.000000in}}%
\pgfpathlineto{\pgfqpoint{0.000000in}{-0.048611in}}%
\pgfusepath{stroke,fill}%
}%
\begin{pgfscope}%
\pgfsys@transformshift{2.107138in}{0.499691in}%
\pgfsys@useobject{currentmarker}{}%
\end{pgfscope}%
\end{pgfscope}%
\begin{pgfscope}%
\definecolor{textcolor}{rgb}{0.000000,0.000000,0.000000}%
\pgfsetstrokecolor{textcolor}%
\pgfsetfillcolor{textcolor}%
\pgftext[x=2.107138in,y=0.402469in,,top]{\color{textcolor}\rmfamily\fontsize{10.000000}{12.000000}\selectfont 3}%
\end{pgfscope}%
\begin{pgfscope}%
\pgfsetbuttcap%
\pgfsetroundjoin%
\definecolor{currentfill}{rgb}{0.000000,0.000000,0.000000}%
\pgfsetfillcolor{currentfill}%
\pgfsetlinewidth{0.803000pt}%
\definecolor{currentstroke}{rgb}{0.000000,0.000000,0.000000}%
\pgfsetstrokecolor{currentstroke}%
\pgfsetdash{}{0pt}%
\pgfsys@defobject{currentmarker}{\pgfqpoint{0.000000in}{-0.048611in}}{\pgfqpoint{0.000000in}{0.000000in}}{%
\pgfpathmoveto{\pgfqpoint{0.000000in}{0.000000in}}%
\pgfpathlineto{\pgfqpoint{0.000000in}{-0.048611in}}%
\pgfusepath{stroke,fill}%
}%
\begin{pgfscope}%
\pgfsys@transformshift{2.688388in}{0.499691in}%
\pgfsys@useobject{currentmarker}{}%
\end{pgfscope}%
\end{pgfscope}%
\begin{pgfscope}%
\definecolor{textcolor}{rgb}{0.000000,0.000000,0.000000}%
\pgfsetstrokecolor{textcolor}%
\pgfsetfillcolor{textcolor}%
\pgftext[x=2.688388in,y=0.402469in,,top]{\color{textcolor}\rmfamily\fontsize{10.000000}{12.000000}\selectfont 4}%
\end{pgfscope}%
\begin{pgfscope}%
\definecolor{textcolor}{rgb}{0.000000,0.000000,0.000000}%
\pgfsetstrokecolor{textcolor}%
\pgfsetfillcolor{textcolor}%
\pgftext[x=1.816513in,y=0.223457in,,top]{\color{textcolor}\rmfamily\fontsize{10.000000}{12.000000}\selectfont Incident Severity}%
\end{pgfscope}%
\begin{pgfscope}%
\pgfsetbuttcap%
\pgfsetroundjoin%
\definecolor{currentfill}{rgb}{0.000000,0.000000,0.000000}%
\pgfsetfillcolor{currentfill}%
\pgfsetlinewidth{0.803000pt}%
\definecolor{currentstroke}{rgb}{0.000000,0.000000,0.000000}%
\pgfsetstrokecolor{currentstroke}%
\pgfsetdash{}{0pt}%
\pgfsys@defobject{currentmarker}{\pgfqpoint{-0.048611in}{0.000000in}}{\pgfqpoint{-0.000000in}{0.000000in}}{%
\pgfpathmoveto{\pgfqpoint{-0.000000in}{0.000000in}}%
\pgfpathlineto{\pgfqpoint{-0.048611in}{0.000000in}}%
\pgfusepath{stroke,fill}%
}%
\begin{pgfscope}%
\pgfsys@transformshift{0.654013in}{0.499691in}%
\pgfsys@useobject{currentmarker}{}%
\end{pgfscope}%
\end{pgfscope}%
\begin{pgfscope}%
\definecolor{textcolor}{rgb}{0.000000,0.000000,0.000000}%
\pgfsetstrokecolor{textcolor}%
\pgfsetfillcolor{textcolor}%
\pgftext[x=0.487346in, y=0.451466in, left, base]{\color{textcolor}\rmfamily\fontsize{10.000000}{12.000000}\selectfont \(\displaystyle {0}\)}%
\end{pgfscope}%
\begin{pgfscope}%
\pgfsetbuttcap%
\pgfsetroundjoin%
\definecolor{currentfill}{rgb}{0.000000,0.000000,0.000000}%
\pgfsetfillcolor{currentfill}%
\pgfsetlinewidth{0.803000pt}%
\definecolor{currentstroke}{rgb}{0.000000,0.000000,0.000000}%
\pgfsetstrokecolor{currentstroke}%
\pgfsetdash{}{0pt}%
\pgfsys@defobject{currentmarker}{\pgfqpoint{-0.048611in}{0.000000in}}{\pgfqpoint{-0.000000in}{0.000000in}}{%
\pgfpathmoveto{\pgfqpoint{-0.000000in}{0.000000in}}%
\pgfpathlineto{\pgfqpoint{-0.048611in}{0.000000in}}%
\pgfusepath{stroke,fill}%
}%
\begin{pgfscope}%
\pgfsys@transformshift{0.654013in}{0.980881in}%
\pgfsys@useobject{currentmarker}{}%
\end{pgfscope}%
\end{pgfscope}%
\begin{pgfscope}%
\definecolor{textcolor}{rgb}{0.000000,0.000000,0.000000}%
\pgfsetstrokecolor{textcolor}%
\pgfsetfillcolor{textcolor}%
\pgftext[x=0.279012in, y=0.932656in, left, base]{\color{textcolor}\rmfamily\fontsize{10.000000}{12.000000}\selectfont \(\displaystyle {1000}\)}%
\end{pgfscope}%
\begin{pgfscope}%
\pgfsetbuttcap%
\pgfsetroundjoin%
\definecolor{currentfill}{rgb}{0.000000,0.000000,0.000000}%
\pgfsetfillcolor{currentfill}%
\pgfsetlinewidth{0.803000pt}%
\definecolor{currentstroke}{rgb}{0.000000,0.000000,0.000000}%
\pgfsetstrokecolor{currentstroke}%
\pgfsetdash{}{0pt}%
\pgfsys@defobject{currentmarker}{\pgfqpoint{-0.048611in}{0.000000in}}{\pgfqpoint{-0.000000in}{0.000000in}}{%
\pgfpathmoveto{\pgfqpoint{-0.000000in}{0.000000in}}%
\pgfpathlineto{\pgfqpoint{-0.048611in}{0.000000in}}%
\pgfusepath{stroke,fill}%
}%
\begin{pgfscope}%
\pgfsys@transformshift{0.654013in}{1.462071in}%
\pgfsys@useobject{currentmarker}{}%
\end{pgfscope}%
\end{pgfscope}%
\begin{pgfscope}%
\definecolor{textcolor}{rgb}{0.000000,0.000000,0.000000}%
\pgfsetstrokecolor{textcolor}%
\pgfsetfillcolor{textcolor}%
\pgftext[x=0.279012in, y=1.413846in, left, base]{\color{textcolor}\rmfamily\fontsize{10.000000}{12.000000}\selectfont \(\displaystyle {2000}\)}%
\end{pgfscope}%
\begin{pgfscope}%
\pgfsetbuttcap%
\pgfsetroundjoin%
\definecolor{currentfill}{rgb}{0.000000,0.000000,0.000000}%
\pgfsetfillcolor{currentfill}%
\pgfsetlinewidth{0.803000pt}%
\definecolor{currentstroke}{rgb}{0.000000,0.000000,0.000000}%
\pgfsetstrokecolor{currentstroke}%
\pgfsetdash{}{0pt}%
\pgfsys@defobject{currentmarker}{\pgfqpoint{-0.048611in}{0.000000in}}{\pgfqpoint{-0.000000in}{0.000000in}}{%
\pgfpathmoveto{\pgfqpoint{-0.000000in}{0.000000in}}%
\pgfpathlineto{\pgfqpoint{-0.048611in}{0.000000in}}%
\pgfusepath{stroke,fill}%
}%
\begin{pgfscope}%
\pgfsys@transformshift{0.654013in}{1.943261in}%
\pgfsys@useobject{currentmarker}{}%
\end{pgfscope}%
\end{pgfscope}%
\begin{pgfscope}%
\definecolor{textcolor}{rgb}{0.000000,0.000000,0.000000}%
\pgfsetstrokecolor{textcolor}%
\pgfsetfillcolor{textcolor}%
\pgftext[x=0.279012in, y=1.895035in, left, base]{\color{textcolor}\rmfamily\fontsize{10.000000}{12.000000}\selectfont \(\displaystyle {3000}\)}%
\end{pgfscope}%
\begin{pgfscope}%
\definecolor{textcolor}{rgb}{0.000000,0.000000,0.000000}%
\pgfsetstrokecolor{textcolor}%
\pgfsetfillcolor{textcolor}%
\pgftext[x=0.223457in,y=1.269691in,,bottom,rotate=90.000000]{\color{textcolor}\rmfamily\fontsize{10.000000}{12.000000}\selectfont Count of TSGs linked to Incidents}%
\end{pgfscope}%
\begin{pgfscope}%
\pgfpathrectangle{\pgfqpoint{0.654013in}{0.499691in}}{\pgfqpoint{2.325000in}{1.540000in}}%
\pgfusepath{clip}%
\pgfsetrectcap%
\pgfsetroundjoin%
\pgfsetlinewidth{2.710125pt}%
\definecolor{currentstroke}{rgb}{0.260000,0.260000,0.260000}%
\pgfsetstrokecolor{currentstroke}%
\pgfsetdash{}{0pt}%
\pgfusepath{stroke}%
\end{pgfscope}%
\begin{pgfscope}%
\pgfpathrectangle{\pgfqpoint{0.654013in}{0.499691in}}{\pgfqpoint{2.325000in}{1.540000in}}%
\pgfusepath{clip}%
\pgfsetrectcap%
\pgfsetroundjoin%
\pgfsetlinewidth{2.710125pt}%
\definecolor{currentstroke}{rgb}{0.260000,0.260000,0.260000}%
\pgfsetstrokecolor{currentstroke}%
\pgfsetdash{}{0pt}%
\pgfusepath{stroke}%
\end{pgfscope}%
\begin{pgfscope}%
\pgfpathrectangle{\pgfqpoint{0.654013in}{0.499691in}}{\pgfqpoint{2.325000in}{1.540000in}}%
\pgfusepath{clip}%
\pgfsetrectcap%
\pgfsetroundjoin%
\pgfsetlinewidth{2.710125pt}%
\definecolor{currentstroke}{rgb}{0.260000,0.260000,0.260000}%
\pgfsetstrokecolor{currentstroke}%
\pgfsetdash{}{0pt}%
\pgfusepath{stroke}%
\end{pgfscope}%
\begin{pgfscope}%
\pgfpathrectangle{\pgfqpoint{0.654013in}{0.499691in}}{\pgfqpoint{2.325000in}{1.540000in}}%
\pgfusepath{clip}%
\pgfsetrectcap%
\pgfsetroundjoin%
\pgfsetlinewidth{2.710125pt}%
\definecolor{currentstroke}{rgb}{0.260000,0.260000,0.260000}%
\pgfsetstrokecolor{currentstroke}%
\pgfsetdash{}{0pt}%
\pgfusepath{stroke}%
\end{pgfscope}%
\begin{pgfscope}%
\pgfsetrectcap%
\pgfsetmiterjoin%
\pgfsetlinewidth{0.803000pt}%
\definecolor{currentstroke}{rgb}{0.000000,0.000000,0.000000}%
\pgfsetstrokecolor{currentstroke}%
\pgfsetdash{}{0pt}%
\pgfpathmoveto{\pgfqpoint{0.654013in}{0.499691in}}%
\pgfpathlineto{\pgfqpoint{0.654013in}{2.039691in}}%
\pgfusepath{stroke}%
\end{pgfscope}%
\begin{pgfscope}%
\pgfsetrectcap%
\pgfsetmiterjoin%
\pgfsetlinewidth{0.803000pt}%
\definecolor{currentstroke}{rgb}{0.000000,0.000000,0.000000}%
\pgfsetstrokecolor{currentstroke}%
\pgfsetdash{}{0pt}%
\pgfpathmoveto{\pgfqpoint{0.654013in}{0.499691in}}%
\pgfpathlineto{\pgfqpoint{2.979013in}{0.499691in}}%
\pgfusepath{stroke}%
\end{pgfscope}%
\begin{pgfscope}%
\definecolor{textcolor}{rgb}{0.000000,0.000000,0.000000}%
\pgfsetstrokecolor{textcolor}%
\pgfsetfillcolor{textcolor}%
\pgftext[x=0.944638in,y=0.514127in,,bottom]{\color{textcolor}\rmfamily\fontsize{10.000000}{12.000000}\selectfont 30}%
\end{pgfscope}%
\begin{pgfscope}%
\definecolor{textcolor}{rgb}{0.000000,0.000000,0.000000}%
\pgfsetstrokecolor{textcolor}%
\pgfsetfillcolor{textcolor}%
\pgftext[x=1.525888in,y=1.410583in,,bottom]{\color{textcolor}\rmfamily\fontsize{10.000000}{12.000000}\selectfont 1893}%
\end{pgfscope}%
\begin{pgfscope}%
\definecolor{textcolor}{rgb}{0.000000,0.000000,0.000000}%
\pgfsetstrokecolor{textcolor}%
\pgfsetfillcolor{textcolor}%
\pgftext[x=2.107138in,y=1.966358in,,bottom]{\color{textcolor}\rmfamily\fontsize{10.000000}{12.000000}\selectfont 3048}%
\end{pgfscope}%
\begin{pgfscope}%
\definecolor{textcolor}{rgb}{0.000000,0.000000,0.000000}%
\pgfsetstrokecolor{textcolor}%
\pgfsetfillcolor{textcolor}%
\pgftext[x=2.688388in,y=0.903891in,,bottom]{\color{textcolor}\rmfamily\fontsize{10.000000}{12.000000}\selectfont 840}%
\end{pgfscope}%
\end{pgfpicture}%
\makeatother%
\endgroup%

%% file: Figures/ttm-severity.pgf
\begingroup%
\makeatletter%
\begin{pgfpicture}%
\pgfpathrectangle{\pgfpointorigin}{\pgfqpoint{2.955556in}{2.139691in}}%
\pgfusepath{use as bounding box, clip}%
\begin{pgfscope}%
\pgfsetbuttcap%
\pgfsetmiterjoin%
\definecolor{currentfill}{rgb}{1.000000,1.000000,1.000000}%
\pgfsetfillcolor{currentfill}%
\pgfsetlinewidth{0.000000pt}%
\definecolor{currentstroke}{rgb}{1.000000,1.000000,1.000000}%
\pgfsetstrokecolor{currentstroke}%
\pgfsetdash{}{0pt}%
\pgfpathmoveto{\pgfqpoint{-0.000000in}{0.000000in}}%
\pgfpathlineto{\pgfqpoint{2.955556in}{0.000000in}}%
\pgfpathlineto{\pgfqpoint{2.955556in}{2.139691in}}%
\pgfpathlineto{\pgfqpoint{-0.000000in}{2.139691in}}%
\pgfpathclose%
\pgfusepath{fill}%
\end{pgfscope}%
\begin{pgfscope}%
\pgfsetbuttcap%
\pgfsetmiterjoin%
\definecolor{currentfill}{rgb}{1.000000,1.000000,1.000000}%
\pgfsetfillcolor{currentfill}%
\pgfsetlinewidth{0.000000pt}%
\definecolor{currentstroke}{rgb}{0.000000,0.000000,0.000000}%
\pgfsetstrokecolor{currentstroke}%
\pgfsetstrokeopacity{0.000000}%
\pgfsetdash{}{0pt}%
\pgfpathmoveto{\pgfqpoint{0.530556in}{0.499691in}}%
\pgfpathlineto{\pgfqpoint{2.855556in}{0.499691in}}%
\pgfpathlineto{\pgfqpoint{2.855556in}{2.039691in}}%
\pgfpathlineto{\pgfqpoint{0.530556in}{2.039691in}}%
\pgfpathclose%
\pgfusepath{fill}%
\end{pgfscope}%
\begin{pgfscope}%
\pgfpathrectangle{\pgfqpoint{0.530556in}{0.499691in}}{\pgfqpoint{2.325000in}{1.540000in}}%
\pgfusepath{clip}%
\pgfsetbuttcap%
\pgfsetmiterjoin%
\definecolor{currentfill}{rgb}{0.962255,0.735784,0.735784}%
\pgfsetfillcolor{currentfill}%
\pgfsetlinewidth{1.003750pt}%
\definecolor{currentstroke}{rgb}{0.000000,0.000000,0.000000}%
\pgfsetstrokecolor{currentstroke}%
\pgfsetdash{}{0pt}%
\pgfpathmoveto{\pgfqpoint{0.646806in}{0.499691in}}%
\pgfpathlineto{\pgfqpoint{1.111806in}{0.499691in}}%
\pgfpathlineto{\pgfqpoint{1.111806in}{1.561129in}}%
\pgfpathlineto{\pgfqpoint{0.646806in}{1.561129in}}%
\pgfpathclose%
\pgfusepath{stroke,fill}%
\end{pgfscope}%
\begin{pgfscope}%
\pgfpathrectangle{\pgfqpoint{0.530556in}{0.499691in}}{\pgfqpoint{2.325000in}{1.540000in}}%
\pgfusepath{clip}%
\pgfsetbuttcap%
\pgfsetmiterjoin%
\definecolor{currentfill}{rgb}{0.962255,0.735784,0.735784}%
\pgfsetfillcolor{currentfill}%
\pgfsetlinewidth{1.003750pt}%
\definecolor{currentstroke}{rgb}{0.000000,0.000000,0.000000}%
\pgfsetstrokecolor{currentstroke}%
\pgfsetdash{}{0pt}%
\pgfpathmoveto{\pgfqpoint{1.809306in}{0.499691in}}%
\pgfpathlineto{\pgfqpoint{2.274306in}{0.499691in}}%
\pgfpathlineto{\pgfqpoint{2.274306in}{1.048836in}}%
\pgfpathlineto{\pgfqpoint{1.809306in}{1.048836in}}%
\pgfpathclose%
\pgfusepath{stroke,fill}%
\end{pgfscope}%
\begin{pgfscope}%
\pgfpathrectangle{\pgfqpoint{0.530556in}{0.499691in}}{\pgfqpoint{2.325000in}{1.540000in}}%
\pgfusepath{clip}%
\pgfsetbuttcap%
\pgfsetmiterjoin%
\definecolor{currentfill}{rgb}{0.735784,0.735784,0.962255}%
\pgfsetfillcolor{currentfill}%
\pgfsetlinewidth{1.003750pt}%
\definecolor{currentstroke}{rgb}{0.000000,0.000000,0.000000}%
\pgfsetstrokecolor{currentstroke}%
\pgfsetdash{}{0pt}%
\pgfpathmoveto{\pgfqpoint{1.111806in}{0.499691in}}%
\pgfpathlineto{\pgfqpoint{1.576806in}{0.499691in}}%
\pgfpathlineto{\pgfqpoint{1.576806in}{0.543183in}}%
\pgfpathlineto{\pgfqpoint{1.111806in}{0.543183in}}%
\pgfpathclose%
\pgfusepath{stroke,fill}%
\end{pgfscope}%
\begin{pgfscope}%
\pgfpathrectangle{\pgfqpoint{0.530556in}{0.499691in}}{\pgfqpoint{2.325000in}{1.540000in}}%
\pgfusepath{clip}%
\pgfsetbuttcap%
\pgfsetmiterjoin%
\definecolor{currentfill}{rgb}{0.735784,0.735784,0.962255}%
\pgfsetfillcolor{currentfill}%
\pgfsetlinewidth{1.003750pt}%
\definecolor{currentstroke}{rgb}{0.000000,0.000000,0.000000}%
\pgfsetstrokecolor{currentstroke}%
\pgfsetdash{}{0pt}%
\pgfpathmoveto{\pgfqpoint{2.274306in}{0.499691in}}%
\pgfpathlineto{\pgfqpoint{2.739306in}{0.499691in}}%
\pgfpathlineto{\pgfqpoint{2.739306in}{0.880142in}}%
\pgfpathlineto{\pgfqpoint{2.274306in}{0.880142in}}%
\pgfpathclose%
\pgfusepath{stroke,fill}%
\end{pgfscope}%
\begin{pgfscope}%
\pgfsetbuttcap%
\pgfsetroundjoin%
\definecolor{currentfill}{rgb}{0.000000,0.000000,0.000000}%
\pgfsetfillcolor{currentfill}%
\pgfsetlinewidth{0.803000pt}%
\definecolor{currentstroke}{rgb}{0.000000,0.000000,0.000000}%
\pgfsetstrokecolor{currentstroke}%
\pgfsetdash{}{0pt}%
\pgfsys@defobject{currentmarker}{\pgfqpoint{0.000000in}{-0.048611in}}{\pgfqpoint{0.000000in}{0.000000in}}{%
\pgfpathmoveto{\pgfqpoint{0.000000in}{0.000000in}}%
\pgfpathlineto{\pgfqpoint{0.000000in}{-0.048611in}}%
\pgfusepath{stroke,fill}%
}%
\begin{pgfscope}%
\pgfsys@transformshift{1.111806in}{0.499691in}%
\pgfsys@useobject{currentmarker}{}%
\end{pgfscope}%
\end{pgfscope}%
\begin{pgfscope}%
\definecolor{textcolor}{rgb}{0.000000,0.000000,0.000000}%
\pgfsetstrokecolor{textcolor}%
\pgfsetfillcolor{textcolor}%
\pgftext[x=1.111806in,y=0.402469in,,top]{\color{textcolor}\rmfamily\fontsize{10.000000}{12.000000}\selectfont 1}%
\end{pgfscope}%
\begin{pgfscope}%
\pgfsetbuttcap%
\pgfsetroundjoin%
\definecolor{currentfill}{rgb}{0.000000,0.000000,0.000000}%
\pgfsetfillcolor{currentfill}%
\pgfsetlinewidth{0.803000pt}%
\definecolor{currentstroke}{rgb}{0.000000,0.000000,0.000000}%
\pgfsetstrokecolor{currentstroke}%
\pgfsetdash{}{0pt}%
\pgfsys@defobject{currentmarker}{\pgfqpoint{0.000000in}{-0.048611in}}{\pgfqpoint{0.000000in}{0.000000in}}{%
\pgfpathmoveto{\pgfqpoint{0.000000in}{0.000000in}}%
\pgfpathlineto{\pgfqpoint{0.000000in}{-0.048611in}}%
\pgfusepath{stroke,fill}%
}%
\begin{pgfscope}%
\pgfsys@transformshift{2.274306in}{0.499691in}%
\pgfsys@useobject{currentmarker}{}%
\end{pgfscope}%
\end{pgfscope}%
\begin{pgfscope}%
\definecolor{textcolor}{rgb}{0.000000,0.000000,0.000000}%
\pgfsetstrokecolor{textcolor}%
\pgfsetfillcolor{textcolor}%
\pgftext[x=2.274306in,y=0.402469in,,top]{\color{textcolor}\rmfamily\fontsize{10.000000}{12.000000}\selectfont 2}%
\end{pgfscope}%
\begin{pgfscope}%
\definecolor{textcolor}{rgb}{0.000000,0.000000,0.000000}%
\pgfsetstrokecolor{textcolor}%
\pgfsetfillcolor{textcolor}%
\pgftext[x=1.693056in,y=0.223457in,,top]{\color{textcolor}\rmfamily\fontsize{10.000000}{12.000000}\selectfont Incident Severity}%
\end{pgfscope}%
\begin{pgfscope}%
\pgfsetbuttcap%
\pgfsetroundjoin%
\definecolor{currentfill}{rgb}{0.000000,0.000000,0.000000}%
\pgfsetfillcolor{currentfill}%
\pgfsetlinewidth{0.803000pt}%
\definecolor{currentstroke}{rgb}{0.000000,0.000000,0.000000}%
\pgfsetstrokecolor{currentstroke}%
\pgfsetdash{}{0pt}%
\pgfsys@defobject{currentmarker}{\pgfqpoint{-0.048611in}{0.000000in}}{\pgfqpoint{-0.000000in}{0.000000in}}{%
\pgfpathmoveto{\pgfqpoint{-0.000000in}{0.000000in}}%
\pgfpathlineto{\pgfqpoint{-0.048611in}{0.000000in}}%
\pgfusepath{stroke,fill}%
}%
\begin{pgfscope}%
\pgfsys@transformshift{0.530556in}{0.499691in}%
\pgfsys@useobject{currentmarker}{}%
\end{pgfscope}%
\end{pgfscope}%
\begin{pgfscope}%
\definecolor{textcolor}{rgb}{0.000000,0.000000,0.000000}%
\pgfsetstrokecolor{textcolor}%
\pgfsetfillcolor{textcolor}%
\pgftext[x=0.363889in, y=0.451466in, left, base]{\color{textcolor}\rmfamily\fontsize{10.000000}{12.000000}\selectfont \(\displaystyle {0}\)}%
\end{pgfscope}%
\begin{pgfscope}%
\pgfsetbuttcap%
\pgfsetroundjoin%
\definecolor{currentfill}{rgb}{0.000000,0.000000,0.000000}%
\pgfsetfillcolor{currentfill}%
\pgfsetlinewidth{0.803000pt}%
\definecolor{currentstroke}{rgb}{0.000000,0.000000,0.000000}%
\pgfsetstrokecolor{currentstroke}%
\pgfsetdash{}{0pt}%
\pgfsys@defobject{currentmarker}{\pgfqpoint{-0.048611in}{0.000000in}}{\pgfqpoint{-0.000000in}{0.000000in}}{%
\pgfpathmoveto{\pgfqpoint{-0.000000in}{0.000000in}}%
\pgfpathlineto{\pgfqpoint{-0.048611in}{0.000000in}}%
\pgfusepath{stroke,fill}%
}%
\begin{pgfscope}%
\pgfsys@transformshift{0.530556in}{1.080626in}%
\pgfsys@useobject{currentmarker}{}%
\end{pgfscope}%
\end{pgfscope}%
\begin{pgfscope}%
\definecolor{textcolor}{rgb}{0.000000,0.000000,0.000000}%
\pgfsetstrokecolor{textcolor}%
\pgfsetfillcolor{textcolor}%
\pgftext[x=0.294444in, y=1.032401in, left, base]{\color{textcolor}\rmfamily\fontsize{10.000000}{12.000000}\selectfont \(\displaystyle {20}\)}%
\end{pgfscope}%
\begin{pgfscope}%
\pgfsetbuttcap%
\pgfsetroundjoin%
\definecolor{currentfill}{rgb}{0.000000,0.000000,0.000000}%
\pgfsetfillcolor{currentfill}%
\pgfsetlinewidth{0.803000pt}%
\definecolor{currentstroke}{rgb}{0.000000,0.000000,0.000000}%
\pgfsetstrokecolor{currentstroke}%
\pgfsetdash{}{0pt}%
\pgfsys@defobject{currentmarker}{\pgfqpoint{-0.048611in}{0.000000in}}{\pgfqpoint{-0.000000in}{0.000000in}}{%
\pgfpathmoveto{\pgfqpoint{-0.000000in}{0.000000in}}%
\pgfpathlineto{\pgfqpoint{-0.048611in}{0.000000in}}%
\pgfusepath{stroke,fill}%
}%
\begin{pgfscope}%
\pgfsys@transformshift{0.530556in}{1.661561in}%
\pgfsys@useobject{currentmarker}{}%
\end{pgfscope}%
\end{pgfscope}%
\begin{pgfscope}%
\definecolor{textcolor}{rgb}{0.000000,0.000000,0.000000}%
\pgfsetstrokecolor{textcolor}%
\pgfsetfillcolor{textcolor}%
\pgftext[x=0.294444in, y=1.613335in, left, base]{\color{textcolor}\rmfamily\fontsize{10.000000}{12.000000}\selectfont \(\displaystyle {40}\)}%
\end{pgfscope}%
\begin{pgfscope}%
\definecolor{textcolor}{rgb}{0.000000,0.000000,0.000000}%
\pgfsetstrokecolor{textcolor}%
\pgfsetfillcolor{textcolor}%
\pgftext[x=0.238889in,y=1.269691in,,bottom,rotate=90.000000]{\color{textcolor}\rmfamily\fontsize{10.000000}{12.000000}\selectfont Mean TTM (hrs)}%
\end{pgfscope}%
\begin{pgfscope}%
\pgfpathrectangle{\pgfqpoint{0.530556in}{0.499691in}}{\pgfqpoint{2.325000in}{1.540000in}}%
\pgfusepath{clip}%
\pgfsetrectcap%
\pgfsetroundjoin%
\pgfsetlinewidth{2.710125pt}%
\definecolor{currentstroke}{rgb}{0.260000,0.260000,0.260000}%
\pgfsetstrokecolor{currentstroke}%
\pgfsetdash{}{0pt}%
\pgfpathmoveto{\pgfqpoint{0.879306in}{1.146065in}}%
\pgfpathlineto{\pgfqpoint{0.879306in}{1.966358in}}%
\pgfusepath{stroke}%
\end{pgfscope}%
\begin{pgfscope}%
\pgfpathrectangle{\pgfqpoint{0.530556in}{0.499691in}}{\pgfqpoint{2.325000in}{1.540000in}}%
\pgfusepath{clip}%
\pgfsetrectcap%
\pgfsetroundjoin%
\pgfsetlinewidth{2.710125pt}%
\definecolor{currentstroke}{rgb}{0.260000,0.260000,0.260000}%
\pgfsetstrokecolor{currentstroke}%
\pgfsetdash{}{0pt}%
\pgfpathmoveto{\pgfqpoint{2.041806in}{1.006073in}}%
\pgfpathlineto{\pgfqpoint{2.041806in}{1.089860in}}%
\pgfusepath{stroke}%
\end{pgfscope}%
\begin{pgfscope}%
\pgfpathrectangle{\pgfqpoint{0.530556in}{0.499691in}}{\pgfqpoint{2.325000in}{1.540000in}}%
\pgfusepath{clip}%
\pgfsetrectcap%
\pgfsetroundjoin%
\pgfsetlinewidth{2.710125pt}%
\definecolor{currentstroke}{rgb}{0.260000,0.260000,0.260000}%
\pgfsetstrokecolor{currentstroke}%
\pgfsetdash{}{0pt}%
\pgfpathmoveto{\pgfqpoint{1.344306in}{0.514578in}}%
\pgfpathlineto{\pgfqpoint{1.344306in}{0.587060in}}%
\pgfusepath{stroke}%
\end{pgfscope}%
\begin{pgfscope}%
\pgfpathrectangle{\pgfqpoint{0.530556in}{0.499691in}}{\pgfqpoint{2.325000in}{1.540000in}}%
\pgfusepath{clip}%
\pgfsetrectcap%
\pgfsetroundjoin%
\pgfsetlinewidth{2.710125pt}%
\definecolor{currentstroke}{rgb}{0.260000,0.260000,0.260000}%
\pgfsetstrokecolor{currentstroke}%
\pgfsetdash{}{0pt}%
\pgfpathmoveto{\pgfqpoint{2.506806in}{0.829146in}}%
\pgfpathlineto{\pgfqpoint{2.506806in}{0.937124in}}%
\pgfusepath{stroke}%
\end{pgfscope}%
\begin{pgfscope}%
\pgfsetrectcap%
\pgfsetmiterjoin%
\pgfsetlinewidth{0.803000pt}%
\definecolor{currentstroke}{rgb}{0.000000,0.000000,0.000000}%
\pgfsetstrokecolor{currentstroke}%
\pgfsetdash{}{0pt}%
\pgfpathmoveto{\pgfqpoint{0.530556in}{0.499691in}}%
\pgfpathlineto{\pgfqpoint{0.530556in}{2.039691in}}%
\pgfusepath{stroke}%
\end{pgfscope}%
\begin{pgfscope}%
\pgfsetrectcap%
\pgfsetmiterjoin%
\pgfsetlinewidth{0.803000pt}%
\definecolor{currentstroke}{rgb}{0.000000,0.000000,0.000000}%
\pgfsetstrokecolor{currentstroke}%
\pgfsetdash{}{0pt}%
\pgfpathmoveto{\pgfqpoint{0.530556in}{0.499691in}}%
\pgfpathlineto{\pgfqpoint{2.855556in}{0.499691in}}%
\pgfusepath{stroke}%
\end{pgfscope}%
\begin{pgfscope}%
\pgfsetbuttcap%
\pgfsetmiterjoin%
\definecolor{currentfill}{rgb}{1.000000,1.000000,1.000000}%
\pgfsetfillcolor{currentfill}%
\pgfsetfillopacity{0.800000}%
\pgfsetlinewidth{1.003750pt}%
\definecolor{currentstroke}{rgb}{0.800000,0.800000,0.800000}%
\pgfsetstrokecolor{currentstroke}%
\pgfsetstrokeopacity{0.800000}%
\pgfsetdash{}{0pt}%
\pgfpathmoveto{\pgfqpoint{1.512769in}{1.348333in}}%
\pgfpathlineto{\pgfqpoint{2.758334in}{1.348333in}}%
\pgfpathquadraticcurveto{\pgfqpoint{2.786112in}{1.348333in}}{\pgfqpoint{2.786112in}{1.376111in}}%
\pgfpathlineto{\pgfqpoint{2.786112in}{1.942469in}}%
\pgfpathquadraticcurveto{\pgfqpoint{2.786112in}{1.970247in}}{\pgfqpoint{2.758334in}{1.970247in}}%
\pgfpathlineto{\pgfqpoint{1.512769in}{1.970247in}}%
\pgfpathquadraticcurveto{\pgfqpoint{1.484991in}{1.970247in}}{\pgfqpoint{1.484991in}{1.942469in}}%
\pgfpathlineto{\pgfqpoint{1.484991in}{1.376111in}}%
\pgfpathquadraticcurveto{\pgfqpoint{1.484991in}{1.348333in}}{\pgfqpoint{1.512769in}{1.348333in}}%
\pgfpathclose%
\pgfusepath{stroke,fill}%
\end{pgfscope}%
\begin{pgfscope}%
\definecolor{textcolor}{rgb}{0.000000,0.000000,0.000000}%
\pgfsetstrokecolor{textcolor}%
\pgfsetfillcolor{textcolor}%
\pgftext[x=1.540547in,y=1.818241in,left,base]{\color{textcolor}\rmfamily\fontsize{10.000000}{12.000000}\selectfont Incident has a TSG}%
\end{pgfscope}%
\begin{pgfscope}%
\pgfsetbuttcap%
\pgfsetmiterjoin%
\definecolor{currentfill}{rgb}{0.962255,0.735784,0.735784}%
\pgfsetfillcolor{currentfill}%
\pgfsetlinewidth{1.003750pt}%
\definecolor{currentstroke}{rgb}{0.000000,0.000000,0.000000}%
\pgfsetstrokecolor{currentstroke}%
\pgfsetdash{}{0pt}%
\pgfpathmoveto{\pgfqpoint{1.789293in}{1.624568in}}%
\pgfpathlineto{\pgfqpoint{2.067071in}{1.624568in}}%
\pgfpathlineto{\pgfqpoint{2.067071in}{1.721790in}}%
\pgfpathlineto{\pgfqpoint{1.789293in}{1.721790in}}%
\pgfpathclose%
\pgfusepath{stroke,fill}%
\end{pgfscope}%
\begin{pgfscope}%
\definecolor{textcolor}{rgb}{0.000000,0.000000,0.000000}%
\pgfsetstrokecolor{textcolor}%
\pgfsetfillcolor{textcolor}%
\pgftext[x=2.178182in,y=1.624568in,left,base]{\color{textcolor}\rmfamily\fontsize{10.000000}{12.000000}\selectfont False}%
\end{pgfscope}%
\begin{pgfscope}%
\pgfsetbuttcap%
\pgfsetmiterjoin%
\definecolor{currentfill}{rgb}{0.735784,0.735784,0.962255}%
\pgfsetfillcolor{currentfill}%
\pgfsetlinewidth{1.003750pt}%
\definecolor{currentstroke}{rgb}{0.000000,0.000000,0.000000}%
\pgfsetstrokecolor{currentstroke}%
\pgfsetdash{}{0pt}%
\pgfpathmoveto{\pgfqpoint{1.789293in}{1.430895in}}%
\pgfpathlineto{\pgfqpoint{2.067071in}{1.430895in}}%
\pgfpathlineto{\pgfqpoint{2.067071in}{1.528117in}}%
\pgfpathlineto{\pgfqpoint{1.789293in}{1.528117in}}%
\pgfpathclose%
\pgfusepath{stroke,fill}%
\end{pgfscope}%
\begin{pgfscope}%
\definecolor{textcolor}{rgb}{0.000000,0.000000,0.000000}%
\pgfsetstrokecolor{textcolor}%
\pgfsetfillcolor{textcolor}%
\pgftext[x=2.178182in,y=1.430895in,left,base]{\color{textcolor}\rmfamily\fontsize{10.000000}{12.000000}\selectfont True}%
\end{pgfscope}%
\end{pgfpicture}%
\makeatother%
\endgroup%

%% file: empirical-study.tex
\section{Empirical Study}
\label{sec:empirical-analysis}

In this section, we empirically study two aspects of TSGs -- \emph{usage} and \emph{quality}.
First, we characterize how TSGs are used for incident mitigation, using factors
like usage frequency, incident severity, and time-to-mitigate (TTM).
We analyze a large dataset of incidents mapped to TSGs by actual click-throughs on
links to the TSG.
We find new insights indicating that TSGs are widely used, and that incidents linked
to TSGs have significantly lower mitigation time.

Then, we perform a large-scale study of 400+ feedback items provided by 100s
of developers on TSGs at \CompanyX.
Our findings indicate significant gaps in quality aspects like \textit{Completeness},
\textit{Correctness}, and \textit{Up-to-dateness}.
Our results also uncover new and granular issues, such as \textit{Broken Links}
and \textit{Empty} TSGs.
Lastly, we discuss the implications of our findings and recommend actions to tackle
these quality challenges.
Further, we propose a new direction -- \textit{TSG Automation} -- that guides our
vision to improve the state of incident troubleshooting.

\input{Tables/feedback-taxonomy}

\subsection{TSG Usage}
\label{subsec:tsg-usage-empirical}

We first collect a large dataset of incidents that have TSGs linked to them.
Here, we use click-through data on links to TSGs to map incidents to corresponding TSGs.
Unlike prior work \cite{jiang2020mitigate} that study recommended TSGs and their effect on reducing effort, this approach captures the actual usage of TSGs by on-call engineers and strengthens our findings.
Consequently, our dataset of incidents to TSG mapping is not static but rather represents actual click-throughs performed on-call for incident mitigation.
Our dataset contains over \iftrue 1000s of \fi 
incidents\footnote{We cannot disclose the number of incidents due to Microsoft Policy.}
mapped to $\approx$\verify{4800} TSGs, collected over a 4 month period.
We then analyze this dataset from various perspectives to answer the \textbf{RQ}: \textit{How are TSGs used for incident management?}

\Paragraph{Incidents per TSG.}
%
To understand how frequently TSGs are used for mitigating incidents, we analyzed the distribution of the number of incidents per TSG in our dataset.
As shown in Figure \ref{fig:emp-1}, in 4 months, $\approx$47\%, 17\%, and 8\% of the TSGs had at least $2$, $5$, and $10$ incidents linked to them, respectively.
We also observed six TSGs linked to 100+ incidents each, the highest being 184.
These results show that TSGs are frequently used to mitigate recurrent incidents.

\Paragraph{TSGs and Incident Severity.}
In Figure \ref{fig:emp-2}, we look at the total number of TSGs linked to incidents of each severity level.
At \CompanyX, incidents are classified into 4 severity levels: (1) Sev 1: Outage, (2) Sev 2: High, (3) Sev 3: Medium, (4) Sev 4: Low.
Here, Sev 1 and 2 are \emph{paging incidents}; i.e., an on-call engineer is alerted as soon as the incident occurs.
Here, we find a small number of TSGs linked to outages (30). This is expected considering they are less frequent and may require deeper analysis and mitigation strategies. Next, we find that 85\% of TSGs are linked to either high (1893) or medium (3048) severity incidents. Lastly, we find a relatively lower 14\% of TSGs linked to low severity (840) incidents that are non-urgent and have no SLA impact. This shows that TSGs are commonly linked to critical incidents that affect multiple services and impact SLAs.

\Paragraph{TSGs and Time-to-mitigate.} Lastly, we analyze the effect of TSGs on efficient incident mitigation. We study the relationship between the time-to-mitigate (TTM) of incidents and whether the incident had a TSG linked (TSG Linkage). Here, we make sure to analyze incidents of Severity 1 and 2 only, since on-call engineers are notified immediately after their occurrence, and hence, TTM, is a reliable proxy for on-call effort.
We find that the mean TTM of incidents without TSGs ($\approx$19 hrs) is significantly higher than that of incidents linked to a TSG ($\approx$13 hrs). Further, in Figure \ref{fig:emp-3}, we analyze the correlation of this finding with incident severity. Here, we observe that severity 1 incidents linked with TSGs ($\approx$36 hrs) had considerably lesser TTM than those without TSGs ($\approx$2 hrs). For severity 2, we see a similar pattern where the mean TTM is reduced from $\approx$18hrs to $\approx$13hrs on linking TSGs to incidents. Overall, our analysis indicates that TSGs tend to significantly reduce effort during incident mitigation.

\subsection{TSG Quality}
\label{subsec:tsg-quality-empirical}

In Section \ref{subsec:tsg-usage-empirical}, we show  that TSGs are key to incident mitigation. However, in an internal survey of on-call experience at \CompanyX, developers picked \textit{TSG Quality \& Coverage} as the top pain-point out of 19 dimensions studied including volume of alerts, timing, and tooling.
Based on developer feedback, we find that TSGs are prone to issues such as missing information, incorrect steps, and outdated content.
In this section, we thus aim to empirically answer the \textbf{RQ}: \textit{How do developers perceive the quality of TSGs?}

\Paragraph{Setup.} In this study, we analyze the quality aspects of TSGs through feedback provided by developers and on-call engineers at \CompanyX. To this end, we first collect a dataset of feedback provided on TSGs over 4 months. Each feedback item contains (1) Thumbs-up/down rating and (2) a free text message to help improve the TSG. Our dataset contains \verify{428} feedback items (\faThumbsOUp:61, \faThumbsODown:367), for \verify{394} unique TSGs. Using this, we first develop a taxonomy for the feedback intent using the open coding approach. We then map each feedback item to an intent category and analyze their distributions.

We start by splitting our dataset into 3 sets: (1) \verify{87} items from month 1, (2) \verify{119} items from month 2, and (3) \verify{222} items from months 3 \& 4. Then, the first two authors of this work used the open coding approach to label the first set. They assigned a label to each feedback item based on what they perceived as the most prominent intent behind it. Subsequently, they discussed the feedback categories and agreed on a common taxonomy. Next, they independently labeled the second set to make sure no new categories emerged. They then had another discussion to settle disagreements and define a common understanding of each category. 

Lastly, they annotated the third set and computed the inter-annotator agreement score using Cohen's kappa \cite{cohen1960coefficient}. The resulting Cohen's kappa score was 0.907 indicating near-perfect agreement. Here, disagreements were mostly because multiple categories applied to certain feedback. Such disagreements were resolved by picking (1) the intent that occurs first in the feedback and (2) the most specific intent. With this approach, the annotators settled all disagreements and created a conclusively labeled dataset, which was used for all further analyses. In Table \ref{tab:feedback-taxonomy}, we show the resulting taxonomy with descriptions of the intent and examples.

\input{Tables/prior-taxonomies}

\Paragraph{Prior Work.} While we focus on understanding TSG quality, there has been prior work investigating different aspects of software documentation. Particularly close to our work, are empirical studies on generic software documentation quality that either (1) survey software practitioners or (2) mine stackoverflow and github data, to manually create a taxonomy of issues. In comparison, we perform a large-scale study of 400+ feedback items provided by 100s of developers, explicitly collected to improve the TSGs. Further, we collect feedback directly where an on-call engineer would visit to use the TSG for mitigation. As a result, we observe unfiltered feedback, enabling an accurate study of developer issues.

Table \ref{tab:prior-taxonomies} shows some taxonomies developed in prior empirical studies \cite{aghajani2019software, aghajani2020software, plosch2014value, garousi2013evaluating} of software documentation quality. Here, we observe some overlap with all prior work, but we also introduce new categories and rename a few others. Particularly, we share significant overlap with the taxonomy defined by Aghajani et al. \cite{aghajani2019software, aghajani2020software}. We retain 4 of their 9 categories -- Correctness, Completeness, Up-to-dateness, and Readability. Further, we rename Usability to User Experience and Usefulness to Relevance, making them more appropriate to the experience of using TSGs to mitigate incidents. We also add specific categories essential to the quality of TSGs -- Empty and Broken Link. Lastly, we do not include the Documentation process and tools categories, as our study focuses on TSG content only.

\input{Tables/component-types}

\Paragraph{Feedback Distribution.} Table \ref{tab:feedback-taxonomy} shows the frequency of feedback categories in our dataset. As shown, \emph{Completeness} (32.24\%) is the most frequent, and includes issues pointing to missing information such as steps, examples, links, and points of contact. The second most frequent category is \emph{Broken Link} (13.32\%). Here, feedback indicated errors while accessing links such as 404-not found, 403-forbidden, page not found, etc. Thirdly, we have \emph{Correctness} (11.21\%), where the feedback indicated misleading/incorrect information such as conflicting steps, incorrect steps, erroneous queries/commands, etc. We observe that these top-3 categories speak to the actual content in the TSG, accounting for a noteworthy total of 56.77\%.


Next, we have \emph{Readability} (10.28\%) and \emph{UX} (10.05\%), that account for $\approx$20\% of the feedback. This shows that a significant portion of quality issues faced by users is associated with how TSGs are presented and used, not just their content. Following that, we have \emph{Empty} (7.24\%) and have \emph{Up-to-dateness} (6.54\%) issues that reveal important maintainability issues with the current state of TSGs. Lastly, we find feedback pointing to TSGs lacking \emph{Relevance} (3.97\%) to the user's needs. This is a critical issue for troubleshooting, as developers follow the steps in TSGs to mitigate incidents.

\begin{figure}[h]
    \centering
    \scalebox{.7}{\input{Figures/feedback-rating-dist.pgf}}
    \caption{Distribution of Ratings for each Category}
    \label{fig:emp-5}
\end{figure}
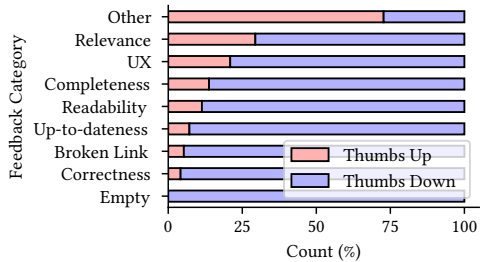

\Paragraph{Rating Distribution.} Each feedback item also contained a thumbs-up (positive) or thumbs-down (negative) rating for the associated TSG. Overall, we find that a majority 367 (85.7\%) items had thumbs-down, and 61 (14.3\%) had thumbs-up ratings. Prior work has shown that this is expected as people tend to give negative feedback far more than positive \cite{Vaish2008NotAE}.
In Figure \ref{fig:emp-5}, we look at the distribution of these ratings within each feedback category.
As shown, all major categories (i.e., leaving the \textit{Other} category) have a majority of negative ratings associated with feedback on TSGs. Further, we looked at the negativity ratio for each category, i.e., the ratio of \faThumbsODown{} to \faThumbsOUp{} votes.

We find that feedback for \textit{Empty} were (naturally) always associated with negative feedback, and hence had the highest negativity ratio. Following that, we have \textit{Correctness} (23:1), \textit{Broken Link} (18:1), and \textit{Up-to-dateness} (13:1) with very high negativity. Next, we have \textit{Readability} (8:1) and \textit{Completeness} (6:1) with high attributed negativity. Then, with relatively lower ratios are \textit{UX} (4:1) and \textit{Relevance} (2:1). Lastly, the \textit{Other} category had the lowest negativity ratio of 0.38:1; i.e., more positive feedback than negative, as shown by examples in Table \ref{tab:feedback-taxonomy}. 
Overall, our results indicate that on-call engineers and developers negatively perceive multiple aspects of TSG quality.

\subsection{Implications \& Recommendations} 

Our TSG usage analysis shows that TSGs are frequently linked to recurring incidents: 14\% TSGs with >5, and six TSGs with 100+ incidents in four months. Secondly, we find that TSGs are most commonly (>85\%) linked to critical severity incidents and significantly reduce mitigation time and effort (avg: 6hrs). Our findings indicate that TSGs are essential and widely used documentation that aid the mitigation of incidents, especially the ones which frequently repeat.

However, our results on the quality of TSGs highlight important issues. We find that the top-3 most frequent feedback categories were \textit{Completeness}, \textit{Broken Link}, and \textit{Correctness}, that account for 56\% of all feedback. TSG users have to deal with missing steps, lack of explanations, incomplete examples, invalid links, incorrect commands, etc. Next, we find that presentation and UX also strongly affect quality; i.e., 20\% of feedback points to \textit{Readability} or \textit{UX} issues stemming from verbose language, poor formatting, and organization. In context, next, we provide recommendations to alleviate these issues from our industry experiences (indicated by \faFlash).

\Paragraph{\faFlash{} Maintenance \& Testing:} A common solution to mitigate the Completeness, Broken Link, and Correctness issues can be to introduce maintenance and testing for TSGs. This is challenging considering TSGs are mostly siloed text documents (Word, OneNote, etc.) and the executables in them need to be parsed. Hence, software research should invest in automated tools that can help parse, review, test, and maintain software documentation, similar to source code.

\Paragraph{\faFlash{} Centralization \& Standardization:} To solve issues with Readability and User Experience, we make 2 recommendations that are being enforced at \CompanyX{}: (1) Converting all TSGs to a standard format (e.g., Markdown) with clear formatting guidelines. (2) Adopting a unifying platform for organizing TSGs, with an accompanying search engine. We find that these recommendations are consistent with those provided by prior work like Aghajani et al. \cite{aghajani2019software}. 

\Paragraph{\faFlash{} User-in-the-loop Review:} Additionally, assuming an enforced TSG review pipeline, we believe that involving users of TSGs in the review process is essential. This can solve readability issues where TSG authors assume clarity, but the users find it difficult to read and use the TSG. From our experience at \CompanyX, teams can adopt this since most TSG users are fellow developers in the team who take up on-call rotation at a specified frequency (e.g., once a month).

\Paragraph{How do we get there?} While we provide some insights and recommendations to improve TSG quality, we note significant research challenges. For instance, as previously stated, it is non-trivial to apply reviewing, testing, and maintenance techniques to semi-structured and informal text documentation like TSGs. However, we observe that the software engineering domain has effectively dealt with improving quality issues in source code. That brings us to the question: \textit{Can we automate TSGs and bring them closer to source code?} We observe that this change can in-turn introduce the recommendations we make into the world of TSGs, such as code review, regression testing, and version control. With this, we envision TSGs of the future to be verified semi-automated (like jupyter notebooks) or fully-automated workflows, that can be executed with minimal manual touches. Such automated TSGs would reduce manual toil, minimize human errors, and also improve DRI health.

%% file: Tables/feedback-taxonomy.tex
\begin{table*}[t]
\small
    \caption{Taxonomy of intents for feedback on TSGs and their frequencies}
    \label{tab:feedback-taxonomy}
    \def\arraystretch{1.05}
    \begin{tabular}[t]{l p{6cm} p{7cm} r}
    \toprule
         \textbf{Feedback Intent} & \textbf{Description} & \textbf{Examples} & \textbf{Frequency} \\
    \midrule
    Completeness & TSG is missing information like examples, links, etc. & \small{`unknown impact and mitigation', `please provide examples'} & $32.24\%$ \\
    
    Broken Link & TSG has broken or invalid links. & \small{`use cases links are broken.', `link leads to "404 - not found"'} & $13.32\%$ \\
    
    Correctness & TSG has incorrect or misleading information. & \small{`information is wrong', `steps didn't work'} & $11.21\%$ \\
    
    Readability & Feedback on TSG clarity, conciseness, grammar, etc. & \small{`too much info, not organized', `confusing terminology'} & $10.28\%$\\
    
    User Exp. (UX) & TSG accessibility, navigation, formatting, etc. & \small{`how to execute those code cells?', `badly formatted query'} & $10.05\%$ \\
    
    Empty & TSG is empty or has dummy content. & \small{`this page is empty', `fill in the TSG, currently just has TODO'} & $7.24\%$ \\
    
    Up-to-dateness & TSG content is outdated. & \small{`out of date: still using visualstudio instead of ado', `deprecated'} & $6.54\%$\\
    
    Relevance & Whether TSG is relevant to the user's issue. & \small{`doesn't tell me how to renew my cert', `nothing useful here'} & $3.97\%$\\
    
    Other & Unclear intent or outside the scope of TSG quality. & \small{`nice page', `loved this page`} & $5.14\%$ \\
    
    \bottomrule
    \end{tabular}
\end{table*}

%% file: Tables/prior-taxonomies.tex
\begin{table}[t]
\small
	\caption{Taxonomies of documentation quality in prior work}
	\label{tab:prior-taxonomies}
	\begin{tabular}{|p{1.3cm}|p{6.25cm}|}
	    \hline
	    \textbf{Study} & \textbf{Taxonomy} \\\hline
	    
		Aghajani \textit{et al.} \cite{aghajani2019software, aghajani2020software}
		& \footnotesize{Correctness, Completeness, Up-to-dateness, Maintainability, Readability, Usability, Usefulness, Doc. process, Doc. tools} \\\hline
		
		Plösch \textit{et al.} \cite{plosch2014value} & \footnotesize{Accuracy, Clarity, Consistency, Readability, Structuredness, Understandability, Completeness, Conciseness, Concreteness, Modifiabiality, Objectivity, Writing Style, Retrievability, Task Orientation, Traceability, Visual Effectiveness} \\\hline
		
		Garousi \textit{et al.} \cite{garousi2013evaluating} & \footnotesize{Completeness, Organization, Including visual models, Relevance, Preciseness, Readability, Accuracy, Consistency, Up-to-date, Examples} \\\hline
	\end{tabular}
\end{table}

%% file: Tables/component-types.tex

        
        
        
                
        
        


\begin{table*}[t]
    \caption{Types of Components in TSGs}
    \label{tab:component-types}
    \def\arraystretch{1.1}
    \begin{tabular}[t]{l l}
    \toprule
         \textbf{Component}: Description & \textbf{Example} \\
    \midrule

        \textbf{\texttt{ADF}} \cite{adf}: Link to an Azure Data Factory, a data integration service. & \small{\seqsplit{http://adf.azure.com/factory=resourceGroup/y/.../factories/z}} \\
        
        \textbf{\texttt{Jarvis}} \cite{azuremonitor}: Link to a Jarvis dashboard, an internal telemetry platform. & \small{https://jarvis.msft.net/dashboard/share/xxx} \\
        
        \textbf{\texttt{Kusto}} \cite{kusto}: Database query in Kusto Query Language (KQL). & {\lstinline!StormEvents | where State == "FLORIDA" | count!} \\
        
        \textbf{\texttt{Powershell}} \cite{powershell}: Statements from CLI scripts built on .NET Runtime. & {\lstinline!\$tenant = "<your tenant id/name>"!} \\
                
        \textbf{\texttt{Torus}}: Secure Powershell scripts to manage Azure resources and datacenters. & {\lstinline!\$rules = Get-TransportRule -Organization \$org!} \\
        
        \textbf{\texttt{Merlin}}: Custom Powershell scripts to diagnose and fix Sharepoint issues. & {\lstinline!Update-GridTenantProvisioningStamp \$TenantId!} \\
        
        \textbf{\texttt{Natural Language}}: Other natural language instructions. & \textit{If the status is green, the problem is self-resolved.} \\

    \bottomrule
    \end{tabular}
\end{table*}

%% file: Figures/feedback-rating-dist.pgf
\begingroup%
\makeatletter%
\begin{pgfpicture}%
\pgfpathrectangle{\pgfpointorigin}{\pgfqpoint{3.709415in}{2.125123in}}%
\pgfusepath{use as bounding box, clip}%
\begin{pgfscope}%
\pgfsetbuttcap%
\pgfsetmiterjoin%
\pgfsetlinewidth{0.000000pt}%
\definecolor{currentstroke}{rgb}{1.000000,1.000000,1.000000}%
\pgfsetstrokecolor{currentstroke}%
\pgfsetstrokeopacity{0.000000}%
\pgfsetdash{}{0pt}%
\pgfpathmoveto{\pgfqpoint{0.000000in}{0.000000in}}%
\pgfpathlineto{\pgfqpoint{3.709415in}{0.000000in}}%
\pgfpathlineto{\pgfqpoint{3.709415in}{2.125123in}}%
\pgfpathlineto{\pgfqpoint{0.000000in}{2.125123in}}%
\pgfpathclose%
\pgfusepath{}%
\end{pgfscope}%
\begin{pgfscope}%
\pgfsetbuttcap%
\pgfsetmiterjoin%
\definecolor{currentfill}{rgb}{1.000000,1.000000,1.000000}%
\pgfsetfillcolor{currentfill}%
\pgfsetlinewidth{0.000000pt}%
\definecolor{currentstroke}{rgb}{0.000000,0.000000,0.000000}%
\pgfsetstrokecolor{currentstroke}%
\pgfsetstrokeopacity{0.000000}%
\pgfsetdash{}{0pt}%
\pgfpathmoveto{\pgfqpoint{1.284415in}{0.515123in}}%
\pgfpathlineto{\pgfqpoint{3.609415in}{0.515123in}}%
\pgfpathlineto{\pgfqpoint{3.609415in}{2.025123in}}%
\pgfpathlineto{\pgfqpoint{1.284415in}{2.025123in}}%
\pgfpathclose%
\pgfusepath{fill}%
\end{pgfscope}%
\begin{pgfscope}%
\pgfpathrectangle{\pgfqpoint{1.284415in}{0.515123in}}{\pgfqpoint{2.325000in}{1.510000in}}%
\pgfusepath{clip}%
\pgfsetbuttcap%
\pgfsetmiterjoin%
\definecolor{currentfill}{rgb}{1.000000,0.698039,0.698039}%
\pgfsetfillcolor{currentfill}%
\pgfsetlinewidth{1.003750pt}%
\definecolor{currentstroke}{rgb}{0.000000,0.000000,0.000000}%
\pgfsetstrokecolor{currentstroke}%
\pgfsetdash{}{0pt}%
\pgfpathmoveto{\pgfqpoint{1.284415in}{0.557068in}}%
\pgfpathlineto{\pgfqpoint{1.284415in}{0.557068in}}%
\pgfpathlineto{\pgfqpoint{1.284415in}{0.640957in}}%
\pgfpathlineto{\pgfqpoint{1.284415in}{0.640957in}}%
\pgfpathclose%
\pgfusepath{stroke,fill}%
\end{pgfscope}%
\begin{pgfscope}%
\pgfpathrectangle{\pgfqpoint{1.284415in}{0.515123in}}{\pgfqpoint{2.325000in}{1.510000in}}%
\pgfusepath{clip}%
\pgfsetbuttcap%
\pgfsetmiterjoin%
\definecolor{currentfill}{rgb}{1.000000,0.698039,0.698039}%
\pgfsetfillcolor{currentfill}%
\pgfsetlinewidth{1.003750pt}%
\definecolor{currentstroke}{rgb}{0.000000,0.000000,0.000000}%
\pgfsetstrokecolor{currentstroke}%
\pgfsetdash{}{0pt}%
\pgfpathmoveto{\pgfqpoint{1.284415in}{0.724846in}}%
\pgfpathlineto{\pgfqpoint{1.376677in}{0.724846in}}%
\pgfpathlineto{\pgfqpoint{1.376677in}{0.808734in}}%
\pgfpathlineto{\pgfqpoint{1.284415in}{0.808734in}}%
\pgfpathclose%
\pgfusepath{stroke,fill}%
\end{pgfscope}%
\begin{pgfscope}%
\pgfpathrectangle{\pgfqpoint{1.284415in}{0.515123in}}{\pgfqpoint{2.325000in}{1.510000in}}%
\pgfusepath{clip}%
\pgfsetbuttcap%
\pgfsetmiterjoin%
\definecolor{currentfill}{rgb}{1.000000,0.698039,0.698039}%
\pgfsetfillcolor{currentfill}%
\pgfsetlinewidth{1.003750pt}%
\definecolor{currentstroke}{rgb}{0.000000,0.000000,0.000000}%
\pgfsetstrokecolor{currentstroke}%
\pgfsetdash{}{0pt}%
\pgfpathmoveto{\pgfqpoint{1.284415in}{0.892623in}}%
\pgfpathlineto{\pgfqpoint{1.400956in}{0.892623in}}%
\pgfpathlineto{\pgfqpoint{1.400956in}{0.976512in}}%
\pgfpathlineto{\pgfqpoint{1.284415in}{0.976512in}}%
\pgfpathclose%
\pgfusepath{stroke,fill}%
\end{pgfscope}%
\begin{pgfscope}%
\pgfpathrectangle{\pgfqpoint{1.284415in}{0.515123in}}{\pgfqpoint{2.325000in}{1.510000in}}%
\pgfusepath{clip}%
\pgfsetbuttcap%
\pgfsetmiterjoin%
\definecolor{currentfill}{rgb}{1.000000,0.698039,0.698039}%
\pgfsetfillcolor{currentfill}%
\pgfsetlinewidth{1.003750pt}%
\definecolor{currentstroke}{rgb}{0.000000,0.000000,0.000000}%
\pgfsetstrokecolor{currentstroke}%
\pgfsetdash{}{0pt}%
\pgfpathmoveto{\pgfqpoint{1.284415in}{1.060401in}}%
\pgfpathlineto{\pgfqpoint{1.442578in}{1.060401in}}%
\pgfpathlineto{\pgfqpoint{1.442578in}{1.144290in}}%
\pgfpathlineto{\pgfqpoint{1.284415in}{1.144290in}}%
\pgfpathclose%
\pgfusepath{stroke,fill}%
\end{pgfscope}%
\begin{pgfscope}%
\pgfpathrectangle{\pgfqpoint{1.284415in}{0.515123in}}{\pgfqpoint{2.325000in}{1.510000in}}%
\pgfusepath{clip}%
\pgfsetbuttcap%
\pgfsetmiterjoin%
\definecolor{currentfill}{rgb}{1.000000,0.698039,0.698039}%
\pgfsetfillcolor{currentfill}%
\pgfsetlinewidth{1.003750pt}%
\definecolor{currentstroke}{rgb}{0.000000,0.000000,0.000000}%
\pgfsetstrokecolor{currentstroke}%
\pgfsetdash{}{0pt}%
\pgfpathmoveto{\pgfqpoint{1.284415in}{1.228179in}}%
\pgfpathlineto{\pgfqpoint{1.536038in}{1.228179in}}%
\pgfpathlineto{\pgfqpoint{1.536038in}{1.312068in}}%
\pgfpathlineto{\pgfqpoint{1.284415in}{1.312068in}}%
\pgfpathclose%
\pgfusepath{stroke,fill}%
\end{pgfscope}%
\begin{pgfscope}%
\pgfpathrectangle{\pgfqpoint{1.284415in}{0.515123in}}{\pgfqpoint{2.325000in}{1.510000in}}%
\pgfusepath{clip}%
\pgfsetbuttcap%
\pgfsetmiterjoin%
\definecolor{currentfill}{rgb}{1.000000,0.698039,0.698039}%
\pgfsetfillcolor{currentfill}%
\pgfsetlinewidth{1.003750pt}%
\definecolor{currentstroke}{rgb}{0.000000,0.000000,0.000000}%
\pgfsetstrokecolor{currentstroke}%
\pgfsetdash{}{0pt}%
\pgfpathmoveto{\pgfqpoint{1.284415in}{1.395957in}}%
\pgfpathlineto{\pgfqpoint{1.589280in}{1.395957in}}%
\pgfpathlineto{\pgfqpoint{1.589280in}{1.479846in}}%
\pgfpathlineto{\pgfqpoint{1.284415in}{1.479846in}}%
\pgfpathclose%
\pgfusepath{stroke,fill}%
\end{pgfscope}%
\begin{pgfscope}%
\pgfpathrectangle{\pgfqpoint{1.284415in}{0.515123in}}{\pgfqpoint{2.325000in}{1.510000in}}%
\pgfusepath{clip}%
\pgfsetbuttcap%
\pgfsetmiterjoin%
\definecolor{currentfill}{rgb}{1.000000,0.698039,0.698039}%
\pgfsetfillcolor{currentfill}%
\pgfsetlinewidth{1.003750pt}%
\definecolor{currentstroke}{rgb}{0.000000,0.000000,0.000000}%
\pgfsetstrokecolor{currentstroke}%
\pgfsetdash{}{0pt}%
\pgfpathmoveto{\pgfqpoint{1.284415in}{1.563734in}}%
\pgfpathlineto{\pgfqpoint{1.747870in}{1.563734in}}%
\pgfpathlineto{\pgfqpoint{1.747870in}{1.647623in}}%
\pgfpathlineto{\pgfqpoint{1.284415in}{1.647623in}}%
\pgfpathclose%
\pgfusepath{stroke,fill}%
\end{pgfscope}%
\begin{pgfscope}%
\pgfpathrectangle{\pgfqpoint{1.284415in}{0.515123in}}{\pgfqpoint{2.325000in}{1.510000in}}%
\pgfusepath{clip}%
\pgfsetbuttcap%
\pgfsetmiterjoin%
\definecolor{currentfill}{rgb}{1.000000,0.698039,0.698039}%
\pgfsetfillcolor{currentfill}%
\pgfsetlinewidth{1.003750pt}%
\definecolor{currentstroke}{rgb}{0.000000,0.000000,0.000000}%
\pgfsetstrokecolor{currentstroke}%
\pgfsetdash{}{0pt}%
\pgfpathmoveto{\pgfqpoint{1.284415in}{1.731512in}}%
\pgfpathlineto{\pgfqpoint{1.935675in}{1.731512in}}%
\pgfpathlineto{\pgfqpoint{1.935675in}{1.815401in}}%
\pgfpathlineto{\pgfqpoint{1.284415in}{1.815401in}}%
\pgfpathclose%
\pgfusepath{stroke,fill}%
\end{pgfscope}%
\begin{pgfscope}%
\pgfpathrectangle{\pgfqpoint{1.284415in}{0.515123in}}{\pgfqpoint{2.325000in}{1.510000in}}%
\pgfusepath{clip}%
\pgfsetbuttcap%
\pgfsetmiterjoin%
\definecolor{currentfill}{rgb}{1.000000,0.698039,0.698039}%
\pgfsetfillcolor{currentfill}%
\pgfsetlinewidth{1.003750pt}%
\definecolor{currentstroke}{rgb}{0.000000,0.000000,0.000000}%
\pgfsetstrokecolor{currentstroke}%
\pgfsetdash{}{0pt}%
\pgfpathmoveto{\pgfqpoint{1.284415in}{1.899290in}}%
\pgfpathlineto{\pgfqpoint{2.894804in}{1.899290in}}%
\pgfpathlineto{\pgfqpoint{2.894804in}{1.983179in}}%
\pgfpathlineto{\pgfqpoint{1.284415in}{1.983179in}}%
\pgfpathclose%
\pgfusepath{stroke,fill}%
\end{pgfscope}%
\begin{pgfscope}%
\pgfpathrectangle{\pgfqpoint{1.284415in}{0.515123in}}{\pgfqpoint{2.325000in}{1.510000in}}%
\pgfusepath{clip}%
\pgfsetbuttcap%
\pgfsetmiterjoin%
\definecolor{currentfill}{rgb}{0.698039,0.698039,1.000000}%
\pgfsetfillcolor{currentfill}%
\pgfsetlinewidth{1.003750pt}%
\definecolor{currentstroke}{rgb}{0.000000,0.000000,0.000000}%
\pgfsetstrokecolor{currentstroke}%
\pgfsetdash{}{0pt}%
\pgfpathmoveto{\pgfqpoint{1.284415in}{0.557068in}}%
\pgfpathlineto{\pgfqpoint{3.498701in}{0.557068in}}%
\pgfpathlineto{\pgfqpoint{3.498701in}{0.640957in}}%
\pgfpathlineto{\pgfqpoint{1.284415in}{0.640957in}}%
\pgfpathclose%
\pgfusepath{stroke,fill}%
\end{pgfscope}%
\begin{pgfscope}%
\pgfpathrectangle{\pgfqpoint{1.284415in}{0.515123in}}{\pgfqpoint{2.325000in}{1.510000in}}%
\pgfusepath{clip}%
\pgfsetbuttcap%
\pgfsetmiterjoin%
\definecolor{currentfill}{rgb}{0.698039,0.698039,1.000000}%
\pgfsetfillcolor{currentfill}%
\pgfsetlinewidth{1.003750pt}%
\definecolor{currentstroke}{rgb}{0.000000,0.000000,0.000000}%
\pgfsetstrokecolor{currentstroke}%
\pgfsetdash{}{0pt}%
\pgfpathmoveto{\pgfqpoint{1.376677in}{0.724846in}}%
\pgfpathlineto{\pgfqpoint{3.498701in}{0.724846in}}%
\pgfpathlineto{\pgfqpoint{3.498701in}{0.808734in}}%
\pgfpathlineto{\pgfqpoint{1.376677in}{0.808734in}}%
\pgfpathclose%
\pgfusepath{stroke,fill}%
\end{pgfscope}%
\begin{pgfscope}%
\pgfpathrectangle{\pgfqpoint{1.284415in}{0.515123in}}{\pgfqpoint{2.325000in}{1.510000in}}%
\pgfusepath{clip}%
\pgfsetbuttcap%
\pgfsetmiterjoin%
\definecolor{currentfill}{rgb}{0.698039,0.698039,1.000000}%
\pgfsetfillcolor{currentfill}%
\pgfsetlinewidth{1.003750pt}%
\definecolor{currentstroke}{rgb}{0.000000,0.000000,0.000000}%
\pgfsetstrokecolor{currentstroke}%
\pgfsetdash{}{0pt}%
\pgfpathmoveto{\pgfqpoint{1.400956in}{0.892623in}}%
\pgfpathlineto{\pgfqpoint{3.498701in}{0.892623in}}%
\pgfpathlineto{\pgfqpoint{3.498701in}{0.976512in}}%
\pgfpathlineto{\pgfqpoint{1.400956in}{0.976512in}}%
\pgfpathclose%
\pgfusepath{stroke,fill}%
\end{pgfscope}%
\begin{pgfscope}%
\pgfpathrectangle{\pgfqpoint{1.284415in}{0.515123in}}{\pgfqpoint{2.325000in}{1.510000in}}%
\pgfusepath{clip}%
\pgfsetbuttcap%
\pgfsetmiterjoin%
\definecolor{currentfill}{rgb}{0.698039,0.698039,1.000000}%
\pgfsetfillcolor{currentfill}%
\pgfsetlinewidth{1.003750pt}%
\definecolor{currentstroke}{rgb}{0.000000,0.000000,0.000000}%
\pgfsetstrokecolor{currentstroke}%
\pgfsetdash{}{0pt}%
\pgfpathmoveto{\pgfqpoint{1.442578in}{1.060401in}}%
\pgfpathlineto{\pgfqpoint{3.498701in}{1.060401in}}%
\pgfpathlineto{\pgfqpoint{3.498701in}{1.144290in}}%
\pgfpathlineto{\pgfqpoint{1.442578in}{1.144290in}}%
\pgfpathclose%
\pgfusepath{stroke,fill}%
\end{pgfscope}%
\begin{pgfscope}%
\pgfpathrectangle{\pgfqpoint{1.284415in}{0.515123in}}{\pgfqpoint{2.325000in}{1.510000in}}%
\pgfusepath{clip}%
\pgfsetbuttcap%
\pgfsetmiterjoin%
\definecolor{currentfill}{rgb}{0.698039,0.698039,1.000000}%
\pgfsetfillcolor{currentfill}%
\pgfsetlinewidth{1.003750pt}%
\definecolor{currentstroke}{rgb}{0.000000,0.000000,0.000000}%
\pgfsetstrokecolor{currentstroke}%
\pgfsetdash{}{0pt}%
\pgfpathmoveto{\pgfqpoint{1.536038in}{1.228179in}}%
\pgfpathlineto{\pgfqpoint{3.498701in}{1.228179in}}%
\pgfpathlineto{\pgfqpoint{3.498701in}{1.312068in}}%
\pgfpathlineto{\pgfqpoint{1.536038in}{1.312068in}}%
\pgfpathclose%
\pgfusepath{stroke,fill}%
\end{pgfscope}%
\begin{pgfscope}%
\pgfpathrectangle{\pgfqpoint{1.284415in}{0.515123in}}{\pgfqpoint{2.325000in}{1.510000in}}%
\pgfusepath{clip}%
\pgfsetbuttcap%
\pgfsetmiterjoin%
\definecolor{currentfill}{rgb}{0.698039,0.698039,1.000000}%
\pgfsetfillcolor{currentfill}%
\pgfsetlinewidth{1.003750pt}%
\definecolor{currentstroke}{rgb}{0.000000,0.000000,0.000000}%
\pgfsetstrokecolor{currentstroke}%
\pgfsetdash{}{0pt}%
\pgfpathmoveto{\pgfqpoint{1.589280in}{1.395957in}}%
\pgfpathlineto{\pgfqpoint{3.498701in}{1.395957in}}%
\pgfpathlineto{\pgfqpoint{3.498701in}{1.479846in}}%
\pgfpathlineto{\pgfqpoint{1.589280in}{1.479846in}}%
\pgfpathclose%
\pgfusepath{stroke,fill}%
\end{pgfscope}%
\begin{pgfscope}%
\pgfpathrectangle{\pgfqpoint{1.284415in}{0.515123in}}{\pgfqpoint{2.325000in}{1.510000in}}%
\pgfusepath{clip}%
\pgfsetbuttcap%
\pgfsetmiterjoin%
\definecolor{currentfill}{rgb}{0.698039,0.698039,1.000000}%
\pgfsetfillcolor{currentfill}%
\pgfsetlinewidth{1.003750pt}%
\definecolor{currentstroke}{rgb}{0.000000,0.000000,0.000000}%
\pgfsetstrokecolor{currentstroke}%
\pgfsetdash{}{0pt}%
\pgfpathmoveto{\pgfqpoint{1.747870in}{1.563734in}}%
\pgfpathlineto{\pgfqpoint{3.498701in}{1.563734in}}%
\pgfpathlineto{\pgfqpoint{3.498701in}{1.647623in}}%
\pgfpathlineto{\pgfqpoint{1.747870in}{1.647623in}}%
\pgfpathclose%
\pgfusepath{stroke,fill}%
\end{pgfscope}%
\begin{pgfscope}%
\pgfpathrectangle{\pgfqpoint{1.284415in}{0.515123in}}{\pgfqpoint{2.325000in}{1.510000in}}%
\pgfusepath{clip}%
\pgfsetbuttcap%
\pgfsetmiterjoin%
\definecolor{currentfill}{rgb}{0.698039,0.698039,1.000000}%
\pgfsetfillcolor{currentfill}%
\pgfsetlinewidth{1.003750pt}%
\definecolor{currentstroke}{rgb}{0.000000,0.000000,0.000000}%
\pgfsetstrokecolor{currentstroke}%
\pgfsetdash{}{0pt}%
\pgfpathmoveto{\pgfqpoint{1.935675in}{1.731512in}}%
\pgfpathlineto{\pgfqpoint{3.498701in}{1.731512in}}%
\pgfpathlineto{\pgfqpoint{3.498701in}{1.815401in}}%
\pgfpathlineto{\pgfqpoint{1.935675in}{1.815401in}}%
\pgfpathclose%
\pgfusepath{stroke,fill}%
\end{pgfscope}%
\begin{pgfscope}%
\pgfpathrectangle{\pgfqpoint{1.284415in}{0.515123in}}{\pgfqpoint{2.325000in}{1.510000in}}%
\pgfusepath{clip}%
\pgfsetbuttcap%
\pgfsetmiterjoin%
\definecolor{currentfill}{rgb}{0.698039,0.698039,1.000000}%
\pgfsetfillcolor{currentfill}%
\pgfsetlinewidth{1.003750pt}%
\definecolor{currentstroke}{rgb}{0.000000,0.000000,0.000000}%
\pgfsetstrokecolor{currentstroke}%
\pgfsetdash{}{0pt}%
\pgfpathmoveto{\pgfqpoint{2.894804in}{1.899290in}}%
\pgfpathlineto{\pgfqpoint{3.498701in}{1.899290in}}%
\pgfpathlineto{\pgfqpoint{3.498701in}{1.983179in}}%
\pgfpathlineto{\pgfqpoint{2.894804in}{1.983179in}}%
\pgfpathclose%
\pgfusepath{stroke,fill}%
\end{pgfscope}%
\begin{pgfscope}%
\pgfsetbuttcap%
\pgfsetroundjoin%
\definecolor{currentfill}{rgb}{0.000000,0.000000,0.000000}%
\pgfsetfillcolor{currentfill}%
\pgfsetlinewidth{0.803000pt}%
\definecolor{currentstroke}{rgb}{0.000000,0.000000,0.000000}%
\pgfsetstrokecolor{currentstroke}%
\pgfsetdash{}{0pt}%
\pgfsys@defobject{currentmarker}{\pgfqpoint{0.000000in}{-0.048611in}}{\pgfqpoint{0.000000in}{0.000000in}}{%
\pgfpathmoveto{\pgfqpoint{0.000000in}{0.000000in}}%
\pgfpathlineto{\pgfqpoint{0.000000in}{-0.048611in}}%
\pgfusepath{stroke,fill}%
}%
\begin{pgfscope}%
\pgfsys@transformshift{1.284415in}{0.515123in}%
\pgfsys@useobject{currentmarker}{}%
\end{pgfscope}%
\end{pgfscope}%
\begin{pgfscope}%
\definecolor{textcolor}{rgb}{0.000000,0.000000,0.000000}%
\pgfsetstrokecolor{textcolor}%
\pgfsetfillcolor{textcolor}%
\pgftext[x=1.284415in,y=0.417901in,,top]{\color{textcolor}\rmfamily\fontsize{10.000000}{12.000000}\selectfont \(\displaystyle {0}\)}%
\end{pgfscope}%
\begin{pgfscope}%
\pgfsetbuttcap%
\pgfsetroundjoin%
\definecolor{currentfill}{rgb}{0.000000,0.000000,0.000000}%
\pgfsetfillcolor{currentfill}%
\pgfsetlinewidth{0.803000pt}%
\definecolor{currentstroke}{rgb}{0.000000,0.000000,0.000000}%
\pgfsetstrokecolor{currentstroke}%
\pgfsetdash{}{0pt}%
\pgfsys@defobject{currentmarker}{\pgfqpoint{0.000000in}{-0.048611in}}{\pgfqpoint{0.000000in}{0.000000in}}{%
\pgfpathmoveto{\pgfqpoint{0.000000in}{0.000000in}}%
\pgfpathlineto{\pgfqpoint{0.000000in}{-0.048611in}}%
\pgfusepath{stroke,fill}%
}%
\begin{pgfscope}%
\pgfsys@transformshift{1.837986in}{0.515123in}%
\pgfsys@useobject{currentmarker}{}%
\end{pgfscope}%
\end{pgfscope}%
\begin{pgfscope}%
\definecolor{textcolor}{rgb}{0.000000,0.000000,0.000000}%
\pgfsetstrokecolor{textcolor}%
\pgfsetfillcolor{textcolor}%
\pgftext[x=1.837986in,y=0.417901in,,top]{\color{textcolor}\rmfamily\fontsize{10.000000}{12.000000}\selectfont \(\displaystyle {25}\)}%
\end{pgfscope}%
\begin{pgfscope}%
\pgfsetbuttcap%
\pgfsetroundjoin%
\definecolor{currentfill}{rgb}{0.000000,0.000000,0.000000}%
\pgfsetfillcolor{currentfill}%
\pgfsetlinewidth{0.803000pt}%
\definecolor{currentstroke}{rgb}{0.000000,0.000000,0.000000}%
\pgfsetstrokecolor{currentstroke}%
\pgfsetdash{}{0pt}%
\pgfsys@defobject{currentmarker}{\pgfqpoint{0.000000in}{-0.048611in}}{\pgfqpoint{0.000000in}{0.000000in}}{%
\pgfpathmoveto{\pgfqpoint{0.000000in}{0.000000in}}%
\pgfpathlineto{\pgfqpoint{0.000000in}{-0.048611in}}%
\pgfusepath{stroke,fill}%
}%
\begin{pgfscope}%
\pgfsys@transformshift{2.391558in}{0.515123in}%
\pgfsys@useobject{currentmarker}{}%
\end{pgfscope}%
\end{pgfscope}%
\begin{pgfscope}%
\definecolor{textcolor}{rgb}{0.000000,0.000000,0.000000}%
\pgfsetstrokecolor{textcolor}%
\pgfsetfillcolor{textcolor}%
\pgftext[x=2.391558in,y=0.417901in,,top]{\color{textcolor}\rmfamily\fontsize{10.000000}{12.000000}\selectfont \(\displaystyle {50}\)}%
\end{pgfscope}%
\begin{pgfscope}%
\pgfsetbuttcap%
\pgfsetroundjoin%
\definecolor{currentfill}{rgb}{0.000000,0.000000,0.000000}%
\pgfsetfillcolor{currentfill}%
\pgfsetlinewidth{0.803000pt}%
\definecolor{currentstroke}{rgb}{0.000000,0.000000,0.000000}%
\pgfsetstrokecolor{currentstroke}%
\pgfsetdash{}{0pt}%
\pgfsys@defobject{currentmarker}{\pgfqpoint{0.000000in}{-0.048611in}}{\pgfqpoint{0.000000in}{0.000000in}}{%
\pgfpathmoveto{\pgfqpoint{0.000000in}{0.000000in}}%
\pgfpathlineto{\pgfqpoint{0.000000in}{-0.048611in}}%
\pgfusepath{stroke,fill}%
}%
\begin{pgfscope}%
\pgfsys@transformshift{2.945129in}{0.515123in}%
\pgfsys@useobject{currentmarker}{}%
\end{pgfscope}%
\end{pgfscope}%
\begin{pgfscope}%
\definecolor{textcolor}{rgb}{0.000000,0.000000,0.000000}%
\pgfsetstrokecolor{textcolor}%
\pgfsetfillcolor{textcolor}%
\pgftext[x=2.945129in,y=0.417901in,,top]{\color{textcolor}\rmfamily\fontsize{10.000000}{12.000000}\selectfont \(\displaystyle {75}\)}%
\end{pgfscope}%
\begin{pgfscope}%
\pgfsetbuttcap%
\pgfsetroundjoin%
\definecolor{currentfill}{rgb}{0.000000,0.000000,0.000000}%
\pgfsetfillcolor{currentfill}%
\pgfsetlinewidth{0.803000pt}%
\definecolor{currentstroke}{rgb}{0.000000,0.000000,0.000000}%
\pgfsetstrokecolor{currentstroke}%
\pgfsetdash{}{0pt}%
\pgfsys@defobject{currentmarker}{\pgfqpoint{0.000000in}{-0.048611in}}{\pgfqpoint{0.000000in}{0.000000in}}{%
\pgfpathmoveto{\pgfqpoint{0.000000in}{0.000000in}}%
\pgfpathlineto{\pgfqpoint{0.000000in}{-0.048611in}}%
\pgfusepath{stroke,fill}%
}%
\begin{pgfscope}%
\pgfsys@transformshift{3.498701in}{0.515123in}%
\pgfsys@useobject{currentmarker}{}%
\end{pgfscope}%
\end{pgfscope}%
\begin{pgfscope}%
\definecolor{textcolor}{rgb}{0.000000,0.000000,0.000000}%
\pgfsetstrokecolor{textcolor}%
\pgfsetfillcolor{textcolor}%
\pgftext[x=3.498701in,y=0.417901in,,top]{\color{textcolor}\rmfamily\fontsize{10.000000}{12.000000}\selectfont \(\displaystyle {100}\)}%
\end{pgfscope}%
\begin{pgfscope}%
\definecolor{textcolor}{rgb}{0.000000,0.000000,0.000000}%
\pgfsetstrokecolor{textcolor}%
\pgfsetfillcolor{textcolor}%
\pgftext[x=2.446915in,y=0.238889in,,top]{\color{textcolor}\rmfamily\fontsize{10.000000}{12.000000}\selectfont Count (\%)}%
\end{pgfscope}%
\begin{pgfscope}%
\pgfsetbuttcap%
\pgfsetroundjoin%
\definecolor{currentfill}{rgb}{0.000000,0.000000,0.000000}%
\pgfsetfillcolor{currentfill}%
\pgfsetlinewidth{0.803000pt}%
\definecolor{currentstroke}{rgb}{0.000000,0.000000,0.000000}%
\pgfsetstrokecolor{currentstroke}%
\pgfsetdash{}{0pt}%
\pgfsys@defobject{currentmarker}{\pgfqpoint{-0.048611in}{0.000000in}}{\pgfqpoint{-0.000000in}{0.000000in}}{%
\pgfpathmoveto{\pgfqpoint{-0.000000in}{0.000000in}}%
\pgfpathlineto{\pgfqpoint{-0.048611in}{0.000000in}}%
\pgfusepath{stroke,fill}%
}%
\begin{pgfscope}%
\pgfsys@transformshift{1.284415in}{0.599012in}%
\pgfsys@useobject{currentmarker}{}%
\end{pgfscope}%
\end{pgfscope}%
\begin{pgfscope}%
\definecolor{textcolor}{rgb}{0.000000,0.000000,0.000000}%
\pgfsetstrokecolor{textcolor}%
\pgfsetfillcolor{textcolor}%
\pgftext[x=0.776312in, y=0.550787in, left, base]{\color{textcolor}\rmfamily\fontsize{10.000000}{12.000000}\selectfont Empty}%
\end{pgfscope}%
\begin{pgfscope}%
\pgfsetbuttcap%
\pgfsetroundjoin%
\definecolor{currentfill}{rgb}{0.000000,0.000000,0.000000}%
\pgfsetfillcolor{currentfill}%
\pgfsetlinewidth{0.803000pt}%
\definecolor{currentstroke}{rgb}{0.000000,0.000000,0.000000}%
\pgfsetstrokecolor{currentstroke}%
\pgfsetdash{}{0pt}%
\pgfsys@defobject{currentmarker}{\pgfqpoint{-0.048611in}{0.000000in}}{\pgfqpoint{-0.000000in}{0.000000in}}{%
\pgfpathmoveto{\pgfqpoint{-0.000000in}{0.000000in}}%
\pgfpathlineto{\pgfqpoint{-0.048611in}{0.000000in}}%
\pgfusepath{stroke,fill}%
}%
\begin{pgfscope}%
\pgfsys@transformshift{1.284415in}{0.766790in}%
\pgfsys@useobject{currentmarker}{}%
\end{pgfscope}%
\end{pgfscope}%
\begin{pgfscope}%
\definecolor{textcolor}{rgb}{0.000000,0.000000,0.000000}%
\pgfsetstrokecolor{textcolor}%
\pgfsetfillcolor{textcolor}%
\pgftext[x=0.482717in, y=0.718565in, left, base]{\color{textcolor}\rmfamily\fontsize{10.000000}{12.000000}\selectfont Correctness}%
\end{pgfscope}%
\begin{pgfscope}%
\pgfsetbuttcap%
\pgfsetroundjoin%
\definecolor{currentfill}{rgb}{0.000000,0.000000,0.000000}%
\pgfsetfillcolor{currentfill}%
\pgfsetlinewidth{0.803000pt}%
\definecolor{currentstroke}{rgb}{0.000000,0.000000,0.000000}%
\pgfsetstrokecolor{currentstroke}%
\pgfsetdash{}{0pt}%
\pgfsys@defobject{currentmarker}{\pgfqpoint{-0.048611in}{0.000000in}}{\pgfqpoint{-0.000000in}{0.000000in}}{%
\pgfpathmoveto{\pgfqpoint{-0.000000in}{0.000000in}}%
\pgfpathlineto{\pgfqpoint{-0.048611in}{0.000000in}}%
\pgfusepath{stroke,fill}%
}%
\begin{pgfscope}%
\pgfsys@transformshift{1.284415in}{0.934568in}%
\pgfsys@useobject{currentmarker}{}%
\end{pgfscope}%
\end{pgfscope}%
\begin{pgfscope}%
\definecolor{textcolor}{rgb}{0.000000,0.000000,0.000000}%
\pgfsetstrokecolor{textcolor}%
\pgfsetfillcolor{textcolor}%
\pgftext[x=0.434490in, y=0.886343in, left, base]{\color{textcolor}\rmfamily\fontsize{10.000000}{12.000000}\selectfont Broken Link}%
\end{pgfscope}%
\begin{pgfscope}%
\pgfsetbuttcap%
\pgfsetroundjoin%
\definecolor{currentfill}{rgb}{0.000000,0.000000,0.000000}%
\pgfsetfillcolor{currentfill}%
\pgfsetlinewidth{0.803000pt}%
\definecolor{currentstroke}{rgb}{0.000000,0.000000,0.000000}%
\pgfsetstrokecolor{currentstroke}%
\pgfsetdash{}{0pt}%
\pgfsys@defobject{currentmarker}{\pgfqpoint{-0.048611in}{0.000000in}}{\pgfqpoint{-0.000000in}{0.000000in}}{%
\pgfpathmoveto{\pgfqpoint{-0.000000in}{0.000000in}}%
\pgfpathlineto{\pgfqpoint{-0.048611in}{0.000000in}}%
\pgfusepath{stroke,fill}%
}%
\begin{pgfscope}%
\pgfsys@transformshift{1.284415in}{1.102346in}%
\pgfsys@useobject{currentmarker}{}%
\end{pgfscope}%
\end{pgfscope}%
\begin{pgfscope}%
\definecolor{textcolor}{rgb}{0.000000,0.000000,0.000000}%
\pgfsetstrokecolor{textcolor}%
\pgfsetfillcolor{textcolor}%
\pgftext[x=0.279012in, y=1.054120in, left, base]{\color{textcolor}\rmfamily\fontsize{10.000000}{12.000000}\selectfont Up-to-dateness}%
\end{pgfscope}%
\begin{pgfscope}%
\pgfsetbuttcap%
\pgfsetroundjoin%
\definecolor{currentfill}{rgb}{0.000000,0.000000,0.000000}%
\pgfsetfillcolor{currentfill}%
\pgfsetlinewidth{0.803000pt}%
\definecolor{currentstroke}{rgb}{0.000000,0.000000,0.000000}%
\pgfsetstrokecolor{currentstroke}%
\pgfsetdash{}{0pt}%
\pgfsys@defobject{currentmarker}{\pgfqpoint{-0.048611in}{0.000000in}}{\pgfqpoint{-0.000000in}{0.000000in}}{%
\pgfpathmoveto{\pgfqpoint{-0.000000in}{0.000000in}}%
\pgfpathlineto{\pgfqpoint{-0.048611in}{0.000000in}}%
\pgfusepath{stroke,fill}%
}%
\begin{pgfscope}%
\pgfsys@transformshift{1.284415in}{1.270123in}%
\pgfsys@useobject{currentmarker}{}%
\end{pgfscope}%
\end{pgfscope}%
\begin{pgfscope}%
\definecolor{textcolor}{rgb}{0.000000,0.000000,0.000000}%
\pgfsetstrokecolor{textcolor}%
\pgfsetfillcolor{textcolor}%
\pgftext[x=0.490818in, y=1.221898in, left, base]{\color{textcolor}\rmfamily\fontsize{10.000000}{12.000000}\selectfont Readability}%
\end{pgfscope}%
\begin{pgfscope}%
\pgfsetbuttcap%
\pgfsetroundjoin%
\definecolor{currentfill}{rgb}{0.000000,0.000000,0.000000}%
\pgfsetfillcolor{currentfill}%
\pgfsetlinewidth{0.803000pt}%
\definecolor{currentstroke}{rgb}{0.000000,0.000000,0.000000}%
\pgfsetstrokecolor{currentstroke}%
\pgfsetdash{}{0pt}%
\pgfsys@defobject{currentmarker}{\pgfqpoint{-0.048611in}{0.000000in}}{\pgfqpoint{-0.000000in}{0.000000in}}{%
\pgfpathmoveto{\pgfqpoint{-0.000000in}{0.000000in}}%
\pgfpathlineto{\pgfqpoint{-0.048611in}{0.000000in}}%
\pgfusepath{stroke,fill}%
}%
\begin{pgfscope}%
\pgfsys@transformshift{1.284415in}{1.437901in}%
\pgfsys@useobject{currentmarker}{}%
\end{pgfscope}%
\end{pgfscope}%
\begin{pgfscope}%
\definecolor{textcolor}{rgb}{0.000000,0.000000,0.000000}%
\pgfsetstrokecolor{textcolor}%
\pgfsetfillcolor{textcolor}%
\pgftext[x=0.360031in, y=1.389676in, left, base]{\color{textcolor}\rmfamily\fontsize{10.000000}{12.000000}\selectfont Completeness}%
\end{pgfscope}%
\begin{pgfscope}%
\pgfsetbuttcap%
\pgfsetroundjoin%
\definecolor{currentfill}{rgb}{0.000000,0.000000,0.000000}%
\pgfsetfillcolor{currentfill}%
\pgfsetlinewidth{0.803000pt}%
\definecolor{currentstroke}{rgb}{0.000000,0.000000,0.000000}%
\pgfsetstrokecolor{currentstroke}%
\pgfsetdash{}{0pt}%
\pgfsys@defobject{currentmarker}{\pgfqpoint{-0.048611in}{0.000000in}}{\pgfqpoint{-0.000000in}{0.000000in}}{%
\pgfpathmoveto{\pgfqpoint{-0.000000in}{0.000000in}}%
\pgfpathlineto{\pgfqpoint{-0.048611in}{0.000000in}}%
\pgfusepath{stroke,fill}%
}%
\begin{pgfscope}%
\pgfsys@transformshift{1.284415in}{1.605679in}%
\pgfsys@useobject{currentmarker}{}%
\end{pgfscope}%
\end{pgfscope}%
\begin{pgfscope}%
\definecolor{textcolor}{rgb}{0.000000,0.000000,0.000000}%
\pgfsetstrokecolor{textcolor}%
\pgfsetfillcolor{textcolor}%
\pgftext[x=0.978859in, y=1.557454in, left, base]{\color{textcolor}\rmfamily\fontsize{10.000000}{12.000000}\selectfont UX}%
\end{pgfscope}%
\begin{pgfscope}%
\pgfsetbuttcap%
\pgfsetroundjoin%
\definecolor{currentfill}{rgb}{0.000000,0.000000,0.000000}%
\pgfsetfillcolor{currentfill}%
\pgfsetlinewidth{0.803000pt}%
\definecolor{currentstroke}{rgb}{0.000000,0.000000,0.000000}%
\pgfsetstrokecolor{currentstroke}%
\pgfsetdash{}{0pt}%
\pgfsys@defobject{currentmarker}{\pgfqpoint{-0.048611in}{0.000000in}}{\pgfqpoint{-0.000000in}{0.000000in}}{%
\pgfpathmoveto{\pgfqpoint{-0.000000in}{0.000000in}}%
\pgfpathlineto{\pgfqpoint{-0.048611in}{0.000000in}}%
\pgfusepath{stroke,fill}%
}%
\begin{pgfscope}%
\pgfsys@transformshift{1.284415in}{1.773457in}%
\pgfsys@useobject{currentmarker}{}%
\end{pgfscope}%
\end{pgfscope}%
\begin{pgfscope}%
\definecolor{textcolor}{rgb}{0.000000,0.000000,0.000000}%
\pgfsetstrokecolor{textcolor}%
\pgfsetfillcolor{textcolor}%
\pgftext[x=0.587269in, y=1.725231in, left, base]{\color{textcolor}\rmfamily\fontsize{10.000000}{12.000000}\selectfont Relevance}%
\end{pgfscope}%
\begin{pgfscope}%
\pgfsetbuttcap%
\pgfsetroundjoin%
\definecolor{currentfill}{rgb}{0.000000,0.000000,0.000000}%
\pgfsetfillcolor{currentfill}%
\pgfsetlinewidth{0.803000pt}%
\definecolor{currentstroke}{rgb}{0.000000,0.000000,0.000000}%
\pgfsetstrokecolor{currentstroke}%
\pgfsetdash{}{0pt}%
\pgfsys@defobject{currentmarker}{\pgfqpoint{-0.048611in}{0.000000in}}{\pgfqpoint{-0.000000in}{0.000000in}}{%
\pgfpathmoveto{\pgfqpoint{-0.000000in}{0.000000in}}%
\pgfpathlineto{\pgfqpoint{-0.048611in}{0.000000in}}%
\pgfusepath{stroke,fill}%
}%
\begin{pgfscope}%
\pgfsys@transformshift{1.284415in}{1.941234in}%
\pgfsys@useobject{currentmarker}{}%
\end{pgfscope}%
\end{pgfscope}%
\begin{pgfscope}%
\definecolor{textcolor}{rgb}{0.000000,0.000000,0.000000}%
\pgfsetstrokecolor{textcolor}%
\pgfsetfillcolor{textcolor}%
\pgftext[x=0.831868in, y=1.893009in, left, base]{\color{textcolor}\rmfamily\fontsize{10.000000}{12.000000}\selectfont Other}%
\end{pgfscope}%
\begin{pgfscope}%
\definecolor{textcolor}{rgb}{0.000000,0.000000,0.000000}%
\pgfsetstrokecolor{textcolor}%
\pgfsetfillcolor{textcolor}%
\pgftext[x=0.223457in,y=1.270123in,,bottom,rotate=90.000000]{\color{textcolor}\rmfamily\fontsize{10.000000}{12.000000}\selectfont Feedback Category}%
\end{pgfscope}%
\begin{pgfscope}%
\pgfsetrectcap%
\pgfsetmiterjoin%
\pgfsetlinewidth{0.803000pt}%
\definecolor{currentstroke}{rgb}{0.000000,0.000000,0.000000}%
\pgfsetstrokecolor{currentstroke}%
\pgfsetdash{}{0pt}%
\pgfpathmoveto{\pgfqpoint{1.284415in}{0.515123in}}%
\pgfpathlineto{\pgfqpoint{1.284415in}{2.025123in}}%
\pgfusepath{stroke}%
\end{pgfscope}%
\begin{pgfscope}%
\pgfsetrectcap%
\pgfsetmiterjoin%
\pgfsetlinewidth{0.803000pt}%
\definecolor{currentstroke}{rgb}{0.000000,0.000000,0.000000}%
\pgfsetstrokecolor{currentstroke}%
\pgfsetdash{}{0pt}%
\pgfpathmoveto{\pgfqpoint{1.284415in}{0.515123in}}%
\pgfpathlineto{\pgfqpoint{3.609415in}{0.515123in}}%
\pgfusepath{stroke}%
\end{pgfscope}%
\begin{pgfscope}%
\pgfsetbuttcap%
\pgfsetmiterjoin%
\definecolor{currentfill}{rgb}{1.000000,1.000000,1.000000}%
\pgfsetfillcolor{currentfill}%
\pgfsetfillopacity{0.800000}%
\pgfsetlinewidth{1.003750pt}%
\definecolor{currentstroke}{rgb}{0.800000,0.800000,0.800000}%
\pgfsetstrokecolor{currentstroke}%
\pgfsetstrokeopacity{0.800000}%
\pgfsetdash{}{0pt}%
\pgfpathmoveto{\pgfqpoint{2.177700in}{0.584568in}}%
\pgfpathlineto{\pgfqpoint{3.512193in}{0.584568in}}%
\pgfpathquadraticcurveto{\pgfqpoint{3.539970in}{0.584568in}}{\pgfqpoint{3.539970in}{0.612346in}}%
\pgfpathlineto{\pgfqpoint{3.539970in}{0.985802in}}%
\pgfpathquadraticcurveto{\pgfqpoint{3.539970in}{1.013580in}}{\pgfqpoint{3.512193in}{1.013580in}}%
\pgfpathlineto{\pgfqpoint{2.177700in}{1.013580in}}%
\pgfpathquadraticcurveto{\pgfqpoint{2.149923in}{1.013580in}}{\pgfqpoint{2.149923in}{0.985802in}}%
\pgfpathlineto{\pgfqpoint{2.149923in}{0.612346in}}%
\pgfpathquadraticcurveto{\pgfqpoint{2.149923in}{0.584568in}}{\pgfqpoint{2.177700in}{0.584568in}}%
\pgfpathclose%
\pgfusepath{stroke,fill}%
\end{pgfscope}%
\begin{pgfscope}%
\pgfsetbuttcap%
\pgfsetmiterjoin%
\definecolor{currentfill}{rgb}{1.000000,0.698039,0.698039}%
\pgfsetfillcolor{currentfill}%
\pgfsetlinewidth{1.003750pt}%
\definecolor{currentstroke}{rgb}{0.000000,0.000000,0.000000}%
\pgfsetstrokecolor{currentstroke}%
\pgfsetdash{}{0pt}%
\pgfpathmoveto{\pgfqpoint{2.205478in}{0.860802in}}%
\pgfpathlineto{\pgfqpoint{2.483256in}{0.860802in}}%
\pgfpathlineto{\pgfqpoint{2.483256in}{0.958024in}}%
\pgfpathlineto{\pgfqpoint{2.205478in}{0.958024in}}%
\pgfpathclose%
\pgfusepath{stroke,fill}%
\end{pgfscope}%
\begin{pgfscope}%
\definecolor{textcolor}{rgb}{0.000000,0.000000,0.000000}%
\pgfsetstrokecolor{textcolor}%
\pgfsetfillcolor{textcolor}%
\pgftext[x=2.594367in,y=0.860802in,left,base]{\color{textcolor}\rmfamily\fontsize{10.000000}{12.000000}\selectfont Thumbs Up}%
\end{pgfscope}%
\begin{pgfscope}%
\pgfsetbuttcap%
\pgfsetmiterjoin%
\definecolor{currentfill}{rgb}{0.698039,0.698039,1.000000}%
\pgfsetfillcolor{currentfill}%
\pgfsetlinewidth{1.003750pt}%
\definecolor{currentstroke}{rgb}{0.000000,0.000000,0.000000}%
\pgfsetstrokecolor{currentstroke}%
\pgfsetdash{}{0pt}%
\pgfpathmoveto{\pgfqpoint{2.205478in}{0.667129in}}%
\pgfpathlineto{\pgfqpoint{2.483256in}{0.667129in}}%
\pgfpathlineto{\pgfqpoint{2.483256in}{0.764352in}}%
\pgfpathlineto{\pgfqpoint{2.205478in}{0.764352in}}%
\pgfpathclose%
\pgfusepath{stroke,fill}%
\end{pgfscope}%
\begin{pgfscope}%
\definecolor{textcolor}{rgb}{0.000000,0.000000,0.000000}%
\pgfsetstrokecolor{textcolor}%
\pgfsetfillcolor{textcolor}%
\pgftext[x=2.594367in,y=0.667129in,left,base]{\color{textcolor}\rmfamily\fontsize{10.000000}{12.000000}\selectfont Thumbs Down}%
\end{pgfscope}%
\end{pgfpicture}%
\makeatother%
\endgroup%

%% file: tool.tex
\begin{figure*}[t]
    \centering
    \includegraphics[trim= 43 390 120 30, clip, keepaspectratio=true, scale=0.8]{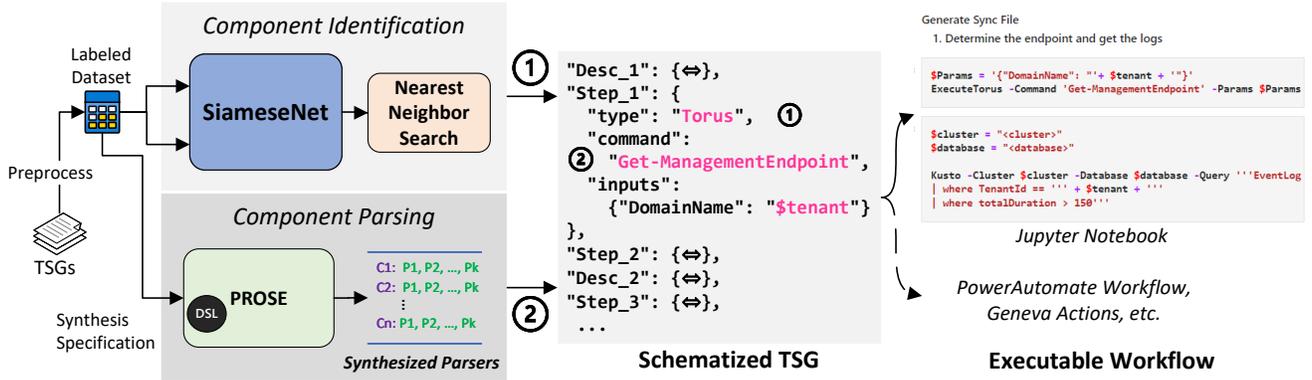}
    \caption{Overview of \tool{} pipeline}
    \label{fig:pipeline}
\end{figure*}

\section{\NoCaseChange{\tool{}}: Towards TSG Automation}

Towards this vision of automated TSGs, in this section, we introduce \textbf{\tool{}}, a tool to aid the automation of manual TSGs to executable workflows. Particularly, \tool{}'s design is guided by three unique observations about TSGs:

\begin{enumerate}
    \item \textbf{TSGs contain components} that are distinct pieces of information. As shown in Table \ref{tab:component-types}, they are commands, database queries for logs, dashboard links, instructions, etc.
    \item \textbf{TSGs have control flow} resembling a decision tree. For e.g., if conditions mentioned in natural language.
    \item \textbf{Components contain constituents} that are parts of a component expected to be parsed for execution. For e.g., command name and parameters for a Powershell command.
\end{enumerate}

These observations guide us to design a two-phase framework for automating troubleshooting guides that first identifies TSG components (Component Identification) and then parses them to extract constituents necessary for execution (Component Parsing). There has been significant research on text/code identification and parsing in prior work \cite{Leclair2018, Ugurel2002what}. However, applying them directly to TSGs requires addressing some unique challenges:

\Paragraph{(1) How do we identify components?} While heuristic-based methods lack coverage, most supervised learning methods need 1000s of labeled examples to train accurate classifiers. In the context of TSGs, this means manually labeling 1000s of examples for 10s of component types. This is laborious and limits the scalability of our tool to new component types. Hence, we need appropriate models that can learn from a limited set of labeled examples.
    
\Paragraph{(2) How do we parse components?} One approach is to hand-craft parsers to extract each constituent of a component. But it assumes significant domain expertise and manual effort. On the other hand, ML based parsers require large training datasets and are stochastic. Hence, we need to also automatically learn verifiable parsing programs from a small set of examples.

In the rest of the section, we describe the design of the \tool{} framework, as shown in Fig. \ref{fig:pipeline}, that address these challenges.



\input{component-identification}

\input{component-parsing}

%% file: component-identification.tex
\subsection{Component Identification}
\label{subsec:component-identification}

\Paragraph{Overview.} Toward TSG automation, we need to first extract and identify individual statements in TSGs that need to be automated -- \textit{Components} -- such as commands, database queries, dashboard links, and also natural language instructions. Here, for automation, it is not only important to extract these statements but also to identify the component type. Table \ref{tab:component-types} shows examples of some popular components used in TSGs at \CompanyX.

Intuitively, we can view component identification as a supervised classification problem -- given a statement, classify it into a category. However, most supervised learning techniques require thousands of labeled examples for training. In the context of TSGs, this implies manually labeling 1000s of examples for 10s of component types.
Further, this poses challenges to the scalability of \tool{} to new components types with very few examples. Hence, we solve this using a few-shot learning setup, where we aim to learn models with a minimal amount of training examples.

But, there are challenges with learning models from very few examples, such as overfitting, robustness, and generalizability. We alleviate this by moving away from the classical learning framework (learning to classify) to a meta-learning framework, where we \textit{learn how to learn} to classify. First, we train a Siamese neural network on a meta-task -- learning how similar or different components are. Then, we use a nearest-neighbor search approach to identify component types. In the following subsections, we describe our approach to component identification in detail.

\subsubsection{\textbf{Preprocessing}} 
Today TSGs are decentralized and can be in various formats such as Word, Markdown, OneNote, etc. Therefore, we first use the \texttt{pandoc} library \cite{pandoc} to convert all TSGs to a single format that is parsable programmatically -- Markdown. Next, we clean the converted TSG using regexes by pruning information that is non-trivial to parse, such as images and tables. Here, we plan to extend \tool{} to capture such multi-modal information with tools such as optical character recognition (OCR). Then, we segment the TSG into statements using newline characters. Here, we use some simple heuristics to handle multi-line commands and queries. For instance, we remove indentation around the `|' character for \texttt{Kusto} queries and around the `\{' and `\}' (braces) for command-line scripts. 
Lastly, we tokenize TSG statements into tokens. Here, we use a custom implemented tokenizer to handle camel-case, URLs, command names, and multi-line database queries.

\subsubsection{\textbf{Meta-Learning Framework}} As previously stated, in this setting, we need to learn a component classifier from a minimal set of examples, i.e., a few-shot learning setup. To enable few-shot learning, we use meta-learning \cite{schmidhuber1987evolutionary} as a framework to simplify the component identification task to a meta-task. Meta-learning \cite{schmidhuber1987evolutionary}, commonly understood as \textit{learning to learn}, refers to an outer (meta) algorithm updating an inner learning algorithm such that the model it learns improves an outer objective (meta-task). 

Here, we choose our meta-task as \textit{"Given a pair of statements, predict the probability that they belong to the same component type"}. 
The intuition here is that one way to learn classification is to differentiate between component types. For instance, to learn to classify Powershell commands and SQL queries, we can learn what makes Powershell and SQL similar or dissimilar. Then, based on learnt properties, we can decide which component type is the closest match.

In meta-learning literature, this translates to metric-based meta-learning \cite{koch2015siamese}. The core idea in metric-based meta-learning is similar to nearest neighbors algorithms (e.g., k-NN classifier \cite{altman1992introduction} and k-means clustering \cite{lloyd1982least}), where the predicted probability $P$ over a set of labels $y$ is a weighted sum of labels of support set samples $S$. The weight is generated by a kernel function $k_{\theta}$, measuring the similarity between two data samples.

\begin{equation}
    P_{\theta}(\boldsymbol{y|x,S}) = \sum_{(x_i, y_i) \in S} k_{\theta}(x, x_i)y_i
\end{equation}

Our aim here is to learn a kernel $k_{\theta}$ that is aligned with learning a similarity metric over component types. Hence, we learn to map TSG components to a latent (embedding) space, where different components are well separated and similar components are close.

\subsubsection{\textbf{Siamese Network}}
\label{subsubsec:siamese-net}

To learn our meta-learning task, we use a Siamese Network (SiameseNet) architecture proposed by Koch et al. \cite{koch2015siamese}. The Siamese Network architecture contains two twin networks and a distance metric, that are jointly trained to learn the relationship between pairs of input data samples. The twin networks are identical and share the same weights and parameters. 
The SiameseNet accepts two inputs $\mathbf{x}_a$ and $\mathbf{x}_b$, which are featurized inputs of the same or different component types. For featurization, we use the simple yet effective bag-of-words \cite{harris1954distributional} approach which creates one-hot features. Next, a convolutional neural network \cite{6795724} learns to encode the 2 resultant vectors via an embedding function $f_{\theta}$. Here, $f_{\theta}$ contains a $100$ dim. embedding layer, 2 convolutional layers \cite{6795724}, 2 max pooling operations \cite{10.5555/3104322.3104338}, and a $128$ dim. dense layer. 

The $L_1$ distance between the two resultant embeddings is $||f_{\theta}(\mathbf{x}_a) - f_{\theta}(\mathbf{x}_b)||_1$. But, to decide whether the two inputs are drawn from the same component type, we need to convert this unbounded distance to a probability $p$. We do that by computing the exponent of the negative L1 norm (equation \ref{eq:manhattan-dist-prob}). Finally, as shown in equation \ref{eq:loss-function}, we train our SiameseNet on a binary cross-entropy loss function as the network label $\mathbf{y}$ is binary. Here $\mathbf{y} = 1$ whenever $\mathbf{x}_a$ and $\mathbf{x}_b$ are of the same component type and $\mathbf{y} = 0$ otherwise.

\begin{equation}
    \label{eq:manhattan-dist-prob}
    p(\mathbf{x}_a, \mathbf{x}_b) = exp(-||f_{\theta}(x_a) - f_{\theta}(x_b)||_1)
\end{equation}

\begin{multline}
    \label{eq:loss-function}
    \mathcal{L}= \sum_{\forall (\mathbf{x}_a, \mathbf{x}_b,\mathbf{y})}
        \mathbf{y}\log p(\mathbf{x}_a, \mathbf{x}_b) +  (1-\mathbf{y})\log (1-p(\mathbf{x}_a, \mathbf{x}_b))
\end{multline}

\subsubsection{\textbf{Nearest neighbour Search}}

With a trained SiameseNet based meta-learner, we can now identify the probability of a pair of statements belonging to the same component. In other words, we can now use the SiameseNet as a comparator that returns the similarity between a pair of TSG statements. With this, a naive approach to performing component identification would be to compare a given TSG statement against every training example. Then, we can label the TSG statement with the component type of the most similar training example -- the nearest neighbor.

But, performing this comparison for each TSG statement with every training example is quite inefficient (O(\#training-examples $\times$ \#statements)), especially for longer TSGs. To mitigate this, for each component type $c \in C$ we pre-compute a \textit{prototype} $\mathbf{prot}_c$ as the mean of the embedded training examples $S_c$ of that component type:

\begin{equation}
    \label{eq:prototype-vector}
    \mathbf{prot}_c = \frac{1}{|S_c|} \sum_{(\mathbf{x}_i, y_i) \in S_c} f_\theta(\mathbf{x}_i)
\end{equation}

Now, for a TSG statement we only compare against each prototype and identify the label as the component type associated with the nearest neighboring prototype. This improves the efficiency of our nearest-neighbor search (O(\#component-types $\times$ \#statements)).

%% file: component-parsing.tex
\subsection{Component Parsing}
\label{subsec:component-parsing}

From the techniques of Section~\ref{subsec:component-identification}, we obtain a list of components in a TSG and their respective component types.
In this section, we learn \emph{component parsers} for each component type to extract the constituents of the component.
For example, given a Powershell command, the component parser will extract the command name and the parameters.

The main challenge here is handling the wide variety of components that commonly occur in TSGs.
Further, large companies and organizations often use components written in custom scripting languages that are unique to only the organization.
Hence, the diversity in component types and the presence of custom component languages make hand-crafting a generic set of parsers infeasible.

Instead of hand-crafting a set of component parsers, \tool{} uses program synthesis techniques to learn parsers from examples.
Specifically, we use programming-by-example techniques to learn parsers from a small number of user-provided examples.
Here, for each component type, the user provides example components along with their expected constituents, and the program synthesis engine produces a parser to extract these constituents in the form of a python program.
Programming-by-example techniques are specifically suitable in this scenario as they are able to learn from few (between $1-5$) examples, and can produce efficient, deterministic, and user-readable parsing programs.
In comparison to machine learning techniques, this both avoids the need for a large amount of training data and further allows the user to edit the produced parser program to fix any minor issues.

\begin{table*}[t]
\def\arraystretch{1.3}
\caption{Description of some component parsers synthesized from example}
\label{table:programs}
\footnotesize
  \begin{tabular}{l|l|l|l|l}
    \toprule
    \textbf{Component} & \textbf{Constituent} & \textbf{E.g. input} & \textbf{E.g. output} & \textbf{Description of synthesized parsing program} \\
    \midrule
      PowerShell
    & $1^{st}$ Param
    & \texttt{Test-PolicyDistributionStatus -Org nybc ...}
    & \texttt{-Org}
    & Extract between first two whitespace spans
    \\
      ADF
    & Subscription
    & \texttt{https://.../subsc/SUB1/resourceGroups/...}
    & \texttt{SUB1}
    & Extract alphanumeric span preceding \texttt{/resourceGroups/}
    \\
      \multirow{3}{*}{Kusto}
    & \multirow{3}{*}{TableName}
    & \texttt{Tba | where ...}
    & \texttt{Tba}
    & \multirow{3}{*}{
      Condition program with 3 branches
    }
    \\
      
    & 
    & \texttt{cluster(...).db(...).AutoTriage | sort ...}
    & \texttt{AutoTriage}
    &
    \\
      
    &
    & \texttt{let result = newUser | where ...}
    & \texttt{newUser}
    &
    \\
      NL Conditional
    & Condition
    & \texttt{If it returns True, you can ...}
    & \texttt{it ... True}
    & Extract between tags \texttt{<CL1>} \& \texttt{</CL1>} after constituency tagging
    \\
  \end{tabular}
\end{table*}

\Paragraph{Programming-by-Example in \tool{}.}
We do not describe the full program synthesis procedure and instead provide a
high-level overview of the PROSE program synthesis library and how \tool{} uses
PROSE.
The reader is referred to~\cite{gulwani2017program} for a literature survey on program synthesis.
PROSE implements an extension of the FlashFill~\cite{gulwani2011automating}
programming-by-example technique.
%
Given a set of examples of the form $\mathsf{in}_i \mapsto \mathsf{out}_i$ where
$\mathsf{in}_i$ and $\mathsf{out}_i$ are strings, PROSE produces a program $P$ (and its
translation to Python) such that $P(\mathsf{in}_i) = \mathsf{out}_i$ for all $i$.
The programs produced by PROSE are of two kinds:
(a) single-branch expressions that are concatenations of constant strings and substrings
of the input string, and
(b) conditional expressions over single-branch expressions.
The conditions in the conditional expressions and the substrings for single-branch expressions
are built-up using standard string and regular expression operators such as \texttt{StartsWith},
\texttt{EndsWith}, \texttt{IndexOf}, \texttt{Regex.Find}, etc.
The reader is referred to the documentation of PROSE for the exact class of programs that can be
synthesized~\cite{prose}.

%

\begin{example}[Single-branch programs]
  Consider the Powershell assignment statements below.
  \begin{lstlisting}[frame=single, columns=flexible]
$mb = Get-Mailbox senderOrRecipientMailbox
$tenant = "<your tenant id/name>"
EOP: $rulePackage = Get-DlpSensitiveInformation -Org ...\end{lstlisting}
  Note that the assignment statement (3) has a label associated with it (\str{EOP}).
  For this component type, one constituent of interest is the left-hand side of
  the assignment, i.e., the variable being assigned to.
  To synthesize the parser for this constituent, the user provides $3$ examples
  of the form
  $(1) \mapsto \str{\$mb}$,
  $(2) \mapsto \str{\$tenant}$, and
  $(3) \mapsto \str{\$rulePackage}$,
  where $(i)$ represents the corresponding assignment statement from above.
  Given these examples, \tool{} uses the program synthesis library
  PROSE~\cite{prose} to synthesize the following python program.
  \begin{lstlisting}
def prog0(self, s):
  idx1 = s.index("$")
  idx2 = re.search(r"[$][\p{L}0-9]+", s).end()
  return s[idx1:idx2] \end{lstlisting}
  This program slices the input assignment statement between the first
  occurrence of a dollar symbol up to the end of the first sequence of
  alphanumeric characters preceded by a dollar symbol.
\end{example}

While the above example can be handled using simple regular expressions, the
synthesizer is able to produce more complex parsers that involve conditionals
and other sophisticated operations.

\begin{example}
  \label{ex:multi-branch}
  Consider the list of Kusto query components below.
\begin{lstlisting}
TbaFilteringException | where time > ago(1d) | ...
cluster('Aznwautotriage').database('autotriage').AutoTriageIcmNer | sort by IncidentId desc
\end{lstlisting}
  From these components, we are interested in extracting the constituent representing
  the table name, i.e., \str{TbaFilteringException} and \str{AutoTriageIcmNer}, respectively.
  Given the examples $(1) \mapsto \str{TbaFilteringException}$ and
  $(2) \mapsto \str{AutoTriageIcmNew}$, PROSE generates a conditional program with
  two branches, each handling one example.
\begin{lstlisting}
def prog0(self, s):
  if re.match("^cluster", s):
    idx1 = s.rindex(".") + 1
    idx2 = re.search(r"\p{Zs}+", _0).start()
  else:
    idx1 = re.search(r"[-.\p{L}0-9]+", _0).start()
    idx2 = re.search(r"\p{Zs}+", _0).start()
  return s[idx1:idx2] \end{lstlisting}
  This program handles both formats of Kusto query components gracefully.
  In practice, there are several more variations of the Kusto query component that occur
  in TSGs and the program generated by PROSE has more branches to deal with them--we present
  this two branch version here for simplicity.
\end{example}

Table~\ref{table:programs} presents some input and output pairs for different
components and corresponding constituents, along with the description of the
parsing program produced by PROSE.

\Paragraph{Special component types.}
While most component types and constituents can be handled using the above techniques, we discuss $2$ special cases which require additional procedures.

\Subparagraph{Handling natural language.}
Traditional program synthesis techniques are not designed to handle the complexities of natural language.
For example, consider the component \emph{``If you need to force the file sync, you can use ForceSync parameter"}.
From this component, we are interested in extracting the \emph{condition clause} constituent (``you need to force the file sync'') and the \emph{action clause} constituent (``can use ForceSync parameter'').
%
%
Using PROSE directly will force us to rely on fragile punctuation based parsers such as ``extract between the word If and the first comma'' which would then fail on differently punctuated statements like \emph{``If it is due to any other error contact the reporting team"}.

To avoid learning such fragile rules, we first annotate the clauses in the input component using a constituency tagger (see, for example, \cite{parsing2009speech, kitaev-etal-2019-multilingual}) to annotate $3$ kinds of constructs: 
%
simple declarative clauses, subordinate clauses, and verb phrases.
The tagged version of the component is \emph{``If <CL1>you need to force the file sync</CL1>, <CL2>you can use ForceSync parameter</CL2>''}.
With this tagged component, the program synthesizer is able to synthesize a simple parser that relies on searching for the anchor points \texttt{<CL1>},
\texttt{</CL1>}, \texttt{<CL2>}, and \texttt{</CL2>}. 
This parser would also handle components with missing punctuation correctly as the constituency tagger natively understands natural language and does not rely on punctuation.

\Subparagraph{Iterative constituent extractions.}
In certain component types, some constituents appear repeatedly.
For example, in a Powershell command, parameter name and value constituents occur
as many times as there are parameters in the command.
In the command \sloppy\lstinline{Test-PolicyDistributionStatus -Org nybc.com -PolicyId 8dbdfce9 -Verbose True}, there are $3$ parameter names (\str{-Org},
\str{-PolicyId}, and \str{Verbose}) and $3$ parameter values
(\str{nybc.com}, \str{8dbdfce9}, and \str{True}).

One potential way of handling this scenario is to learn a different parser for the $i^{th}$ parameter for each $i$.
While sound, this strategy would repeat work learning a similar parser for each $i$.
Instead, we follow an iterative strategy in \tool{}: we only learn parsers for the $1^{st}$ constituents.
During extraction, we
  (a) first extract the constituents for the first parameter from the
    component obtaining \str{-Org} and \str{nybc.com},
  (b) delete the extracted constituents from the component to obtain the new
    component \sloppy\lstinline{Test-PolicyDistributionStatus -PolicyId 8dbdfce9 -Verbose True}, and
  (c) repeat the steps as long as there are more constituents to be extracted
    obtaining in sequence \str{-PolicyId}, \str{8dbdfce9} and \str{-Verbose}, \str{True}.

%% file: evaluation.tex
\input{Tables/classification-eval}

\section{Evaluation}

\subsection{Component Identification Evaluation}
\label{subsec:classification-eval}

\Paragraph{Setup.} To evaluate \tool{}'s component identification model, we use a manually labeled dataset. We begin by first extracting \verify{1902} statements from \verify{50} TSGs from various services at \CompanyX. We then manually classify these sentences into their respective component types, using a combination of domain expertise and existing command databases. For evaluation, we choose \verify{7} component types (shown in Table \ref{tab:component-types}), based on their frequency of occurrence in TSGs as reported by domain experts at \CompanyX{}.

To mitigate the explosion of data and class imbalance, caused by a large number of natural language instructions in TSGs and our pair-wise sampling approach during meta-learning,  we limit \texttt{Natural Language} to \verify{200} random examples. With this, we create a dataset of \verify{661} labeled examples.
We then compare \tool{}'s few-shot SiameseNet model (Section \ref{subsubsec:siamese-net}) against multiple baselines, in a  5-fold cross-validation setting that ensures models do not overfit training data. In Table \ref{tab:component-identification}, we report the 5-fold cross-validation precision, recall, and F1 scores, for each component type. We also report overall aggregated metrics, including the accuracy of our models.

\Paragraph{Baselines.} First we have \texttt{KNN\_BoW} -- a K-nearest-neighbor \cite{altman1992introduction} model using a Bag-of-words \cite{harris1954distributional} as features. 
Next, with \texttt{RF\_BoW}, we introduce a Random Forest \cite{breiman2001random} model, while keeping Bag-of-words as features. Here, we specifically choose these two models as they are simpler, yet learn classification similar to our \texttt{SiameseNet} -- by separating classes from each other.
Next, we have \texttt{KNN\_W2V} and \texttt{RF\_W2V}, where we retain the models, but update the Bag-of-words feature space to Word2Vec \cite{mikolov2013efficient}. For Word2Vec models, we finetuned pre-trained models to a corpus of 3000+ TSG sentences, using \texttt{sentencepiece} \cite{kudo2018sentencepiece} and \texttt{gensim} \cite{rehurek_lrec}.


\Paragraph{Baseline Results.} From Table \ref{tab:component-identification}, we see that all baselines perform quite well overall. \texttt{RF\_BoW} is the best, with an overall F1 of 0.81 and an accuracy of 0.78. However, we see that these models perform well for certain components, but poorly for others.
For instance, in Table \ref{tab:component-identification}, we observe high F1 scores (0.71--0.93) for \texttt{ADF}, \texttt{Jarvis}, and \texttt{Kusto}. Table \ref{tab:component-types} shows that these components have distinct structure and syntax (e.g., `https://adf.', `https://jarvis-', `| where'), making them easier to identify.
However, we find poor F1 scores (0.40--0.63) for components that are harder to distinguish, such as \texttt{Torus}, \texttt{Merlin}, and \texttt{Powershell}. 
This can be attributed to these components sharing syntax/structure, but having variations in vocabulary/semantics (refer Table \ref{tab:component-types}). This makes distinguishing these components non-trivial and our baselines fail to capture these semantic variations.

\Paragraph{SiameseNet.} Lastly, we evaluate our proposed \texttt{SiameseNet} approach, which incorporates meta-learning to first learn a meta-task -- distinguishing between components. We then utilize a nearest neighbor search approach to classify sentences into components in the embedding space. As shown in Table \ref{tab:component-identification}, our approach achieves an average accuracy of 0.89, which is significantly better than our baselines. We also observe that it reaches high F1 scores (0.72--0.98) across all component types. Particularly, unlike the baselines, we see strong results for components that are harder to distinguish like \texttt{Torus}, \texttt{Merlin}, and \texttt{Powershell}. Hence, with \tool{}'s meta-learning approach capturing syntax and semantics of components, we are able to outperform multiple strong baselines.

\input{Tables/parsing-evaluation}


\subsection{Component Parsing Evaluation}
\label{subsec:parsing-eval}

\Paragraph{Setup.} Next, we evaluate \tool{}'s effectiveness to parse and extract constituents of components using synthesized programs from PROSE. As described in Section \ref{subsec:component-parsing}, \tool{} uses programming-by-examples to synthesizes parsers, for each constituent of a component (e.g., \texttt{parameters} of a \texttt{Powershell} command). 

We then test these synthesized programs on the \verify{1902} sentence labeled dataset described in Section \ref{subsec:classification-eval} and collect the parsed outputs. Here, we ensure that there is no overlap between the specification examples and the test dataset. Then, we manually validate the precision (i.e., constituents parsed are correct) and recall (i.e., all constituents of the input are parsed) of each parsed output.

\Paragraph{Results.} Table \ref{tab:component-parsing} shows the average precision and recall of parsing, aggregated for each component. 
As shown, \tool{}'s parser synthesizer can generate accurate parsers that can be learned from 5-10 specification examples. For instance, with just 8 examples, we can learn parsers for \texttt{Powershell} statements with a precision of 0.9 and recall of 0.8. Also, with just 10 examples of conditional statements, we can learn parsers to extract conditions (\hlc[lightcornflowerblue]{\thickspace\thickspace}) and actions (\hlc[lightgreen]{\thickspace\thickspace}) accurately.
More importantly, these parsers have high recall; i.e., are robust to variations in conditional statements such as 
\textit{``If \hlc[lightcornflowerblue]{command returns True}, then \hlc[lightgreen]{create an incident}''}, 
\textit{``If the \hlc[lightcornflowerblue]{status is False} \hlc[lightgreen]{delete the resource}''}, 
\textit{``If \hlc[lightcornflowerblue]{average latency is > 300 ms}''}, etc. Overall, we find that our parsers have a high average precision of 0.94 and recall of 0.91.

Further, we looked at some common test examples that were incorrectly extracted. For instance \texttt{Torus} parsers incorrectly parsed the statement: \lstinline{\$m = Get-Mailbox -Arbitrate -Identity \$identity} and returned \lstinline{\{variable: '\$m', command: 'Get-Mailbox', parameters: [('Arbitrate', '-Identity')]\}}. 
As shown, the parsers incorrectly identified \sloppy\lstinline{Arbitrate} as a parameter whose value is \lstinline{-Identity}. This is due to the usage of flag parameters like \lstinline{-Arbitrate} in between commands, that were unseen in the synthesis specification. We find that these kinds of errors can be fixed by learning programs from a larger set of specification examples that cover these variations. 
In another example of a \texttt{Powershell} statement:  \lstinline{\$m | Format-List \$db}, we observed that the parsers returned empty results. This is because of the usage of pipe: | in a conjunction of statements, which was unseen in the specification. We find that these kinds of errors can be fixed with some preprocessing, such as splitting the command on pipes and iteratively calling the parsers. 

Thus, our analysis shows that precise component parsers for TSGs can be effectively learned from a small set of examples through program synthesis -- creating a scalable solution to expand \tool{} to other kinds of components with minimal manual effort.

\subsection{TSG Coverage Evaluation}
\label{subsec:tsg-coverage-eval}

\Paragraph{Setup.} In this section, we look at the overall coverage of \tool{} for TSG automation. For this, we select the top \verify{15} most frequently used TSGs for incident mitigation. We then run \tool{} on these TSGs and validate the returned results for each line in a TSG -- both the component type and parsed output. First, we mark all lines in the TSG that are \texttt{notAutomatable}; i.e., lines that cannot be converted to an executable. Next, for every automatable line, we verify if the result is \texttt{Valid} or \texttt{Invalid} for automation; i.e., whether the line was correctly identified and parsed by \tool{}. Finally, we report the coverage of \tool{} as the percentage of valid automatable lines.

\begin{figure}[t]
    \centering
    \includegraphics[]{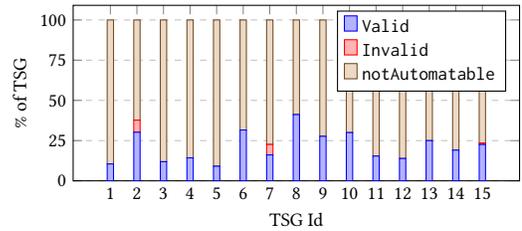}
	\caption{Coverage for TSG Automation}
	\label{fig:coverage-analysis}
\end{figure}

\Paragraph{Results.} Figure \ref{fig:coverage-analysis} shows the results of this evaluation. First, we observe that TSGs have majority of \texttt{notAutomatable} lines. This is expected as much of a manual TSG is made up of statements that are either not executable or not required for automation. For example, in TSGs \#1, \#3, \#4, and \#5, we observe statements such as section headers, dates, links to other TSGs, author name, step descriptions, comments, and points of contact. While we accurately identify these statements as \texttt{Natural Language}, we cannot parse them into executables, hence, making them \texttt{notAutomatable}.

Next, we find that on average 21.24\% of all lines in TSG is correctly automatable (\texttt{Valid}) using \tool{}. Also, for 6 TSGs, more than 25\% of all lines are correctly automatable, with a maximum coverage of 41.2\% for TSG\#8. Overall, we observe that for 12/15 TSGs, 100\% of automatable lines in a TSG were correctly identified and parsed by \tool{} showing that \tool{} is highly effective at accurately identifying and parsing necessary information to automate a TSG into executable workflows.

\subsection{\tool{} User Study}
\label{subsec:user-study-eval}

So far, we have shown that \tool{} has very high accuracy and coverage. Next, we perform a user study to understand the importance of TSG automation and the usefulness of \tool{}.

\Paragraph{Setup.} In order to select the study participants, randomly sampled 30 on-call engineers at \CompanyX{} who have mitigated or resolved incidents in the last six months and invited them for interviews. With a response rate of 33\%, 10 out of the 30 invitees agreed for an interview. These 10 participants ($P_1$-$P_{10}$) work across 8 teams at \CompanyX. They had mitigated or resolved on average 1852 incidents, with the minimum being 41 and maximum 5432. We conducted $\approx$15-minute semi-structured interviews with these 10 participants. 


We began by asking how often the participant uses TSGs for incident mitigation. Next, we recorded verbatim responses on the issues they face with today's TSGs. Then, we introduced a manual TSG and asked if automating the TSG would help.
If the response was yes, we introduced the participant to the results of TSG automation using \tool{} and a human-in-the-loop approach.
First, we showed a schematized TSG (JSON), returned by \tool{}, with components identified and parsed -- as in Fig. \ref{fig:pipeline}.
Then, we showed the final automated TSG (Jupyter notebook) that would be programmatically generated from the schematized TSG.
Given this, we asked the participants to rate the usefulness of \tool{} for TSG automation on a 5-point Likert scale \cite{Robinson2014}. However, if the participant responded that automating the TSG would not help, we collected responses on why they thought so. Next, we summarize the results of the study.

\Paragraph{Q1. How often do you use TSGs for mitigation?} Here, a majority of six participants said they use TSGs every other week,  while three others said they use them every month. Interestingly, one participant said they use TSGs every day. When asked why they use TSGs every day, they said:

\begin{quote}
    $P_2$: \textit{``My usual on-call rotation is bi-weekly. But I end up helping my peers use TSGs every day, because they are bad and confusing.''}
\end{quote}

\Paragraph{Q2. What challenges do you face with TSGs?} Here, we observed that all participants responded with some type of challenge they faced due to quality issues in TSGs. We find the majority of them talked about completeness, readability, and usability issues. Here are some representative verbatim responses:

\begin{quote}
    $P_4$: \textit{``TSGs maybe maintained, but we don't find them elaborative. SME (subject matter expert) thinks this is well defined, but for us or new members additional info should be added.''}
\end{quote}

\begin{quote}
    $P_6$: \textit{``I work on a service with a lot of customization. So when incidents arise, I have to manually verify which steps in a TSG will work for that incident. There is no guaranteed solution.''} 
\end{quote}

\begin{quote}
    $P_7$: \textit{``Mostly outdated TSGs or missing the information I need. So I generally discuss with my team for many of the issues.''} 
\end{quote}

\Paragraph{Q3. Do you think automating TSGs would help?} Here, notably, all participants of the user study said that automating TSGs would help. Apart from reducing on-call load, effort, and human error, the participants identified other positive outcomes of automation:

\begin{quote}
    $P_2$: \textit{``The team being able to review automated TSGs is very valuable. We can check for the safety of steps -- which today is missing.''}
\end{quote}

\Paragraph{Q3.a. If Yes, how useful is \tool{} for TSG Automation?} Here, we look at the distribution of the 5-point Likert scale ratings provided by the participants. We find that the participants gave the usefulness of \tool{} an average rating of 4.2. A majority six participants gave a rating of 4, three participants gave a rating of 5, and one participant gave a neutral rating of 3. Overall, we find that our participants strongly perceived \tool{} to positively aid the automation of TSGs. Some even remarked at how \tool{} motivates an automated on-call experience of the future:

\begin{quote}
    $P_1$: \textit{``This tool also motivates teams to better document their queries, commands, etc., from the get-go in executable \iffalse formats such as jupyter\fi notebooks.''}
\end{quote}

\begin{quote}
    $P_3$: \textit{``When will this be available? This is great! This can drive future data analysis like which TSGs, commands, etc. were run frequently, and help find major issues.''}
\end{quote}

\Paragraph{Q3.b. If No, why is automation not useful?} As stated, no participant responded that automation of TSGs would not be useful.

%% file: Tables/classification-eval.tex
\begin{table*}[t]
\small
    \caption{Evaluation of Component Identification}
    \label{tab:component-identification}
    \begin{tabular}[t]{lccc|ccc|ccc|ccc|cccc}
    \toprule
    
         \textbf{Component}
         
         & \multicolumn{3}{c}{\texttt{KNN\_BoW}} 
         & \multicolumn{3}{c}{\texttt{RF\_BoW}} 
         & \multicolumn{3}{c}{\texttt{KNN\_W2V}} 
         & \multicolumn{3}{c}{\texttt{RF\_W2V}} 
         & \multicolumn{3}{c}{\texttt{SiameseNet}} \\\cmidrule(lr){2-16}
         
         & Pre. & Rec. & F1 & Pre. & Rec. & F1 & Pre. & Rec. & F1 & Pre. & Rec. & F1 & Pre. & Rec. & F1 & Support \\

    \midrule
    
        ADF & 0.65 & 1.00 & 0.75 & 0.90 & 1.00 & \textbf{0.93} & 1.00 & 1.00 & 1.00 & 0.87 & 1.00 & 0.90 & 0.67 & 1.00 & 0.72 & 6\\
        
        Jarvis & 1.00 & 1.00 & 1.00 & 1.00 & 1.00 & 1.00 & 0.69 & 1.00 & 0.78 & 0.82 & 0.77 & 0.75 & 1.00 & 1.00 & \textbf{1.00} & 14\\
        
        Kusto & 1.00 & 0.60 & 0.72 & 0.87 & 0.87 & 0.85 & 0.62 & 0.87 & 0.72 & 0.92 & 0.53 & 0.62 & 0.97 & 0.87 & \textbf{0.91} & 29\\
        
        Merlin & 0.60 & 0.51 & 0.54 & 0.69 & 0.49 & 0.55 & 0.45 & 0.42 & 0.43 & 0.49 & 0.38 & 0.43 & 0.95 & 0.65 & \textbf{0.76} & 106\\
        
        Torus & 0.70 & 0.84 & 0.76 & 0.75 & 0.84 & 0.79 & 0.54 & 0.78 & 0.64 & 0.52 & 0.72 & 0.60 & 0.82 & 0.95 & \textbf{0.87} & 202\\
        
        Powershell & 0.48 & 0.57 & 0.51 & 0.64 & 0.72 & 0.67 & 0.88 & 0.29 & 0.41 & 0.54 & 0.38 & 0.42 & 0.97 & 0.82 & \textbf{0.87} & 104\\
        
        Natural Language & 0.94 & 0.69 & 0.79 & 0.95 & 0.90 & 0.92 & 0.95 & 0.82 & 0.88 & 0.83 & 0.80 & 0.81 & 1.00 & 0.97 & \textbf{0.99} & 200\\
        
    \midrule
        Overall & 0.77 & 0.74 & 0.72 & 0.83 & 0.83 & 0.82 & 0.73 & 0.74 & 0.69 & 0.71 & 0.65 & 0.65 & \textbf{0.91} & \textbf{0.90} & \textbf{0.87} \\
        
        Accuracy & \multicolumn{3}{c}{0.69} & \multicolumn{3}{c}{0.78} & \multicolumn{3}{c}{0.62} & \multicolumn{3}{c}{0.65} & \multicolumn{3}{c}{\textbf{0.89}} \\
            
    \bottomrule
    \end{tabular}
\end{table*}

%% file: Tables/parsing-evaluation.tex
\begin{table}[ht]
\small
    \caption{Evaluation of Component Parsing}
    \label{tab:component-parsing}
    \def\arraystretch{1.1}
    \begin{tabular}[t]{p{5.2cm}p{0.5cm}p{0.5cm}p{0.5cm}}
    \toprule
         \textbf{Component \{Constituents\}} & \textbf{Sup.} & \textbf{Pre.} & \textbf{Rec.}\\
    \midrule
    
    Torus~~\small{\{variable, command, parameters\}} & 202 & 0.93 & 0.81 \\
    
    Merlin~~\small{\{variable, command, parameters\}} & 106 & 0.92 & 0.86 \\
    
    Powershell~~\small{\{variable, command, parameters\}} & 104 & 0.90 & 0.80\\
    
    Kusto~~\small{\{cluster, database, table, query\}} & 29 & 1.00 & 1.00\\
    
    ADF~~\small{\{subscription, resourcegroup, factory\}} & 6 & 1.00 & 1.00\\
    
    Conditionals~~\small{\{condition, action\}} & 127 & 0.87 & 0.97\\
    
    \midrule
    \textbf{Overall} & & \textbf{0.94} & \textbf{0.91} \\
    \bottomrule
    
    \end{tabular}
\end{table}

%% file: related-work.tex
\section{Related Work}

\Paragraph{Incident Management.} Troubleshooting guides are critical for incident management, which has been a popular research direction in software engineering. Recent work has focused on multiple aspects of incident management like triaging \cite{EmpiricalIcMICSE2019, ContinuousTriageASE2019}, mitigation \cite{jiang2020mitigate}, diagnosis \cite{nair2015learning, bansal2019decaf, luo2014correlating}, and more. Particularly close to our work, are efforts that attempt to mine structured knowledge from various artifacts, such as incident reports \cite{kikuchi2015prediction, shetty2021neural, softnerEMSE} and root cause documentation \cite{saha2022mining}. 
However, we tackle an aspect of incident management, that has received relatively lesser attention -- TSGs. Jiang et al. \cite{jiang2020mitigate}, analyzed the usage of troubleshooting guides and proposed a TSG recommendation system to help developers find relevant TSGs. Our empirical study also supports the findings of such prior work. However, different from them, we also study the quality aspects of TSGs, like completeness, correctness, etc., that make them difficult to use. Lastly, our findings motivate the automation of TSGs, that in-turn introduces properties of source code to TSGs. We introduce \tool{} -- a novel framework to aid with the automation of TSGs.

\Paragraph{Software Documentation.} While studies on troubleshooting guides are limited, there have been several efforts to study software documentation in general. These can be classified into 2 categories: (1) tools to generate/recommend documentation and (2) empirical investigation of documentation usage and quality. Regarding automation for documentation, research has focused on either summarization or recommendation for bug reports \cite{lotufo2015modelling, mani2012ausum}, code \cite{cortes2014automatically, jiang2017towards, mcburney2015automatic}, user stories \cite{krasniqi2017tracelab}, API usage examples \cite{holmes2005using, li2018learning, moreno2015can, stylos2006mica}, etc.
Different from these, our tool \tool{} focuses on automation that helps translate manual text documentation to executable workflows. Closer to our work in this space are the empirical studies on documentation. These studies use user surveys to analyze the importance and quality of software documentation \cite{de2005study, chen2009empirical, aghajani2019software, aghajani2020software, plosch2014value, garousi2013evaluating}, but focus on software maintenance in general, unlike our work on the specific task of troubleshooting. Most analogous to our work on TSG quality is the taxonomy of documentation quality proposed by Aghajani et al. \cite{aghajani2019software}, to which we compare our work in detail in Section \ref{subsec:tsg-quality-empirical}.

\Paragraph{Few-shot Learning \& Meta-Learning.} Few-Shot Learning (FSL) \cite{fei2006one, fink2004object} is a type of machine learning problem, where we learn a task from only a limited number of examples for the target. FSL is particularly useful to help reduce the burden of collecting large datasets of supervised information, such as in large-scale image classification \cite{koch2015siamese, tian2020rethinking, zhang2020deepemd}, language modeling \cite{vinyals2016matching}, drug discovery \cite{altae2017low}, robotics \cite{finn2017model, duan2017one}, and more. Unlike these well studied scenarios, in this work, we use FSL in the domain of TSGs, with varying kinds of information -- commands, queries, links, instructions, etc.

To enable FSL, we use meta-learning \cite{schmidhuber1987evolutionary}, commonly known as \textit{learning to learn}. It refers to learning related tasks, and using this to learn new tasks much faster than otherwise possible \cite{vanschoren2018meta}. 
Particularly, in our scenario, we use a flavour of meta-learning called metric-learning. Here, the idea is to learn input representations and a similarity metric during the meta-learning phase \cite{koch2015siamese, vinyals2016matching, snell2017prototypical, sung2018learning}. In this work, we use a Siamese convolutional network \cite{koch2015siamese} to embed TSG statements and separate the final task, component identification, from the neural net, and instead use a fast nearest neighbor search approach for classification, like Snell et al. \cite{snell2017prototypical}.

\Paragraph{Program Synthesis.} Program synthesis techniques, especially programming-by-examples (PBE), have been applied to various domains \cite{lessenich2015balancing, le2014flashextract, meng2011systematic}. Gulwani \cite{gulwani2011automating} introduced FlashFill to synthesize string transformation scripts from examples. 
In the software engineering domain, there have been efforts to apply program synthesis techniques on code related tasks such as learning version update patches \cite{andersen2010generic}, code edit scripts \cite{meng2011systematic}, and merge conflict resolutions \cite{pan2021can}. However, our work targets a new domain that has not been explored using PBE: software documentation. Different from prior work, TSGs contain a multitude of components (commands, queries, natural language), each with a large number of input variations. Using PROSE \cite{prose} and its text transformation DSL (core to the FlashFill system), we show that even under such conditions, robust parsers can be learnt using a minimal set of input-output specifications.

%% file: conclusion.tex
\vspace{-3mm}
\section{Discussion and Conclusion}
In this work, we presented a large-scale empirical study of over 4K+ TSGs mapped to 
\iffalse17K\fi\iftrue 1000s of \fi
incidents. Our analysis indicates that TSGs are very frequently used for incident mitigation and notably help reduce mitigation time and effort. However, on studying feedback provided by 400+ on-call engineers at \CompanyX{}, we uncover significant gaps in TSG quality, such as completeness, maintainability, readability, etc., characterized by our proposed taxonomy of issues. These insights motivate us to investigate the automation of TSGs and propose \tool{} -  a novel framework to aid with the automation of manual TSGs to executable workflows combining machine learning and program synthesis. Our evaluation of \tool{} on 50 TSGs shows the effectiveness of the tool to both identify TSG statements (accuracy 0.89) and parse them for execution (precision 0.94 and recall 0.91). Lastly, with a survey of engineers at \CompanyX{}, we show the usefulness of TSG automation and \tool{} for TSG users.
As next step, we are planning to deploy \tool{} as a self-serve tool at \CompanyX{} with an accompanying user interface and a feedback loop. We envision TSG authors uploading their current TSGs and viewing the automated TSG returned by our tool. Here, the author would edit/fix errors in the results, which we collect as feedback for re-training and improvement. Further, prior research \cite{li2021fighting} has shown that due to highly complex dependencies between services, on-call engineers find it challenging to mitigate incidents. With \tool{} and automated TSGs, we plan to automatically infer of a queue of TSGs to run, from various services, and help mitigate such complex incidents.

\section{Acknowledgements}
\label{sec:acknowledgements}
We would like to acknowledge the invaluable contributions and support of Tarun Sharma, Abhilekh Malhotra, Sunil Singhal, Harinder Pal, Gurpreet Singh, Shalki Aggarwal, Sakshum Sharma, Rahul Mittal, Puneet Kapoor, Saravan Rajmohan, and B. Ashok.